%% file: main.tex
\RequirePackage{xcolor}
\documentclass[journal,twoside,web]{ieeecolor}
\makeatletter
\let\NAT@parse\undefined
\makeatother
\usepackage[colorlinks=true]{hyperref} 
\usepackage{cite}     

\usepackage[pdftex]{graphicx}
\usepackage{generic}
\usepackage{amssymb,amsfonts}
\usepackage{amsthm}
\usepackage{todonotes} 
\input{includes/base}
\input{includes/commands}

\input{includes/symbols}
\usepackage{nameref}

\begin{document}
\bstctlcite{bstctl:etal, bstctl:nodash, bstctl:simpurl}

\pgfdeclarelayer{background}
\pgfdeclarelayer{background1}
\pgfsetlayers{background1,background,main}

\title{Learned Bayesian Cram\'er-Rao Bound \\
for Unknown Measurement Models \\ {Using} Score Neural Networks}
\author{Hai Victor Habi, Hagit~Messer,~Life~Fellow,~IEEE and Yoram~Bresler,~Life~Fellow,~IEEE
\thanks{H.V. Habi and H. Meseer are with the School of Electrical Engineering, Tel Aviv University, Tel Aviv 6139001, Israel (e-mail: haivictorh@mail.tau.ac.il; messer@eng.tau.ac.il).}
\thanks{Y.Bresler is with the Department of Electrical and Computer Engineering and the Coordinated Science Lab, University of Illinois Urbana-Champaign, USA (e-mail: ybresler@illinois.edu).
}}
\date{}

\markboth{February 2025}%
{Habi, Messer, and Bresler: Learned Bayesian Cram\'er-Rao Bound for Unknown Measurement Models  using Score Neural Networks}

\maketitle

\begin{abstract}
    The Bayesian Cramér-Rao bound (BCRB) is a crucial tool in signal processing for assessing the fundamental limitations of any estimation problem as well as benchmarking within a Bayesian frameworks. However, the BCRB cannot be computed without full knowledge of the prior and the measurement distributions. In this work, we propose a fully learned Bayesian Cramér-Rao bound (LBCRB) that learns both the prior and the measurement distributions. Specifically, we suggest two approaches to obtain the LBCRB: the Posterior Approach and the Measurement-Prior Approach. The Posterior Approach provides a simple method to obtain the LBCRB, whereas the Measurement-Prior Approach enables us to incorporate domain knowledge to improve the sample complexity and {interpretability}. To achieve this, we introduce a Physics-encoded score neural network which enables us to easily incorporate such domain knowledge into a neural network. We {study the learning} errors of the two suggested approaches theoretically, and  validate them numerically. We demonstrate the two approaches on several signal processing examples, including a linear measurement problem with unknown mixing and Gaussian noise covariance matrices, frequency estimation, and quantized measurement. In addition, we test our approach on a nonlinear signal processing problem of frequency estimation with real-world underwater ambient noise.
\end{abstract}
\raggedbottom


\section{Introduction}
The Bayesian Cramér-Rao Bound (BCRB) \cite{van2004detection} is a crucial tool in signal processing for {assessing} the fundamental limitations of any estimation problem within a Bayesian framework. For instance, the BCRB has been employed to elucidate various signal processing applications such as localization \cite{10184105,10140073} and MIMO systems \cite{nasir2013phase}, among others \cite{xu2004bayesian,rosentha2024asymptotically,mazor2024limitations}. Besides understanding the intrinsic limits of problems, the BCRB has also been used for system design, including waveform design \cite{huleihel2013optimal,turlapaty2014bayesian,zuo2010conditional,sun2024optimal}. 

However, to obtain the BCRB requires \emph{complete knowledge} of both the prior and the measurement distributions. In addition, in  some cases, even when both prior and measurement distributions are known, the BCRB cannot be computed, because the required integration over the parameter distribution does not have a close-form solution.  

Various methods have been proposed 
to derive a bound from data, {rather than analytically.} Some leverage prior knowledge of the problem combined with a learnable component. For instance, \cite{lutati22_interspeech} presents a non-Bayesian bound for single-channel speech separation. Another category of methods, as those described in \cite{duy2022fisher,6975144}, proposes bounds based on f-divergences. However, these methods require access to a {special} dataset {containing observations where for each parameter value there are also observations with a slightly perturbed parameter value.} 
This condition is generally viable only when one can generate an observation vector {for any desired value of the parameter vector that one aims to estimate}.

Recently, thanks to the success of generative models in modeling complex, high-dimensional data distributions\cite{song2019generative,kobyzev2020normalizing}, a new approach has been introduced that suggested using a learned Generative Cram\'er Rao bound \cite{habi2023learning} when the measurement distribution is \emph{completely unknown,} but a dataset of independent and identically distributed  (i.i.d)  measurement-parameter pairs is available. The  Generative Cram\'er Rao bound \cite{habi2023learning} achieves this by first learning the measurement distribution using a generative model, and then utilizing it to obtain the learned Generative Cram\'er Rao bound. The approach of using a generative model to learn a performance estimation bound from data has been extended to several other non-Bayesian bounds, such as the misspecified CRB\cite{habi2023learned} and the Barankin bound \cite{habi2024learning}.  

Those approaches utilized normalizing flows\cite{kobyzev2020normalizing,papamakarios2021normalizing} which enable the computation of the probability density function of the measurements. This comes with a major limitation: that there exist an invertible mapping between the measurements and some base distribution that is analytically tractable (usually standard Gaussian). An example in which such mappings do not exist is when the measurements are quantized\cite{habi2022generative}. 

In 
the Bayesian setting, recent work \cite{crafts2023bayesian} suggested to learn the prior distribution using score matching \cite{hyvarinen2005estimation}. 
{After learning the prior score ($\nabla_{\p}\log\probt{\p}{\pr}$), obtaining the BCRB using this method requires complete {knowledge} of the Fisher score function ($\nabla_{\p}\log\probt{\x|\p}{\X|\pr}$) and its computation. {Given the known Fisher score function and the learned prior score function,} the BCRB is then approximately calculated using empirical means. A {limitation} of this approach is the need for  full knowledge of the Fisher score and ability to compute it, which can be challenging. For instance, calculating the Fisher score for quantized measurement with correlated noise involves integrating a multidimensional Gaussian with nondiagonal covariance, which must be done numerically. Another example is an application with incomplete knowledge of the Fisher score, such as noise from a physical image sensor \cite{abdelhamed2019noise}, or underwater noise \cite{weiss2023towards,msg0-ag12-22}.}

In this work, we 
introduce the \emph{learned Bayesian Cram'er Rao bound (\name{})}, which learns both the prior and the measurement distributions. Specifically, we suggest two approaches to learn the BCRB. The first, the \emph{Posterior Approach,} is based on the score of the posterior distribution ($\nabla_{\p}\log\probt{\p|\x}{\pr|\X}$). The second, the \emph{Measurement-Prior Approach,} is based on the Fisher score function ($\nabla_{\p}\log\probt{\x|\p}{\X|\pr}$) and the prior score function ($\nabla_{\p}\log\probt{\p}{\pr}$). 

The \emph{Posterior Approach} emphasizes learning a conditional score function of the parameter based on the given observation. This process requires 
{to learn} the posterior score function from a dataset consisting of observation-parameter pairs. Following this, the trained posterior score and the dataset are employed to evaluate the BCRB. The \emph{Posterior} Approach offers the advantage of simplicity in both the learning process and the design of the neural network. However, it is restricted in its capacity to integrate domain knowledge regarding the relationship between the parameter and the observations.  

To utilize such domain knowledge and obtain the benefits of combining with a learnable score (similar to model-based deep learning \cite{shlezinger2022model,shlezinger2023model}, physics-informed neural network \cite{banerjee2024physics}, {physics-encoded neural networks \cite{faroughi2024physics,meinders2024application}}), we propose the \emph{Measurement-Prior} Approach. This technique involves  learning the Fisher and prior score functions independently, and subsequently using them to calculate the learned BCRB. 

We {learn} the Fisher and prior score functions from a dataset consisting of observation-parameter pairs. Although it is possible to learn the prior score function using traditional score matching \cite{hyvarinen2005estimation}, the same approach, {or existing conditional score matching \cite{hyvarinen2005estimation,liu2022estimating,yu2019generalized,yu2022generalized}},
do not work for learning the Fisher score $\nabla_{\p}\log\probt{\x|\p}{\X|\pr}$ since the derivative is w.r.t. {the conditioning variable,} the parameter $\p$, {rather than w.r.t the conditioned variable $\x$, as in conditional  score matching.} To address this issue, we introduce a variation of score matching that enables to learn the Fisher score, which we call \emph{Fisher Score Matching (FSM)}. This {new} score matching process involves first learning the prior score function via standard score matching \cite{hyvarinen2005estimation} and then using the obtained prior score function to learn the Fisher score function. 

The primary advantage of employing the {Measurement-Prior} Approach is the integration of domain {knowledge}. {Inspired by \cite{faroughi2024physics,meinders2024application}, we suggest a \emph{\pe{} Score Neural Network,} a new type of score neural network that encodes the physicals of the problem in the structure of the network. This improves the accuracy of the approximation and {reduces} the sample complexity of learning it.} Moreover, this method enables to evaluate the learned BCRB in scenarios with {any desired number of} i.i.d. measurements, without {having to learn} a new score function.



We apply the \name{} 
to several examples to show its benefits and {study} its behavior. {As a technical aspect in the experiments, neural networks were trained with conditioning variables\cite{mirza2014conditional,abdelhamed2019noise,liu2019conditional,ho2021classifier}, 
such as the signal-to-noise ratio, allowing us to obtain a single neural network for all SNR values.}

Our findings on a linear mixing measurement model with Gaussian noise,  and on its {1-bit} quantized variant illustrate the advantages of directly learning the Fisher score function from data. In these scenarios, the BCRB is known analytically or can be calculated numerically,
{enabling us to assess} the  {deviation of the \name{} from the exact BCRB due to learning error.} {Next, testing} 
 our approach on a non-linear signal processing problem of frequency estimation, highlights the advantages of a \pe{} score neural network {for reducing}
 sample complexity, {and enabling highly-accurate approximation to the BCRB with limited data.}
Lastly, we {study} two signal processing problems in which the BCRB cannot be computed with previous methods: 
measurement with correlated noise and quantization; and frequency estimation with underwater ambient noise. {While the first problem is important in many sensor systems, including sensor arrays, the second is important for vessel identification \cite{erbe2019effects}.}

The {main} contributions of this paper are 
as follows:
\begin{itemize}
    \item We introduce a fully learned Bayesian Cram'er-Rao bound that learns both prior and measurement distributions using score matching.
    
    \item  We introduce Fisher score matching (FSM) that enables to learn the Fisher score function. 
    The Fisher score {
    may be of independent}
    interest, 
    e.g., for the computation of {a learned} non-Bayesian CRB.
    \item  
    We theoretically explore the FSM and demonstrate that {it produces a strongly consistent estimate of the true score.}
    \item  We {quantify theoretically the non-asymptotic, finite sample learning} error of the two suggested approaches, and validate {the theoretical predictions.}  numerically. 
    In addition, we show that the LBCRB is a strongly consistent {estimator of} the true BCRB. 

    \item  {We demonstrate the two new approaches} on several signal processing examples, including linear and non-linear estimation {and} a real-world application  with underwater ambient noise. 
\end{itemize}

In the spirit of reproducible research, we make the code of the Learned Bayesian Cram\'er-Rao bound available online \cite{lbcrb_repo}.

The paper is organized as follows. The background and notation are  in Section~\ref{sec:background}. 
The Learned Bayesian Cram'er-Rao bound method is presented in Sec.~\ref{sec:lbcrb_method}. A detailed derivation of Learned Bayesian Cram'er-Rao bound is present in Sec.~\ref{sec:lbcrb} followed by a {study} of its theoretical properties in Sec.~\ref{sec:theory}.   In Sec.~\ref{sec:example_models} we present a set of parameter estimation examples, including linear and nonlinear estimation problems, and frequency estimation with underwater ambient noise. The experimental results for the \name{} are described in Sec.~\ref{sec:experimental}, with conclusions in
Sec.~\ref{sec:conclusions}.
Proofs of the theoretical results of this paper are  in the appendices, which are included in the online Supplementary Material.

\input{files/background}
\input{files/method_short}
\input{files/learend_bayesian_bound}
\input{files/theortical_results}
\input{files/model_examples}

\input{files/results}

\section{Conclusions}\label{sec:conclusions}
This paper proposes two approaches to determine a learned Bayesian Cramér-Rao Bound (LBCRB): the Posterior Approach, and the Measurement-Prior Approach. Both approaches derive the LBCRB using only a dataset of independent and identically distributed (i.i.d.) measurement-parameter pairs. In the Posterior Approach we learn the score of the parameter conditioned on the measurements and use it to derive the BCRB. For the Measurement-Prior Approach, we introduce a variant of score matching that enables  to learn the Fisher score from i.i.d. samples. In addition, we propose a \pe{} score neural network that integrates domain knowledge into the network. 

We  {studied} both approaches theoreticallyand demonstrated the superiority of the Measurement-Prior Approach  over the Posterior Approach when domain knowledge is available, {or when it is desired to compute the \name{} for a measurement comprising multiple i.i.d. samples}. Furthermore, we showed theoretically that in both approaches the LBCRB converges almost surely to the true BCRB . 

We conduct several numerical experiments that validate our theoretical findings and highlight the advantages of the Measurement-Prior Approach when domain knowledge is present. Finally, we present two scenarios where the BCRB cannot be evaluated but the LBCRB can: one involving correlated quantization measurement and the other involving frequency estimation with real-world underwater noise. These examples highlight the usefulness of LBCRB in real life problems. 


A future research direction is to ensure that LBCRB is
a valid lower bound (rather than a good appropriation to
it) by utilizing methods for error estimation and model
selection \cite{shalev2014understanding}. On the practical side, it will be interesting
to study some of the many real-world applications that
can benefit from this approach, such as inverse problems with poorly characterized sensors. Finally, an addition direction is to use the Fisher score to obtain a learned maximum likelihood estimator.

\bibliographystyle{IEEEtran}
\bibliography{ref}

\clearpage
\appendix
\input{files/appendix/implamentation_deatiles}
\input{files/proofs/proofs}

\input{files/appendix/examples}
\input{files/appendix/prior_fim}

\end{document}

%% file: includes/base.tex
\usepackage{amsthm}
\usepackage{amssymb}
\usepackage{algorithm}
\usepackage{amsmath}
\usepackage{cite}
\usepackage{acronym}

\usepackage{commath,amsmath,amssymb,amsfonts}

\usepackage{changes}
\usepackage{accents}
\usepackage[inkscapearea=page]{svg}

\usepackage{mathtools}
\usepackage{hyperref}
\usepackage{amssymb}
\usepackage{bm}
\usepackage{graphicx}
\usepackage{float}
\usepackage{amsmath}
\usepackage{tikz}
\usepackage{multicol}
\usepackage{dsfont}
\usepackage{tcolorbox}
\usepackage{bbm}
\usepackage{multirow}
\usetikzlibrary{positioning, backgrounds, fit, shapes.arrows}
\usetikzlibrary{decorations.markings}
\usepackage{cuted}
\let\labelindent\relax
\usepackage{enumitem}
\usepackage{caption}
\usepackage{booktabs}
\usepackage{subcaption}
\newtheorem{definition}{Definition}[section]
\newtheorem{theorem}{Theorem}[section]
\newtheorem{corollary}[theorem]{Corollary}
\newtheorem{lemma}[theorem]{Lemma}
\newtheorem{prop}[theorem]{Proposition}
\newtheorem*{problem}{Problem Statement}
\newtheorem{assumption}{Assumption}[section]
\newtheorem{remark}{Remark}

\usepackage{algorithm}
\usepackage{algpseudocode}
\tikzset{block/.style = {draw, fill=white, rectangle,
		minimum height=3em, minimum width=2cm},
	input/.style = {coordinate},
	output/.style = {coordinate},
	pinstyle/.style = {pin edge={to-,t,black}}
	radiation/.style={{decorate,decoration={expanding waves,angle=90,segment   length=4pt}}}
	
}
\usepackage{smartdiagram}

\tikzstyle{block} = [draw, rectangle, minimum height=2em, minimum width=2em]
\tikzstyle{sum} = [draw, circle,minimum width=0.1 cm]
\tikzstyle{input} = [coordinate]
\tikzstyle{output} = [coordinate]
\tikzstyle{dummy} = [coordinate]
\tikzstyle{pinstyle} = [pin edge={to-,thin,black}]
\usetikzlibrary{positioning, fit, arrows.meta}
\usetikzlibrary{positioning}
\usetikzlibrary{shapes,arrows}
\tikzstyle{frame_cyan} = [thick, draw=blue, solid,inner sep=0.3em]
\tikzstyle{frame_red} = [thick, draw=red, solid,inner sep=0.3em]
\tikzstyle{frame_green} = [thick, draw=green, solid,inner sep=0.3em]

\usepackage{tikz}
\usepackage{xcolor}
\definecolor{fc}{HTML}{1E90FF}
\definecolor{h}{HTML}{228B22}
\definecolor{bias}{HTML}{87CEFA}
\definecolor{noise}{HTML}{8B008B}
\definecolor{conv}{HTML}{FFA500}
\definecolor{pool}{HTML}{B22222}
\definecolor{up}{HTML}{B22222}
\definecolor{view}{HTML}{FFFFFF}
\definecolor{bn}{HTML}{FFD700}
\tikzset{fc/.style={black,draw=black,fill=fc,rectangle,minimum height=1cm}}
\tikzset{h/.style={black,draw=black,fill=h,rectangle,minimum height=1cm}}
\tikzset{bias/.style={black,draw=black,fill=bias,rectangle,minimum height=1cm}}
\tikzset{noise/.style={black,draw=black,fill=noise,rectangle,minimum height=1cm}}
\tikzset{conv/.style={black,draw=black,fill=conv,rectangle,minimum height=1cm}}
\tikzset{pool/.style={black,draw=black,fill=pool,rectangle,minimum height=1cm}}
\tikzset{up/.style={black,draw=black,fill=up,rectangle,minimum height=1cm}}
\tikzset{view/.style={black,draw=black,fill=view,rectangle,minimum height=1cm}}
\tikzset{bn/.style={black,draw=black,fill=bn,rectangle,minimum height=1cm}}
 
\usepackage{xspace}

\tikzstyle{dummy} = [coordinate]
\pgfkeys{/pgf/.cd,
  parallelepiped offset x/.initial=2mm,
  parallelepiped offset y/.initial=2mm
}
\pgfdeclareshape{parallelepiped}
{
  \inheritsavedanchors[from=rectangle] 
  \inheritanchorborder[from=rectangle]
  \inheritanchor[from=rectangle]{north}
  \inheritanchor[from=rectangle]{north west}
  \inheritanchor[from=rectangle]{north east}
  \inheritanchor[from=rectangle]{center}
  \inheritanchor[from=rectangle]{west}
  \inheritanchor[from=rectangle]{east}
  \inheritanchor[from=rectangle]{mid}
  \inheritanchor[from=rectangle]{mid west}
  \inheritanchor[from=rectangle]{mid east}
  \inheritanchor[from=rectangle]{base}
  \inheritanchor[from=rectangle]{base west}
  \inheritanchor[from=rectangle]{base east}
  \inheritanchor[from=rectangle]{south}
  \inheritanchor[from=rectangle]{south west}
  \inheritanchor[from=rectangle]{south east}
  \backgroundpath{
    \southwest \pgf@xa=\pgf@x \pgf@ya=\pgf@y
    \northeast \pgf@xb=\pgf@x \pgf@yb=\pgf@y
    \pgfmathsetlength\pgfutil@tempdima{\pgfkeysvalueof{/pgf/parallelepiped offset x}}
    \pgfmathsetlength\pgfutil@tempdimb{\pgfkeysvalueof{/pgf/parallelepiped offset y}}
    \def\ppd@offset{\pgfpoint{\pgfutil@tempdima}{\pgfutil@tempdimb}}
    \pgfpathmoveto{\pgfqpoint{\pgf@xa}{\pgf@ya}}
    \pgfpathlineto{\pgfqpoint{\pgf@xb}{\pgf@ya}}
    \pgfpathlineto{\pgfqpoint{\pgf@xb}{\pgf@yb}}
    \pgfpathlineto{\pgfqpoint{\pgf@xa}{\pgf@yb}}
    \pgfpathclose
    \pgfpathmoveto{\pgfqpoint{\pgf@xb}{\pgf@ya}}
    \pgfpathlineto{\pgfpointadd{\pgfpoint{\pgf@xb}{\pgf@ya}}{\ppd@offset}}
    \pgfpathlineto{\pgfpointadd{\pgfpoint{\pgf@xb}{\pgf@yb}}{\ppd@offset}}
    \pgfpathlineto{\pgfpointadd{\pgfpoint{\pgf@xa}{\pgf@yb}}{\ppd@offset}}
    \pgfpathlineto{\pgfqpoint{\pgf@xa}{\pgf@yb}}
    \pgfpathmoveto{\pgfqpoint{\pgf@xb}{\pgf@yb}}
    \pgfpathlineto{\pgfpointadd{\pgfpoint{\pgf@xb}{\pgf@yb}}{\ppd@offset}}
  }
}
\pgfdeclareshape{document}{
\inheritsavedanchors[from=rectangle] 
\inheritanchorborder[from=rectangle]
\inheritanchor[from=rectangle]{center}
\inheritanchor[from=rectangle]{north}
\inheritanchor[from=rectangle]{north east}
\inheritanchor[from=rectangle]{north west}
\inheritanchor[from=rectangle]{south}
\inheritanchor[from=rectangle]{south east}
\inheritanchor[from=rectangle]{south west}
\inheritanchor[from=rectangle]{west}
\inheritanchor[from=rectangle]{east}
\backgroundpath{%
\southwest \pgf@xa=\pgf@x \pgf@ya=\pgf@y
\northeast \pgf@xb=\pgf@x \pgf@yb=\pgf@y
\pgf@xc=\pgf@xb \advance\pgf@xc by-5pt 
\pgf@yc=\pgf@ya \advance\pgf@yc by5pt
\pgfpathmoveto{\pgfpoint{\pgf@xa}{\pgf@ya}}
\pgfpathlineto{\pgfpoint{\pgf@xa}{\pgf@yb}}
\pgfpathlineto{\pgfpoint{\pgf@xb}{\pgf@yb}}
\pgfpathlineto{\pgfpoint{\pgf@xb}{\pgf@yc}}
\pgfpathlineto{\pgfpoint{\pgf@xc}{\pgf@ya}}
\pgfpathclose
\pgfpathmoveto{\pgfpoint{\pgf@xc}{\pgf@ya}}
\pgfpathlineto{\pgfpoint{\pgf@xc}{\pgf@yc}}
\pgfpathlineto{\pgfpoint{\pgf@xb}{\pgf@yc}}
\pgfpathclose
}
}
\tikzstyle{block} = [draw, fill=white, rectangle, 
    minimum height=3em, minimum width=6em]
    
\usetikzlibrary{backgrounds}
\usepackage{pifont}
\tikzstyle{startstop} = [rectangle, rounded corners, minimum width=2cm, minimum height=0.7cm,text centered, draw=black, fill=lime!30]
\usepackage{stackengine}

\makeatletter
\newcounter{phase}[algorithm]
\newlength{\phaserulewidth}
\newcommand{\setphaserulewidth}{\setlength{\phaserulewidth}}

\makeatother
\setphaserulewidth{.7pt}
\makeatother
\usepackage{colortbl}

%% file: includes/commands.tex
\newcommand\MakeUppercaseGreek[1]{
  \begingroup
    \let\psi\Psi
    \let\omega\Omega
    \let\gamma\Gamma
    \MakeUppercase{#1}
  \endgroup}
\newcommand{\vectorsym}[1]{\bm{#1}}
\newcommand{\randomvec}[1]{\MakeUppercaseGreek{\bm{#1}}}
\newcommand{\expectation}[2]{\mathbb{E}_{#2}\left[#1\right]}

\newcommand{\brackets}[1]{\left(#1\right)}
\newcommand{\normaldis}[2]{\mathcal{N}\brackets{#1,#2}}

\newcommand{\normf}[1]{\left\lVert#1\right\rVert_{\mathrm{F}}}

\newcommand{\matsym}[1]{\mathbf{#1}}
\newcommand{\squareb}[1]{\left[{#1}\right]}

\newcommand{\trace}[1]{\mathrm{Tr}\left(#1\right)}

\newcommand{\probt}[2]{f_{#2}\left(#1\right)}

\newcommand{\divc}[2]{\frac{\partial#1}{\partial#2}}

\newcommand{\at}[2]{\left.#1\right\vert_{#2}}
\newcommand{\eigmax}[1]{\lambda_{\mathrm{max}}\brackets{#1}}
\newcommand{\eigmin}[1]{\lambda_{\mathrm{min}}\brackets{#1}}

\newcommand{\probP}{\text{I\kern-0.15em P}}
\newcommand{\pmft}[2]{{\probP}_{#2}\left(#1\right)}

%% file: includes/symbols.tex
\newcommand{\name}[0]{LBCRB}

\newcommand{\X}[0]{\randomvec{X}}
\newcommand{\Yr}[0]{\randomvec{Y}}

\newcommand{\F}[0]{\matsym{F}}
\newcommand{\Z}[0]{\randomvec{Z}}

\newcommand{\p}[0]{\vectorsym{\theta}}

\newcommand{\z}[0]{\vectorsym{z}}
\newcommand{\x}[0]{\vectorsym{x}}

\newcommand{\Xr}[0]{\randomvec{x}}

\newcommand{\Ps}[0]{\mathcal{T}}
\newcommand{\Psb}[0]{\partial\Ps^{+}}
\newcommand{\pr}[0]{\Theta}
\newcommand{\ds}[0]{\mathcal{D}}
\newcommand{\nds}[0]{N_{\ds}}
\newcommand{\nx}[0]{{d_{\x}}}
\newcommand{\np}[0]{{d_{\p}}}

\newcommand{\xset}[0]{\mathcal{X}}
\newcommand{\xsetr}[0]{\widetilde{\mathcal{X}}}

\newcommand{\e}[0]{\vectorsym{\epsilon}}
\newcommand{\niideval}[0]{n_{iid}}
\newcommand{\niiddata}[0]{m_{iid}}

\newcommand{\postscore}[2]{\vectorsym{s}_{B}\brackets{#1|#2}}
\newcommand{\priorscore}[1]{\vectorsym{s}_{P}\brackets{#1}}
\newcommand{\lscore}[2]{\vectorsym{s}_{F}\brackets{#1|#2}}
\newcommand{\postscores}[2]{\vectorsym{s}_{B}^*\brackets{#1|#2}}
\newcommand{\priorscores}[1]{\vectorsym{s}_{P}^*\brackets{#1}}
\newcommand{\lscores}[2]{\vectorsym{s}_{F}^*\brackets{#1|#2}}

\newcommand{\iscore}[2]{\vectorsym{s}_{I}\brackets{#1|#2}}
\newcommand{\paramp}[0]{\Omega_{P}}
\newcommand{\paramf}[0]{\Omega_{F}}
\newcommand{\pscore}[1]{\vectorsym{s}\brackets{#1}}
\newcommand{\fm}[0]{\F_{M}}

\newcommand{\fb}[0]{\F_{B}}
\newcommand{\fp}[0]{\F_{P}}
\newcommand{\bcrb}[0]{\mathcal{V}}
\newcommand{\nbe}[0]{N_B^{(e)}}
\newcommand{\nbet}[0]{{K}_B^{(e)}}
\newcommand{\nmpe}[0]{N_{MP}^{(e)}}
\newcommand{\nmpet}[0]{{K}_{MP}^{(e)}}

\newcommand{\lbfimlp}[0]{\overline{\matsym{F}}_{MP}}
\newcommand{\lbcrblps}[0]{\widehat{\bcrb}_{MP}}
\newcommand{\lbfimlps}[0]{\widehat{\matsym{F}}_{MP}}
\newcommand{\lbfimb}[0]{\overline{\matsym{F}}_{B}}
\newcommand{\lbcrbbs}[0]{\widehat{\bcrb}_{B}}
\newcommand{\lbfimbs}[0]{\widehat{\matsym{F}}_{B}}
\newcommand{\lpfim}[0]{\widehat{\matsym{F}}_{P}}
\newcommand{\lmfim}[0]{\widehat{\matsym{F}}_{M}}
\newcommand{\tlmfim}[0]{\overline{\matsym{F}}_{M}}
\newcommand{\tlpfim}[0]{\overline{\matsym{F}}_{P}}
\newcommand{\mfle}[0]{\eta_{M}^{(a)}}
\newcommand{\pfle}[0]{\eta_{P}^{(a)}}
\newcommand{\bfle}[0]{\eta_{B}^{(a)}}

\newcommand{\mpfse}[0]{\eta_{MP}^{(e)}}
\newcommand{\mpfle}[0]{\eta_{MP}^{(a)}}

\newcommand{\bfse}[0]{\eta_{B}^{(e)}}
\newcommand{\cmark}{\text{\ding{51}}}%
\newcommand{\lossbsm}[0]{\widehat{\mathcal{L}_{B}}}
\newcommand{\lossfsm}[0]{\widehat{\mathcal{L}_{F}}}
\newcommand{\losspsm}[0]{\widehat{\mathcal{L}_{P}}}
\newcommand{\lossbs}[0]{{\mathcal{L}_{B}^{0}}}
\newcommand{\lossfs}[0]{{\mathcal{L}_{F}^{0}}}
\newcommand{\lossps}[0]{{\mathcal{L}_{P}^{0}}}
\newcommand{\lossbst}[0]{{{\mathcal{L}}_{B}}}
\newcommand{\lossfst}[0]{{{\mathcal{L}}_{F}}}

\newcommand{\intdim}{\mathrm{intdim}}

\newcommand{\cb}{c_B}
\newcommand{\cp}{c_P}
\newcommand{\cm}{c_M}
\newcommand{\dbb}{{\bar{d}_B}}
\newcommand{\dmpb}{{\bar{d}_{MP}}}
\newcommand{\db}{{{d}_B}}
\newcommand{\dm}{{{d}_M}}
\newcommand{\intdp}{{{d}_P}}

\newcommand{\pe}[0]{{Physics-encoded}}
\newcommand{\peac}[0]{{PeSNN}}

%% file: files/background.tex
\section{Notation and Background}\label{sec:background}
\subsection{Notations}

Upper case 
$\randomvec{A}$ indicates a random vector, lower case italics $a$ and boldface $\vectorsym{a}$ indicate a scalar and a vector, respectively, with $\norm{\vectorsym{a}}_2$ denoting 
the $l_2$ norm. The $i$-th element of a vector $\vectorsym{a}$ is indicated by $\squareb{\vectorsym{a}}_i$. Upper case boldface $\matsym{A}$ indicates a matrix, with its trace, determinant, transpose, Frobenius norm, spectral norm (largest singular value), and condition number denoted by $\trace{\matsym{A}}$, $\det{\matsym{A}}$, 
$\matsym{A}^T$, $\normf{\matsym{A}}$, $\norm{\matsym{A}}_2$, and $\kappa(\matsym{A})$,
respectively.  An identity matrix of size $k\times k$ is denoted by $\matsym{I}_k$. 
For symmetric matrix $\matsym{A}$ the notation $\matsym{A} \succ 0$ (or $\matsym{A} \succeq 0$) means that $\matsym{A}$ is positive-definite (or positive semidefinite). 
For symmetric $\matsym{A}$ and $\matsym{B}$ the inequality $A \succ B$ means that $A-B \succ 0$. 
The minimal and maximal eigenvalues of a symmetric matrix $\matsym{A}$ are denoted by $\eigmin{\matsym{A}}$ and $\eigmax{\matsym{A}}$, respectively. {Finally, the $d \times d$ Jacobian matrix of a vector function $\pscore{\p}: \mathbb{R}^d \rightarrow \mathbb{R}^d $ is denoted by $\partial \pscore{\p} / \partial \p$.}

\input{files/background_bcrb}

\subsection{Overview of Score Matching}\label{sec:score_over_view}
Score matching\cite{hyvarinen2005estimation,hyvarinen2007some,liu2022estimating} is a well-known method for learning the score function of a data distribution when only having access to a set of i.i.d. samples. Various scores have been developed over the years. We begin with a brief introduction to classical score matching\cite{hyvarinen2005estimation}. Specifically, the goal of score matching is to learn, using a dataset $\ds$ of $\nds$ i.i.d. samples,
a {model} function {(represented e.g., by a neural network)} 
$\pscore{\p;\Omega}:\mathbb{R}^{\np}\rightarrow\mathbb{R}^{\np}$ 
{paramterized by  vector $\Omega$}, such that $\pscore{\p}=\pscore{\p;\Omega}\approx\nabla_{\p}
    \log\probt{\p}{\p}$. {
    The mismatch between the true score function and the model} is formulated 
    as the 
    objective function
\begin{equation}\label{eq:loss_prior}
    \mathcal{L}^0\brackets{\Omega}=\expectation{\norm{\pscore{\p;\Omega}-\nabla_{\p}
    \log\probt{\p}{\p}}_2^2}{\p},
\end{equation}
{where the expectation over $\p$ can be replaced by an empirical mean over the samples in $\ds$. }

We wish to minimize {the score mismatch} \eqref{eq:loss_prior} w.r.t. $\Omega$. Since we do not have direct access to {the true score} $\nabla_{\p}\log\probt{\p}{\pr}$, only a set of i.i.d. samples $\ds$, we cannot directly minimize the objective 
\eqref{eq:loss_prior}. 
Instead, an {equivalent} objective function has been {derived} in score matching\cite{hyvarinen2005estimation}
{that does not require direct access to $\nabla_{\p}\log\probt{\p}{\pr}$:
\begin{equation}\label{eq:loss_prior_sm}
    {\mathcal{L}}\brackets{\Omega}=\expectation{\norm{\pscore{\pr;\Omega}}_2^2}{\pr}+2\trace{
    {\expectation{\frac{\partial \pscore{\pr;\Omega}}{\partial\p}}{\pr}}}.
\end{equation}
The objective \eqref{eq:loss_prior_sm}  is related to \eqref{eq:loss_prior} via 
    ${\mathcal{L}}^0\brackets{\Omega}={\mathcal{L}}\brackets{\Omega}+C,$
where $C$ is a constant independent of $\Omega$.
The optimum parameters $\Omega^*$ can then be determined by minimizing a sample average version \eqref{eq:score_prior_mean} of the objective  \eqref{eq:loss_prior_sm} over a dataset of i.i.d. measurements.
} 

{For the relation between \eqref{eq:loss_prior} and \eqref{eq:loss_prior_sm} to hold, certain technical conditions need to be satisfied.
First are the} boundary conditions
\begin{equation}\label{eq:boundary_conditions}
    \lim\limits_{\p\rightarrow \Psb}\pscore{\p;\Omega}\probt{\p}{\pr}=0, 
\end{equation}
where $\Psb$ is the boundary of the set $\Ps$ augmented by the points at infinity in $\Ps$ in case $\Ps$ is unbounded. 
{
Second are the following} regularity conditions\cite{hyvarinen2005estimation}. 
\begin{assumption}[Score Matching regularity]\label{ass:score_reg_prior} 
\hspace*{2cm}
\begin{enumerate}[label={\ref*{ass:score_reg_prior}}.\arabic*,labelsep=*, leftmargin=*]

\item The {log-{prior}} ${\log}\probt{\p}{\pr}$ is differentiable w.r.t. $\p$ {at all $\p\in\Ps$ where $\probt{\p}{\pr}>0$} \label{assum:diff_prob_prior}. 
    
\item The score neural network $\pscore{\p ;\Omega}$ is differentiable w.r.t. $\p$. 
\item The expectations $\expectation{\norm{\pscore{\pr;\Omega}}_2^2}{\pr}$ and $\expectation{\norm{\nabla\log\probt{\pr}{\pr}}_2^2}{\pr}$ are finite. \label{assume:expecte_fine_score_matching}
\end{enumerate}
\end{assumption}
In case the boundary conditions \eqref{eq:boundary_conditions} do not hold, 
extensions have been suggested \cite{hyvarinen2007some} for  data  
 $\p$ with non-negative support 
 {$[0,\infty)$ for each component} and 
 {PDF that does not vanish at the boundary,}  or for general truncated domains\cite{liu2022estimating}.

%% file: files/background_bcrb.tex
\subsection{Bayesian Cram\'er Rao Bound}
Let $\X\in\Upsilon\subseteq\mathbb{R}^\nx$ be a random observation vector {that depends on a random parameter vector} $\pr\in\Ps\subseteq\mathbb{R}^\np$, 
and let $\probt{\x,\p}{\X,\pr}$ be their joint {probability density function (PDF).}\footnote{To simplify  notation we write the expressions for continues random vector $\X$;  for discrete or mixed $\X$ they can be written in terms of the joint CDF, or for $\X$ discrete $\probt{\x,\p}{\X,\pr}$ can be replaced by $P_{\X}(\x)\probt{\p | \x}{\pr|\X}$ or by $\probt{\p}{\pr} P_{\X|\pr}(\x|\p)$, where $P_{\X}(\x)$ and $P_{\X|\pr}(\x|\p)$ are the probability mass function (PMF), and conditional PMF of $\X$, respectively.  
}
For convenience we model the parameter set $\Ps$ as a closed set (in the Euclidean metric) with no isolated points. Assume
the following conditions\cite[pages 33-35]{van2007bayesian}, \cite{weinstein1988general}. 
\begin{assumption}[BCRB Regularity Conditions]
    \label{assum:bcrb_reg}
$\probt{\x,\p}{\X,\pr}$ 
satisfies the following conditions:  
\begin{enumerate}[label={\ref*{assum:bcrb_reg}}.\arabic*,labelsep=*, leftmargin=*]

\item\label{sas:derivative} The gradient $\nabla_{\p}\log\probt{\x,\p}{\X,\pr}$ with respect to $\p$  exists and {each of its elements} is absolutely integrable w.r.t $\x$ and $\p$ {on $\Upsilon \times \Ps$}.
\item\label{sas:derivative_prior} {The gradient $\nabla_{\p}\log\probt{\p}{\pr}$ with respect to $\p$  exists.}
\item \label{assume:non_singular}  The matrix  $$\expectation{\nabla_{\pr}\log\probt{\x,\pr}{\X,\pr}\nabla_{\pr}\log\probt{\x,\pr}{\X,\pr}^T}{\X,\pr} 
$$  is {positive-definite} and finite. 
\item The limits $\lim\limits_{\p\rightarrow\partial\Ps{^+}}\p\probt{\x,\p}{\X,\pr}=0$ hold, where $\partial\Ps{^+}$ is the boundary of the set $\Ps$ augmented by the points  of $\Ps$ at infinity  in case $\Ps$ is unbounded. \label{ass:lim}
\item  The conditional expectation of the score  {
vanishes for all $\x$,} $\expectation{\nabla_{\pr}\log\probt{\pr,\x}{\pr,\X}}{\pr|\X}=0$ .\label{ass:score_expection}
\item For all $\p\in\Ps$, the densities $\probt{\x,\p}{\X,\pr}$ have a common support {$\{\x:\probt{\x,\p}{\X,\pr}>0\}{\subseteq \Upsilon}$}  w.r.t. {$\x$} that is independent of $\p$. \label{assume:common_support}
\end{enumerate}
{
\begin{remark}
   Assumption~\ref{assume:non_singular}, although sometimes not stated explicitly (e.g., \cite[pages 33-35]{van2007bayesian}), is, in view of \eqref{eq:bfim}, a natural non-degeneracy assumption for the BCRB.
\end{remark}
}

\begin{remark}
When $\X$ is a discrete random variable, a slightly altered set of
regularity conditions {can be used} \cite{zeitler2012bayesian}.
\end{remark}
  \begin{remark}
      In situations where {one or more of assumptions} ~\ref{ass:lim} 
      and ~\ref{ass:score_expection} is not satisfied, {both can be substituted by} the bias condition $\lim\limits_{\p\rightarrow\partial\Ps{^+}}\int_{\x}\brackets{\hat{\p}\brackets{\x}-\p}\probt{\x,\p}{\X,\pr}d \x =0$, {where $\hat{\p}\brackets{\x}$ is the estimator of $\p$ that is being considered.}
  \end{remark}
\end{assumption}
{Let $\hat{\p}\brackets{{\xset}}$ be an arbitrary estimator of $\p$  using 
{a set of i.i.d observations $\xset\triangleq\set{\X_i}_{i=1}^{\niideval}$ of $\X$}, with estimator error $\e=\p-\hat{\p}\brackets{
{\xset}}$. Then, subject to Assumptions~\ref{assum:bcrb_reg},}
the following lower bound on the mean square error (MSE) matrix holds\cite{van2007bayesian}:
\begin{equation}\label{eq:bcrb}
\mathrm{MSE}\brackets{\e}\triangleq\expectation{\e\e^T}{\X,\pr}\succeq\bcrb\triangleq \fb^{-1},
\end{equation}
where $\bcrb$ is the Bayesian Cram\'er Rao Bound  (BCRB), and
\begin{align}\label{eq:bfim} &\fb\triangleq\expectation{\nabla_{\pr}\log\probt{\xset,\pr}{\xset,\pr}\nabla_{\pr}\log\probt{\xset,\pr}{\xset,\pr}^T}{\xset,\pr},\nonumber\\
        &=\expectation{\nabla_{\pr}\log\probt{\pr|\xset}{\pr|\xset}\nabla_{\pr}\log\probt{\pr|\xset}{\pr|\xset}^T}{\xset,\pr}
\end{align}
is the \emph{Bayesian Fisher Information Matrix} {(Bayesian FIM, or BFIM).}
{The second line} in \eqref{eq:bfim} {follows by the} product rule of probability and 
{the vanishing of} $\nabla_{\p}\log\probt{\x}{\X}=0$.
A different representation of the BFIM decomposes   it \cite{van2007bayesian} into the sum of two parts, the \emph{{measurement} {FIM}} {$\fm$} 
{of a \emph{single} observation vector $\X$} and the \emph{prior {FIM}} {$\fp$}:
\begin{equation}\label{eq:bfim_decomposition_base}
\fb=\niideval\cdot\fm+\fp=\niideval\cdot\expectation{\F\brackets{\pr}}{\pr}+\fp,
\end{equation}
where $\F\brackets{\p}$ is the \emph{non Bayesian {FIM}} {(whose inverse is the non-Bayesian CRB)} 
{of a single observation vector $\X$} 
\begin{align} \label{eq:non-BayesFim}
\F\brackets{\p}\triangleq&\expectation{\nabla_{\p}\log\probt{\X|\p}{\X|\pr}\nabla^T_{\p}\log\probt{\X|\p}{\X|\pr}}{\X|\p}
\end{align}
\begin{equation} \label{eq:PriorFim}
 \text{and} \qquad  \fp\triangleq\expectation{\nabla_{\pr}\log\probt{\pr}{\pr}\nabla_{\pr}\log\probt{\X|\p}{\X|\pr}^T}{\pr} .
\end{equation}


We use the term \emph{Posterior Approach} to describe the computation of the BFIM according to \eqref{eq:bfim}, and the term \emph{Measurement-Prior Approach} to describe the computation 
according to \eqref{eq:bfim_decomposition_base}.

%% file: files/method_short.tex
\section{Learned Bayesian Cram\'er Rao Bound:Overview}\label{sec:lbcrb_method}
We briefly 
overview the problem that the LBCRB address, and the methods to compute it. {The  overview in this section covers all the information that a practitioner would need to apply the proposed techniques. Detailed formulation, derivations and theoretical analysis are in later sections.}
Our goal is to determine the Bayesian Cram\'er-Rao bound \eqref{eq:bcrb}
in scenarios where the prior $\probt{\p}{\pr}$, the measurement $\probt{\x|\p}{\X|\pr}$, or both are either
unknown or partially known. However, a data set 
\begin{equation}\label{eq:dataset_rel}
 \ds=\set{\p_n, \xsetr_n=\set{{\x}_{n,j}}_{j=1}^{\niiddata}   }_{
 {n=1}}^{\nds},
\end{equation}
of $\nds$ parameter-measurement sets pairs is given,
{where  $\p_n$ drawn from $\probt{\pr}{\pr}$ and  each $\x_{n,j}$ drawn from $\probt{\x|\p_n}{\X|\pr}$.}. 
{The measurement set in $\ds$ contains $\niiddata$ i.i.d samples 
{for} the same value of $\p_i$.} 
{It is acceptable that $\niiddata\neq \niideval$, that is, 
$\niiddata$  may differ from $\niideval$ defined earlier in the context of \eqref{eq:bfim_decomposition_base}}.

We 
propose two approaches 
{to learn the \name{} from $\ds$:}
 the \emph{Posterior Approach}; and the \emph{Measurement-Prior Approach}.
 
\subsection{Posterior Approach}
\subsubsection{\textbf{Learning Step}} Define a neural network $\postscore{\p}{\xsetr;\Omega}$ 
 parameterized by $\Omega$ 
{to {model} $\nabla_{\pr}\log\probt{\xsetr,\p}{\xset,\pr}$.} Then minimize the following objective 
 {with respect to $\Omega$}:
\begin{align}\label{eq:score_post_mean}
     &\lossbsm\brackets{\Omega}=\frac{1}{\nds}\sum_{\p,\xsetr\in\ds}\ell_B\brackets{\p,\xsetr;\Omega},\\
     &\ell_B\brackets{\p,\xsetr;\Omega}\triangleq\norm{\postscore{\p}{\xsetr;\Omega}}_2^2+2\trace{{\frac{\partial \postscore{\p}{\xsetr;\Omega}}{\partial\p}}}.\nonumber
 \end{align}
{Denote the minimizer determined in this step by $\Omega^*$.}

\subsubsection{\textbf{Evaluation Step}} 
{Using} 
{
$\postscores{\p}{\xsetr} \triangleq \postscore{\p}{\xsetr;\Omega^*}$
} \footnote{\label{fn:model}%
For conciseness, we drop $\Omega$ or $\Omega^*$ from the notation, whenever the NN parameters are fixed (e.g. after training). }  
compute the Learned Bayesian Fisher Information Matrix 
\begin{equation}\label{eq:mean_fully}
    \lbfimbs\triangleq \frac{1}{\nds}\sum_{\p,\xsetr\in\mathcal{D}}\postscores{\p}{\xsetr}\postscores{\p}{\xsetr}^T.
\end{equation}
Finally, to obtain the LBCRB, invert $\lbfimbs$, which results in ${\bcrb}\approx\lbcrbbs\triangleq\lbfimbs^{-1}$.  
The Posterior Approach is illustrated in Figure~\ref{fig:main_post}. 

{We emphasize that the LBCRB in \eqref{eq:mean_fully} is calculated  {for a measurement that contains  $\niideval$ i.i.d samples $\x_i, i=1, \ldots, \niideval$, where $\niideval= \niiddata$, that is, $\niideval$ coincides with the number $\niiddata = |\xsetr|$ of i.i.d samples available in the training set $\ds$ for each  value of $\p$.  
}
The LBCRB can be determined for $\niideval\leq \niiddata$ {using such a data set,}  but this requires learning a different score function for each desired $\niideval$. {On the other hand, computing 
the LBCRB for $\niideval > \niiddata$ is not possible using the same $\ds$, and would require a data set with a larger $\niiddata$.
 Both of these limitations are overcome by the Measurement-Prior Approach of the next section.
}

We 
will {quantify the approximation of $\bcrb$ by $\lbcrbbs$ and also}
show that if  {the model neural net $\postscore{\p}{\xsetr;\Omega}$} has sufficient capacity, then
{
$
    \lbcrbbs 
    \xrightarrow{\nds\rightarrow \infty} \bcrb \quad\text{
    {a.s}} 
$
(almost surely, i.e., with probability 1). In other words, the approximation $\lbcrbbs$ enjoys the important statistical 
 property of \emph{strong consistency}.
}

\subsection{Measurement-Prior Approach}
\label{subsec:MP}
To improve the sample complexity of the Posterior Approach, we suggest to decompose, {similar to \eqref{eq:bfim_decomposition_base},} the LBCRB into two parts: Prior; and Measurement. As an additional advantage, this enables to introduce domain knowledge. We begin with the prior term.

\subsubsection{\textbf{Learning Prior Step}}  Define a neural network $\priorscore{\p;\paramp}$  parameterized by $\paramp$ 
{to {model} $\nabla_{\p}\log\probt{\p}{\pr}$.} Then
{find a minimizer $\Omega_P^*$ 
of the following objective
}
\begin{align}\label{eq:score_prior_mean}
     &\losspsm\brackets{\paramp}=\frac{1}{\nds}\sum_{\p\in\ds}\ell_P\brackets{\p;\paramp},\\
     &\ell_P\brackets{\p;\paramp}\triangleq\norm{\priorscore{\p;\paramp}}_2^2+2\trace{\frac{\partial \priorscore{\p;\paramp}}{\partial\p}}.\nonumber
 \end{align}
\subsubsection{\textbf{Construct \pe{} Score Neural Network (\peac{})}} \label{subsubsec:MoISNN}
Assume knowledge of a model function $\mathcal{M}\brackets{\p}$ such that {the PDF of a single measurement sample can be expressed as} $\probt{\x|\p}{\X|\pr}=\probt{{\x}|\mathcal{M}\brackets{\p}}{\X|\vectorsym{\tau}}$, and define a neural network $\iscore{\x}{\vectorsym{\tau}; \paramf}$ that is parameterized by same parameter $\paramf$ of $\lscore{\x}{\p;\paramf}$. 
Then, a \emph{\pe{} Score Neural Network} (\peac{}) 
{to {model} the Fisher score $\nabla_{\p}\log\probt{{\x}|\p}{{\X}|\p}$} is given by:
\begin{equation}\label{eq:model_base_score}
\lscore{{\x}}{\p;\paramf}=\divc{\mathcal{M}\brackets{\p}}{\p}^T\at{\iscore{{\x}}{\vectorsym{\tau};\paramf}}{\vectorsym{\tau}=\mathcal{M}\brackets{\p}}.
\end{equation}
The \pe{} Score 
Neural Network is illustrated 
in Figure~\ref{fig:model_inforamed}.

As an example of the application of \peac{}, consider 
{a frequency estimation problem} . 
Let $\squareb{\X}_n=
{\cos\brackets{ \theta n }}+\squareb{\randomvec{W}}_n$,
{$n= 1, \ldots, N$} 
{be the observation, with $\theta$ the frequency to be estimated, and $\randomvec{W}$   random noise with unknown distribution.} Here, $\squareb{\mathcal{M}\brackets{\p}}_n=
{\cos\brackets{\theta n}}$. Using \peac{}, we only need to learn the 
{the score for the PDF of the} noise component, $\randomvec{W}$, {eliminating the need to learn and represent the cosine function}.
\input{files/tikz/flow_two_approch_compare}
\subsubsection{\textbf{Learning Fisher Score Step}} 
{Use the neural network $\lscore{{\x}}{\p;\paramf}$ parameterized by $\paramf$ to 
model {the Fisher score} $\nabla_{\p}\log\probt{\x|\p}{\X|\pr}$ {for a single measurement sample.}
Then 
{find a mininimizer $\paramf^*$}
of the following objective 
($\paramp
{^*}$ is known from the step of learning the Prior).
\begin{align}\label{eq:score_lik_mean}
    &\lossfsm\brackets{\paramf; \paramp
    {^*}}=\frac{1}{\nds\cdot\niiddata}\sum_{\xsetr,\p\in\mathcal{D}}\sum_{{\x}\in\xsetr}{\ell_{F}\brackets{\x,\p;\paramf,\paramp
    {^*}}}{} \nonumber\\
    &\ell_{F}\brackets{\x,\p;\paramf,\paramp
    {^*}}\triangleq \norm{\lscore{{\x}}{\p;\paramf}}_2^2\\
    &+2\lscore{{\x}}{\p;\paramf}^T\priorscore{\p;\paramp
    {^*}}+2\trace{\frac{\partial \lscore{{\x}}{\p;\paramf}}{\partial\p}}\nonumber
\end{align}
\subsubsection{\textbf{Evaluation Step}}  
{Using $\priorscores{\p}=\priorscore{\p; \Omega^*_P}$ and $\lscores{\x}{\p}$ = $\lscore{\x}{\p;\Omega^*_F}$ } compute 
{the learned Measurement {FIM for a single measurement sample} and the Prior FIM}
\begin{equation} \label{eq:mean_efim_likd}
    \lmfim=\frac{1}{\nds\cdot \niiddata}\sum_{\p,\xsetr\in\ds} \sum_{\x\in\xsetr}\lscores{\x}{\p}\lscores{\x}{\p}^T,
\end{equation}
\begin{equation}\label{eq:bprior_mean}
    \lpfim=\frac{1}{\nds} \sum_{\p\in\ds} \priorscores{\p}\priorscores{\p}^T,
\end{equation}
{Then}
the Learned Bayesian FIM for any desired number $\niideval$ of i.i.d samples is given by
\begin{equation}\label{eq:bfim_apx_final}
     \lbfimlps{(\niideval)} =\niideval\cdot  \lmfim+ \lpfim.
\end{equation}
Finally, 
{
the Learned BCRB 
is obtained as }
    $\lbcrblps=\lbfimlps^{-1}.$   
    
    The Measurement-Prior Approach is 
    illustrated in Figure~\ref{fig:main_lik_prior}. 
    
    {This method has important advantages over the Posterior Approach. First,} 
    it provides the same guarantees as those 
    of the Posterior Approach but with 
    lower sample complexity, and with a more interpretable model. 
    {
   Second, because 
    $\niideval$ 
    can be chosen different to $\niiddata$,} 
   the  Measurement-Prior Approach 
   can 
   provide the LBCRB
   for \emph{any} 
   {desired} number of $\niideval$ i.i.d. samples without any additional effort 
   {or additional training data, which the Posterior Approach cannot}.
   
    {We will prove
    that minimizing that objective in \eqref{eq:score_lik_mean} is equivalent to learning the true Fisher score $\nabla_{\p}\log\probt{\x|\p}{\X|\pr}$, and that the Fisher score neural network is consistent estimator of the true Fisher score. Moreover, we will quantify the approximation of $\bcrb$ by $\lbcrblps$ and also show that if the score neural networks $\priorscores{\p}$ and $\lscores{\x}{\p}$ have sufficient capacity, then {
    $\lbcrblps\xrightarrow{\nds\rightarrow\infty} \bcrb$} a.s. (almost surely, i.e., with probability 1). In other words, the approximation $\lbcrblps$ enjoys the important statistical property of \emph{strong consistency}. }

%% file: files/tikz/flow_two_approch_compare.tex
\definecolor{mydarkblue}{RGB}{0,0,139} 
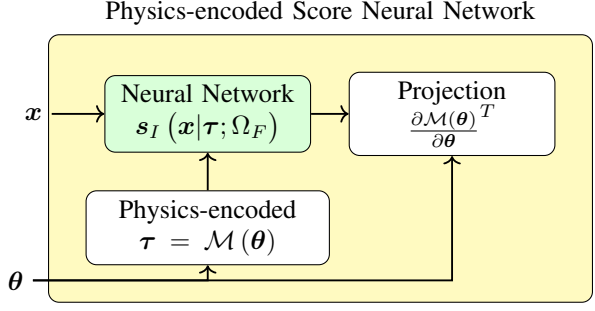
\begin{figure}
    \centering
     \begin{tikzpicture}[every label/.append style={align=center}]
        
        \node[startstop,fill=green!15, text width=2.5cm,align=center] (model_learn) {Neural Network\\   $ \iscore{{\x}}{\vectorsym{\tau};\paramf}$ };
        \node[below=0.5cm of model_learn,startstop,fill=white, text width=3cm,align=center] (model_base) {\pe{}\\   $\vectorsym{\tau}=\mathcal{M}\brackets{\p}$ };
        \node[right=0.5cm of model_learn,startstop,fill=white, text width=2.5cm,align=center] (model_base_porj) {Projection\\   $\divc{\mathcal{M}\brackets{\p}}{\p}^T$ };

        \node[left=0.7 of model_learn ] (x) {${\x}$};
        \node[left=0.7 of model_base,yshift=-0.7cm] (theta) {$\p$};

        \begin{pgfonlayer}{background}
            \node[draw,rounded corners,fit= (model_base) (model_learn) (model_base_porj), fill=yellow!30,inner sep=14pt,label={\pe{} Score Neural Network}](fit2){};

        \end{pgfonlayer}

        \draw[->,line width=0.25mm] (x)  -- ( model_learn);
        \draw[->,line width=0.25mm] (theta)  -| (model_base);
        \draw[->,line width=0.25mm] (theta)  -| (model_base_porj);
        \draw[->,line width=0.25mm] (model_base)  -- (model_learn);
        \draw[->,line width=0.25mm] (model_learn)  -- (model_base_porj);
\end{tikzpicture} 

    \caption{\pe{} Score Neural Network: The white blocks indicate the usage of $\mathcal{M}\brackets{\p}$ (a known function), whereas the green block represents a neural network that is trained during the learning process. 
    }
    \label{fig:model_inforamed}
\end{figure}

%% file: files/learend_bayesian_bound.tex
\section{{Derivation of the \name{}}}\label{sec:lbcrb}
{We turn now to a precise problem statement and derivation of the two proposed methods for the \name{}.}
\begin{tcolorbox}
\begin{problem}\label{problem_one}
 Suppose that $\probt{\x|\p}{\X|\pr}$ and $\probt{\p}{\pr}$ satisfy Assumptions \ref{assum:bcrb_reg} and 
 score matching Assumptions \ref{ass:score_cond_reg} or \ref{ass:score_reg} and \ref{ass:score_reg_prior}. 
 {Let $\bcrb$ be the}  
Bayesian Cram\'er-Rao lower bound (BCRB) \eqref{eq:bcrb} on the estimation error of parameter $\pr\in\Ps$ from a 
measurement {$\xset = \set{{\X}_{i}}_{i=1}^{\niideval}$ containing $\niideval$ i.i.d. measurements ${\X}_i \sim  \probt{\X|\pr_n}{\X|\pr}$.}
{Assume that $\probt{\x|\p}{\X|\pr}$ and $\probt{\p}{\pr}$ are completely or partially unknown.}
Given a data set 
        {\begin{equation}\label{eq:dataset}
          \ds=\set{\pr_n, \mathcal{\X}_n=\set{{\X}_{n,j}}_{j=1}^{\niiddata}   }_{n=1}^{\nds} 
        \end{equation}}
\noindent 
of  {parameter-measurement} samples that are independent and identically-distributed (i.i.d) 
as    %
{${\X}_{n,j} \sim  \probt{\X|\pr_n}{\X|\pr}, \pr_n\sim \probt{\pr}{\pr}$, }
obtain {a learned approximation  \name{} $\hat{\bcrb}(\ds)$ to  $\bcrb$} satisfying:
\begin{equation}
     {\hat{\bcrb}}\brackets{\ds}
     \xrightarrow{\nds \rightarrow \infty}\bcrb\quad\text{%
     {a.s}}.
\end{equation}

\end{problem}
\end{tcolorbox}
\begin{remark}
    The data set in \eqref{eq:dataset_rel} represents an instance {(realization)} of the data set defined in Problem~\ref{problem_one} (\eqref{eq:dataset}). For theoretical analysis concerning the stochastic nature of the \name{}, we employ the random data set specified in \eqref{eq:dataset}.
\end{remark}

To address Problem~\ref{problem_one}, we propose two approaches: 
(A) the Posterior Approach, which 
uses conditional score matching to learn the posterior score;  
and (B) the Measurement-Prior Approach, which learns two score functions -- one for the prior, and another for the measurement distribution.
 The Measurement-Prior Approach facilitates the incorporation of domain knowledge into the score neural network {improving the representation and learning of the true score}. 
Similar to previous works \cite{habi2023learned,habi2023learning, crafts2023bayesian}, both approaches comprise two stages - of learning, and evaluation. {However, as discussed in the Introduction, they differ from the previous works in most other key aspects.}   
The two approaches 
are illustrated in Figure~\ref{fig:main}.

\subsection{Posterior Approach}\label{sec:post_learn}
In the Posterior Approach {( Fig.~\ref{fig:main_post}),} we use the BFIM in \eqref{eq:bcrb}, which only requires the posterior score $\nabla_{\p}\log\probt{\p|\xsetr}{\p|\xset}$, 
a conditional score of $\p$ given a measurement $\X$. Then we use the learned conditional score to evaluate the \name{} by replacing the expectation with an empirical mean. 
\input{files/tikz/apporch_compare}

\subsubsection{Score Learning}\label{sec:score_learning_post}
 To learn $\nabla_{\p}\log\probt{\p|\xsetr}{\pr|\xset}$, we suggest to use conditional score matching \cite{hyvarinen2005estimation,song2019generative,song2020score}. The goal is to {to fit} a model $\postscore{\p}{\xsetr; \Omega}$ parameterized by $\Omega$ (usually implemented as a neural network) to 
 the true score function $\nabla_{\p}\log\probt{\p|\xsetr}{\pr|\xset}$ {by minimizing w.r.t. $\Omega$ the discrepancy between the two expressed by} the objective
\begin{equation}
\label{eq:score_matching_objective}
    \lossbs\brackets{\Omega}=
    \mathbb{E}_{\pr, \xset} \norm{\postscore{\pr}{\xset;\Omega}-\nabla_{\pr}\log\probt{\pr|\xset}{\pr|\xset}}_2^2
\end{equation}
Since we do not have direct access to $\nabla_{\p}\log\probt{\p|\xsetr}{\pr|\xset}$, only a set of i.i.d. samples $\ds$, we cannot directly minimize the objective 
\eqref{eq:score_matching_objective}. 
Similar 
to standard score matching, an 
equivalent objective function is used {that does not require direct access to $\nabla_{\p}\log\probt{\p|\xsetr}{\pr|\xset}$:
\begin{align}\label{eq:conditional_score_matching}
&\lossbst\brackets{\Omega}=\expectation{\ell_B\brackets{\X,\pr;\Omega}}{\X,\pr}=\\
&\expectation{\norm{\postscore{\pr}{\xset;\Omega}}_2^2}{\pr,\xset}+2\trace{\expectation{\frac{\partial \postscore{\pr}{\xset;\Omega}}{\partial\pr}}{\xset,\pr}}.\nonumber
\end{align}

Objective \eqref{eq:conditional_score_matching} is equivalent to \eqref{eq:score_matching_objective} for the purpose of finding the minimizer $\Omega^*$,
in the sense that the two only differ by a constant $C$ that is independent of $\Omega$, i.e., $\lossbs\brackets{\Omega}=\lossbst\brackets{\Omega}+C$, provided that the following conditions hold.
First,} the boundary condition
\begin{equation}\label{eq:boundary_conditions_post}
    \lim\limits_{\p\rightarrow \Psb}\postscore{\p}{\xsetr}\probt{\p|\xsetr}{\pr|\xset}=0,\quad\forall\xsetr,
\end{equation}
{
Second, the following} regularity conditions. 
\begin{assumption}\label{ass:score_cond_reg}$\,$

\begin{enumerate}[label={\ref*{ass:score_cond_reg}}.\arabic*,labelsep=*, leftmargin=*]
\item \label{assum:diff_prob_post} The {log-posterior} $\log\probt{\p|\xsetr}{\pr|\xset}$ is differentiable w.r.t. $\p$ {at all $\xsetr$ and $\p\in\Ps$  where $\probt{\p,\xsetr}{\pr,\xset} > 0$}. 

\item The expectation $\expectation{\norm{\nabla\log\probt{\pr|\xset}{\pr|\xset}}_2^2}{\pr,\xset}$ is finite. \label{assum:bound_expection_post}
\item The score neural network $\postscore{\p}{\xsetr}$ is differentiable w.r.t. $\p$. \label{assum:net_cond_score}
\item The expectation $\expectation{\norm{\postscore{\pr}{\xset}}_2^2}{\pr,\xset}$  is finite. \label{assum:expected_cond_score}

\end{enumerate}
\end{assumption}
\begin{remark}\label{remark:condtion_hold_post}
    {Assumptions~\ref{assum:diff_prob_post} and \ref{assum:bound_expection_post} 
are implied by the regularity Assumptions~\ref{sas:derivative} and \ref{assume:non_singular} of the BCRB, 
respectively.}
{On the other hand,}
Assumptions~\ref{assum:net_cond_score}, \ref{assum:expected_cond_score} and the boundary condition \eqref{eq:boundary_conditions_post} can be inherently satisfied by selecting an appropriate neural network architecture and non-linear activation function.
\end{remark}
The objective \eqref{eq:conditional_score_matching} enables 
{ to determine the optimum parameters $\Omega^*$ by minimizing a sample average version \eqref{eq:score_post_mean} using} 
a dataset $\ds$ of i.i.d. measurements.
{ 
This yields} an approximation of the Bayesian score $\postscores{\p}{\xsetr} \triangleq \postscore{\p}{\xsetr;\Omega^*}\approx\nabla_{\p}\log\probt{\p|\xsetr}{\pr|\xset}$. 
\subsubsection{Evaluation of the LBCRB}\label{sec:lbcrb_eval_post}
We replace the Bayesian score in \eqref{eq:bfim} with the learned one.
\footnote{%
{ By Assumption~\ref{ass:score_expection} we have $\expectation{\nabla_{\pr}\log\probt{{\X},\pr}{{\X},\pr}}{\X,\pr}=0$.  This suggests that the mean of the learned score function can be subtracted to eliminate any bias introduced by it. However, in our numerical experiments this mean subtraction had a negligible effect 
for moderate-size training sets (e.g., $\nds=600$.), and was not used. %
{
It may prove useful though} for situations with small $\nds$.     }} 
This results  in the learned Bayesian Fisher Information Matrix
\begin{equation}\label{eq:f_score}
\lbfimb\triangleq\expectation{\postscores{\pr}{\xset}\postscores{\pr}{\xset}^T}{\xset,\pr}.
\end{equation}
Since we cannot evaluate the expectation {in \eqref{eq:f_score}}, we replace the expectation with an empirical mean over the entire dataset $\mathcal{D}$ as shown in \eqref{eq:mean_fully}, 
producing $\lbfimbs$.
Finally, to obtain the LBCRB, we invert $\lbfimbs$, which results in ${\bcrb}\approx\lbcrbbs\triangleq\lbfimbs^{-1}$.

{The Posterior Approach addresses Problem~\ref{problem_one} in a simple way, requiring the training of only one model network for the posterior conditional score. However, this simplicity comes at the cost of two main drawbacks.} 
{First, because
}
the \emph{Posterior} Approach involves directly learning of $\nabla_{\p}\log\probt{\p|\xsetr}{\pr|\xset}$, it focuses only on the score of $\xsetr$ without considering that $\xsetr$ may consist of a set of i.i.d. measurements.  This 
requires learning  {a score model with a  conditioning input $\xsetr$ of high dimension $\niiddata$, increasing the complexity of the neural network and the sample complexity for training it}. {Second}, if one has {some domain}  knowledge about the problem, it is unclear how to incorporate it into the score function model. In Section~\ref{sec:lik_learn}, we propose the Measurement-Prior Approach, which
{ overcomes both of these limitations, while providing additional advantages.}
\subsection{Measurement-Prior Approach}\label{sec:lik_learn}


%
{
In this approach, illustrated
in Fig.~\ref{fig:main_lik_prior},
}
we employ the decomposition of the BFIM 
in \eqref{eq:bfim_decomposition_base} 
into two components: the prior Fisher Information Matrix (FIM), which requires only the score $\nabla_{\p}\log\probt{\p}{\pr}$, and the measurement FIM, which relies on the Fisher score $\nabla_{\p}\log\probt{{\x}|\p}{{\X}|\pr}$.
To learn of the prior score we use the standard score matching presented in Section~\ref{sec:background}, whereas for learning the measurement score, we derive a new Fisher Score Matching (FSM) objective.

{
The separation of the Bayesian score into two learned components 
provides several important} 
advantages, which we highlight here, {and discuss in detail later in this section.

}
First, it enables to determine
bounds 
{for a measurement with} 
an arbitrary number of 
independently and identically distributed 
i.i.d samples without requiring additional data or training.

{Second, it enables to learn and use a model for the Fisher score for only a \emph{single measurement sample}, reducing the complexity of the model and facilitating its training. 
Third, it enables to}
to incorporate domain knowledge about the estimation problem 
into the score neural network{, further reducing its complexity and facilitating its training.
}

\subsubsection{Score learning}\label{sec:score_learning_lik}
We  learn two distinct score functions, parametrized by $\paramf$ and $\paramp$, respectively: the Fisher score function $\lscore{\x}{\p; \paramf}$ for a single measurement sample;  and the prior score function $\priorscore{\p;\paramp}$.
We begin by learning the prior score $\priorscore{\p {;\paramp}}\approx\nabla_{\p}\log\probt{\p}{\pr}$ using the conventional score matching method detailed in Section~\ref{sec:score_over_view}. To apply the objective 
in \eqref{eq:loss_prior_sm}, 
we assume that 
the boundary condition 
\eqref{eq:boundary_conditions} 
and Assumptions~\ref{ass:score_reg_prior} hold  
for 
reasons analogous to {those given in Remark~\ref{remark:condtion_hold_post}} for the {posterior} score matching. 


Next we address the learning of the Fisher score function. We wish to minimize the mismatch between the true and model scores in the objective
{
\begin{align}
        &\lossfs\brackets{\paramf}\nonumber\\
        &=\expectation{\norm{\lscore{{\X}}{\p;\paramf}-\nabla_{\pr}\log\probt{{\X}|\pr}{{\X}|\pr}}_2^2}{{\X},\pr}.
\end{align}
}
However, in this case, we cannot use the standard {(conditional)} score matching {technique, because unlike the latter, where the gradient is with respect to the conditioned variable, the gradient in }
{$\nabla_{\p}\log\probt{{\x}|\p}{{\X}|\pr}$} {is with respect to the conditioning variable $\p$.} 
To {handle this fundamentally different scenario}, we 
{derive a new Fisher score matching objective. It requires the boundary condition \begin{equation}\label{eq:boundary_condtions_gen_direct}
    \lim\limits_{\p\rightarrow\Psb}  \lscore{{\x}}{\p}\probt{{\x},\p}{{\X},\p} =0,\quad \forall {\x}\in\widetilde{\Upsilon}
\end{equation} and the following regularity conditions.
}
\begin{assumption}[Fisher Score Matching Regularity]\label{ass:score_reg}$\,$
    \begin{enumerate}[label=\text{\ref{ass:score_reg}}.\arabic*,labelsep=*, leftmargin=*]
    \item The 
    {measurement {log-likelihood}} 
    {$\log\probt{{\x}|\p}{{\X}|\pr}$} is differentiable w.r.t. $\p$ {at all $\x\in\Upsilon$ and $\p\in\Ps$  where $\probt{{\x},\p}{{\X},\pr} > 0$} \label{assum:diff_prob}. 

    \item\label{ass:bound_expection} The expectation 
    {$\expectation{\norm{\nabla_{\pr}\log\probt{{\X}|\pr}{{\X}|\pr}}_2^2}{{\X},\pr}$} is finite. 
    \item The score neural network 
    {$\lscore{{\x}}{\p}$} is differentiable w.r.t. $\p$. \label{assum:diff_net} 
    \item\label{ass:bound_expection_v2} The expectation   
    {$\expectation{\norm{\lscore{{\X}}{\pr}}_2^2}{{\X},\pr}$} is finite. 
\end{enumerate}
\end{assumption}

\begin{remark}\label{remark:fisher_score}
Assumptions  \ref{assum:diff_net}, \ref{ass:bound_expection_v2}, and the boundary conditions  \eqref{eq:boundary_condtions_gen_direct} hold by the arguments 
in Remark~\ref{remark:condtion_hold_post}.  {As for Assumption~\ref{assum:diff_prob} it implied by Assumptions~\ref{sas:derivative} and \ref{sas:derivative_prior}.}

Similarly, by \eqref{eq:bfim_decomposition_base} and \eqref{eq:non-BayesFim}, 
Assumption~\ref{ass:bound_expection} is implied by the non-degeneracy BCRB regularity Assumption~\ref{assume:non_singular}.

\end{remark}
\begin{theorem}[Fisher Score Matching]\label{thm:liklihood} 
{Suppose that the boundary condition \eqref{eq:boundary_condtions_gen_direct} and the
}
regularity conditions Assumptions~\ref{ass:score_reg}) hold.
{Define}\begin{align}\label{eq:score_match_param_gen_direct}
    &\lossfst\brackets{\paramf}=2\expectation{\lscore{{\X}}{\pr;\paramf}^T\nabla_{\p}\log\probt{\pr}{\pr}}{{\X},\pr} +\\
    &2\trace{\expectation{\frac{\partial \lscore{\x}{\pr;\paramf}}{\partial\pr}}{\X,\pr}}+\expectation{\norm{\lscore{{\X}}{\pr;\paramf}}_2^2}{{\X},\pr}.\nonumber 
\end{align}
{Then $\lossfs=\lossfst+C$, where $C$ is a constant independent of $\paramf$.}
\end{theorem}
Theorem~\ref{thm:liklihood} is proved 
in Appendix~\ref{sec:lik_score_proof}. {Comparing the new Fisher Score Matching objective \eqref{eq:score_match_param_gen_direct}
with the standard conditional score matching objective (e.g., \eqref{eq:conditional_score_matching}), we notice that the difference is the added new first term in \eqref{eq:score_match_param_gen_direct}. This term involves the expectation of the inner product of the model conditional score of the conditioned variable $\X$ (i.e., the Fisher score $\lscore{{\X}}{\pr}$) with the true score of the conditioning variable $\pr$ (i.e, the true score of the prior, $\nabla_{\p}\log\probt{\p}{\pr}$).}

To employ Theorem~\ref{thm:liklihood}, 
requires the true prior score function. 
Instead, we substitute the prior score function with the learned version $\priorscore{\p}\approx \nabla_{\p}\log\probt{\p}{\pr}$. This allows us to leverage a dataset of i.i.d. measurements and parameter pairs to learn the Fisher score 
using the objective 
\eqref{eq:score_lik_mean}.
\subsubsection{\pe{} Score Neural Network}\label{sec:model_informed}
{The separation of the modeling and learning of the prior score from that of the Fisher (measurement) score in the Measurement-Prior Approach opens a new opportunity to} integrate domain-knowledge (e.g., knowledge of the physics of the measurement process) into the {Fisher score model. This reduces the NN model complexity, and improves sample complexity.}

{Specifically, } let us assume that we know some deterministic physical model $\mathcal{M}\brackets{\p}$ that relates the parameter vector $\p$ to the measurement $\X$ 
{in the following manner}:
\begin{equation}\label{eq:physical_model_score}
    \probt{{\x}|\p}{{\X}|\pr}=\probt{{\x}|\mathcal{M}\brackets{\p}}{{\X}|\vectorsym{\tau}},
\end{equation}
where $\probt{{\x}|\vectorsym{\tau}}{{\X}|\vectorsym{\tau}}$ is an unknown PDF 
parameterized by the known physical model $\mathcal{M}$. %
{
Such a representation can fit several signal processing problems as demonstrated in {Sec.~\ref{subsubsec:MoISNN} and further} in Sec.~\ref{sec:example_models}. }
The score function of the model presented in \eqref{eq:physical_model_score} is given by:
    $\nabla_{\p}\log\probt{{\x}|\pr}{{\X}|\p}=\divc{\mathcal{M}\brackets{\p}}{\p}^T\at{\nabla_{\vectorsym{\tau}}\log\probt{{\x}|\vectorsym{\tau}}{{\X}|\vectorsym{\tau}}}{\vectorsym{\tau}=\mathcal{M}\brackets{\p}}.$
Since $\probt{{\x}|\vectorsym{\tau}}{{\X}|\vectorsym{\tau}}$ is unknown we replace it 
by a neural network model $\iscore{\x}{\vectorsym{\tau}}$.  {The combined model for the Fisher score then becomes} \eqref{eq:model_base_score}.
The \pe{} Score 
Neural Network is illustrated 
in Figure~\ref{fig:model_inforamed}.
\begin{remark}\label{remark:opt_mi}
    To optimize the parameters %
    {$\paramf$ defining} $\vectorsym{s}_{I}$, we substitute \eqref{eq:model_base_score} into \eqref{eq:score_lik_mean}, implying that the optimization process relies on a dataset consisting of 
    {measurement-parameter ($\x,\p$) rather than $(\x,\vectorsym{\tau})$ pairs}. 
    Importantly, $\vectorsym{\tau}=\mathcal{M}\brackets{\p}$ is calculated during the optimization for every given value of $\p$.
\end{remark}
\subsubsection{Evaluation of the LBCRB}\label{sec:lbcrb_eval_lik}
The Prior Score and Fisher Score {models} learned in Section~\ref{sec:lbcrb_eval_lik} 
are used in the evaluation phase of the \emph{Measurement-Prior} Approach. %
{Specifically, let $\lscores{\x}{\p}=\lscore{\x}{\p;\paramf^*}\approx\nabla_{\p}\log\probt{\x|\p}{\X|\pr}$ and $\priorscores{\x}=\priorscore{\p;\paramp^*}\approx\nabla_{\p}\log\probt{\p}{\pr}$ be the learned Fisher and prior scores, where $\paramf^*$ and $\paramp$ denote the minimizer's of \eqref{eq:score_lik_mean} and \eqref{eq:score_prior_mean}, respectively.  }
We use the learned Prior and Fisher scores to {express} their FIM's 
\begin{subequations}\label{eq:split_fim_exp}
\begin{equation}
    \tlmfim\triangleq\expectation{\lscores{{\X}}{\pr}\lscores{{\X}}{\pr}^T}{{\X},\pr},
\end{equation}
\begin{equation}
\tlpfim\triangleq\expectation{\priorscores{\pr}\priorscores{\pr}^T}{\pr}.
\end{equation}
\end{subequations}
Substituting into the decomposition of the Bayesian FIM in \eqref{eq:bfim_decomposition_base} 
yields {an approximation for the Bayesian FIM for a measurement contatining $\niideval$ i.i.d samples,}
\begin{equation}\label{eq:f_score_decomposition}
    \fb\approx\lbfimlp\brackets{\niideval}\triangleq\niideval\cdot\tlmfim+\tlpfim.
\end{equation}
To compute \eqref{eq:split_fim_exp}, we replace the expectation with an empirical mean over the training dataset  $\ds$, 
in \eqref{eq:mean_efim_likd} and \eqref{eq:bprior_mean}. 
{Combining the two learned FIMs as in \eqref{eq:f_score_decomposition} yields
the final learned Bayesian FIM (LBFIM) \eqref{eq:bfim_apx_final} of the Measurement-Prior Approach,
with the corresponding LBCRB $\lbcrblps(\niideval)=\lbfimlps^{-1}(\niideval).$}
 
We would like to emphasize that \eqref{eq:f_score_decomposition} and \eqref{eq:bfim_apx_final} enable us to compute a {learned} BCRB for a measurement containing any desired number $\niideval$ of i.i.d. samples, {regardless of the number $\niiddata$ of i.i.d measurement samples available in the training data set $\ds$ for each value of $\p$.} 

{The explicit use of the statistical independence of the $\niideval$ measurement samples in the Measurement-Prior Approach provides two additional advantages:} (1) 
the bound generalizes to any number $\niideval$ of i.i.d. samples {in the measurement,} without  training a new model; and (2) {
the model $\lscore{\x}{\p;\paramf}$ for the score function for a single measurement sample is far simpler than that for multiple samples, making training easier and reducing the number of training samples required.}
 


%% file: files/tikz/apporch_compare.tex
\definecolor{mydarkblue}{RGB}{0,0,139} 

\begin{figure*}[ht]
    \centering
    \begin{subfigure}[t]{1.0\textwidth}
        \centering
        \input{files/tikz/fig_post}
        \caption{Posterior Approach}
        \label{fig:main_post}
    \end{subfigure}%
    
    \begin{subfigure}[t]{1.0\textwidth}
        \centering
        \input{files/tikz/fig_lik_prior}    
        \caption{Measurement-Prior Approach}
        \label{fig:main_lik_prior}
    \end{subfigure}
    
    \caption{ {
    \name{}: Overview. 
    Fig.~\ref{fig:main_post}: 
    Posterior Approach. 
    The learning phase consists of a single training step for the posterior score; this neural net model is subsequently used in the evaluation phase. 
    Fig.~\ref{fig:main_lik_prior}: Measurement-Prior Approach. 
    The learning phase comprises two consecutive training steps: 
    (i) training the prior score; and (ii)} training the measurement score—yielding two distinct NN models: the Prior and the Fisher scores. These models are then applied in the evaluation phase. }
    \label{fig:main}
\end{figure*}

%% file: files/tikz/fig_post.tex
     \begin{tikzpicture}[font=\small,every label/.append style={font=\small,align=center}]



        \node[startstop,fill=yellow!15, text width=5.0cm,align=center,yshift=0cm] (loss_post) { Conditional Score Matching \eqref{eq:score_post_mean} \\
       $\Omega^* =\arg\min\limits_{{\Omega_{}}}\lossbsm\brackets{{\Omega}}$};


        \node[right=0.3 cm of loss_post, startstop,fill=green!15, text width=5.5cm,align=center,yshift=+3pt] (post_score) { Posterior Score \\$\nabla_{\p}\log\probt{\p|\x}{\pr|\X} \approx \postscore{\p}{\x;\Omega^*}$};
        


        
        \node[right=0.3cm of post_score,startstop,fill=white, text width=2cm,align=center,fill=red!15] (eval_bfim_post) {   Compute $\lbfimbs$\\ using \eqref{eq:mean_fully}};

        \node[right=0.3cm of eval_bfim_post,startstop,fill=white, text width=2.5cm,align=center,fill=red!15] (inv_post) {   Invert $\lbfimbs$\\ $\lbcrbbs\triangleq\lbfimbs^{-1}$ };
        
        \begin{pgfonlayer}{background}



            \node[draw,rounded corners,fit= (post_score), fill=white,inner sep=2pt,label={Score Model}](fit11){};

            \node[draw,rounded corners,fit= (loss_post), fill=white,inner sep=2pt,label={Learning Stage}](fit12){};

                        \node[draw,rounded corners,fit= (inv_post)  (eval_bfim_post), fill=white,inner sep=2pt,label=Evaluation Stage](fit_eval2){};
            
        \end{pgfonlayer}

        


        \draw[->,line width=0.25mm] (loss_post)  -- ( post_score);
        \draw[->,line width=0.25mm] (post_score)  -- (eval_bfim_post);
        \draw[->,line width=0.25mm] (eval_bfim_post)  -- (inv_post);
\end{tikzpicture} 

%% file: files/tikz/fig_lik_prior.tex
     \begin{tikzpicture}[font=\small,every label/.append style={font=\small,align=center}]

        \node[startstop,fill=white, text width=5cm,align=center,yshift=-0.0cm,fill=yellow!15] (loss_prior) {  Prior Score Matching \eqref{eq:score_prior_mean}\\ 
        $\paramp^* =\arg\min\limits_{{\paramp}}\losspsm\brackets{{\paramp}}$};

        \node[below=0.4cm of loss_prior,startstop,fill=cyan, text width=5cm,align=center,yshift=0cm] (loss_dmle) { Fisher Score Matching \eqref{eq:score_lik_mean} \\
        $\paramf^* =\arg\min\limits_{{\paramf}}\lossfsm\brackets{{\paramf};\paramp^*}$};


        \node[right=0.3 cm of loss_prior, startstop,fill=green!15, text width=5.5cm,align=center,yshift=+3pt] (prior_score) { Prior Score \\$\nabla_{\p}\log\probt{\p}{\pr}\approx \priorscore{\p;\paramp^*}$};
		\node[ right=0.3cm of loss_dmle,startstop,fill=green!15, text width=5.5cm,align=center,yshift=+3pt] (dmle) { Fisher Score\\ $\nabla_{\p}\log\probt{\x|\p}{\X|\pr}\approx$ $\lscore{\x}{\p;\paramf^*}$};

        
        \node[right=0.3cm of prior_score,startstop,fill=white, text width=2cm,align=center,fill=red!15] (eval_prior) {  Compute $\lpfim$\\ using \eqref{eq:bprior_mean} };

        \node[right=0.3cm of dmle,startstop,fill=white,fill=red!15 , text width=2cm,align=center] (eval_fim) {   Compute $\lmfim$\\ using \eqref{eq:mean_efim_likd}};

                \node[right=0.3cm of eval_prior,startstop,fill=white, text width=3.5cm,align=center,yshift=-4pt,fill=red!15] (eval_bfim) {  Compute the LBFIM \\
                $\lbfimlps =\niideval\cdot  \lmfim+ \lpfim$\\
                and Invert LBFIM to obtain LBCRB\\
                $\lbcrblps=\lbfimlps^{-1}$
                };
        

        
        \begin{pgfonlayer}{background}
            \node[draw,rounded corners,fit= (prior_score) (dmle), fill=white,inner sep=2pt,label={Score Models}](fit1){};

            \node[draw,rounded corners,fit= (loss_dmle) (loss_prior), fill=white,inner sep=2pt,label={Learning Stage}](fit2){};

            \node[draw,rounded corners,fit= (eval_bfim) (eval_fim)  (eval_prior), fill=white,inner sep=2pt,label=Evaluation Stage](fit_eval){};



            
        \end{pgfonlayer}

        

        \draw[->,line width=0.25mm] (loss_prior)  -- (prior_score);
        \draw[->,dashed,line width=0.25mm,color=red,label=$\Omega_p^*$] (loss_prior)  -- node[left] {$\Omega_p^*$} (loss_dmle);
        \draw[->,line width=0.25mm] (loss_dmle)  -- (dmle);
        \draw[->,line width=0.25mm] (dmle)  -- (eval_fim);
        \draw[->,line width=0.25mm] (prior_score)  -- (eval_prior);
        \draw[->,line width=0.25mm] (eval_prior)  -- (eval_bfim);
        \draw[->,line width=0.25mm] (eval_fim)  -| (eval_bfim);

\end{tikzpicture} 

%% file: files/theortical_results.tex
\section{Theoretical Results}\label{sec:theory}
Here, we 
investigate the errors of the \name{} due to learning error. %
{Following the standard approach in } learning theory \cite[Chapter 5]{shalev2014understanding}, we divide the learning error into two components: the \emph{approximation error} and the \emph{empirical-mean error}\footnote{In {learning theory} \cite[Chapter 5]{shalev2014understanding}, the empirical mean error is 
known as the "estimation error". However, to avoid confusion with the estimator error 
in \eqref{eq:bcrb}, we 
refer to it here as the "empirical-mean error".} . We then examine each of these errors individually and their collective impact.
 In the last part, we demonstrate that the {learned score neural network models are strongly consistent approximations of}  
 the true scores, and that \name{} converges {with probability 1 to the BCRB as the size of the training data set $\ds$ increases,} providing conditions to achieve an accurate approximation.


{The results} in this section
use the  concept of \emph{intrinsic dimension of a matrix}\cite{tropp2015introduction,ipsen2024stable} as a measure of its {\emph{effective}} dimension: 
\begin{definition}[Intrinsic Dimension of a Matrix]
    Let $\matsym{A}\in\mathbb{R}^{d\times d}$ be a non-zero positive semi-definite square matrix. Then its intrinsic dimension is defined as:
    \begin{equation*}
        \mathrm{intdim}\brackets{\matsym{A}}\triangleq\frac{\trace{\matsym{A}}}{\norm{\matsym{A}}_2}.
    \end{equation*}
\end{definition}
The intrinsic dimension of a matrix can be understood as quantifying the number of dimensions
{by accounting for} the spectral intensity over all dimensions. 
From 
the definition of intrinsic dimension and the inequality $1\leq \mathrm{intdim}\brackets{\matsym{A}} \leq \mathrm{rank}\brackets{\matsym{A}}\leq d,$
we observe that $\mathrm{intdim}\brackets{\matsym{A}}=d$ when all 
eigenvalues are identical and nonzero, whereas $\mathrm{intdim}\brackets{\matsym{A}}=1$ when $\matsym{A}$ has rank one. In addition, the intrinsic dimension is more robust to small perturbations than the matrix rank \cite{ipsen2024stable}. 

\subsection{Approximation error}
{We evaluate the approximation error of both approaches, demonstrating that the 
}
score matching objective {that expresses the score mismatch} provides an upper bound on the {learned} Fisher Information Matrix (FIM) errors.
\begin{theorem}[\name{} %
{Approximation Error:} 
{Posterior} Approach]\label{thm:lrn:direct}
    Let $\mathrm{sRE}_B\triangleq\frac{\lossbs}{\trace{\fb}}$ be the %
    {relative %
    {approximation error} of the} Bayesian score 
    {and let $\db=\mathrm{intdim}\brackets{\F_B}$.} Suppose that Assumptions~\ref{assum:bcrb_reg}, \ref{ass:score_cond_reg} hold and $\db{\cdot \mathrm{sRE}_B}\leq {0.16}$. Then:
    \begin{align}\label{eq:error_learn_post_approch}
        {\mathrm{RE}_{B}^{(a)}}=\frac{\norm{\lbfimb-\fb}_2}{\norm{\fb}_2} &\leq\bfle \triangleq {2.4}\sqrt{\db\cdot\mathrm{sRE}_B}.
    \end{align}
\end{theorem}

\begin{theorem}[\name{} %
{Approximation Error}: Measurement-Prior Approach]\label{thm:lrn:lik_prior}
    Let $\mathrm{sRE}_M\triangleq\frac{\lossfs}{\trace{\fm}}$ and $\mathrm{sRE}_P\triangleq\frac{\lossps}{\trace{\fp}}$ be the {relative errors in learning the Fisher and prior scores}, respectively and let $\dm=\mathrm{intdim}\brackets{\fm}$ and $\intdp=\mathrm{intdim}\brackets{\fm}$. 
    Suppose that Assumptions~\ref{assum:bcrb_reg}, \ref{ass:score_reg_prior}, and \ref{ass:score_reg} hold and ${\dm\cdot \mathrm{sRE}_M}\leq {0.16}$, ${\intdp\cdot \mathrm{sRE}_P}\leq {0.16}$. 
    Then:
    \begin{align}\label{eq:error_learn_lik_prior_approch}
        &%
        {\mathrm{RE}_{MP}^{(a)}}=\frac{\norm{\lbfimlp-\fb}_2}{\norm{\fb}_2} \leq  \mpfle\\
    &\mpfle\triangleq2.4\brackets{\frac{\norm{%
    {\niideval}\fm}_2}{\norm{\fb}_2}\sqrt{\dm\cdot\mathrm{sRE}_M}+\frac{\norm{\fp}_2}{\norm{\fb}_2}\sqrt{ \intdp\cdot\mathrm{sRE}_P}}\nonumber
    \end{align}
\end{theorem}

Theorems~\ref{thm:lrn:direct} and \ref{thm:lrn:lik_prior} %
{(proved in Appendix C)} 
{quantify the} relationship between the %
{score matching} optimization objective, %
{(or equivalently, the relative error in %
{approximating} the score vector),} and the  %
{approximation error} 
{of the corresponding FIM}. They 
{show how a} reduction in the objective (smaller score %
{approximation error}) 
{translates to} a smaller learning error %
{for the FIM}.  
Moreover, the tightest bound for a selected neural network architecture is achieved when %
{$\Omega
=\Omega^*$ --  the NN parameters that minimize the objective $\mathcal{L}$.} 
Finally, the assumptions $\mathrm{intdim}\brackets{\fb}\frac{\lossbs}{\norm{\fb}_2}\leq 0.16$, $\mathrm{intdim}\brackets{\fm}\frac{\lossfs}{\norm{\fm}_2}\leq 0.16$ and $\mathrm{intdim}\brackets{\fp}\frac{\lossps}{\trace{\fp}}\leq 0.16$ 
introduced in Theorems~\ref{thm:lrn:direct} and \ref{thm:lrn:lik_prior} can be satisfied  by minimizing the score 
{objectives} $\lossbs, \lossfs, \text{and } \lossps$.

{Comparing between the theoretical results on the {approximation error} of the two approaches reveals 
that the Measurement-Prior Approach will 
have an advantage in cases where the measurement FIM is the dominant part, and some additional information can be used in the Fisher (Measurement) score model,
e.g. the reuse of i.i.d. samples,  or a model-informed score neural network. This is also demonstrated numerically in Fig.~\ref{fig:k_study} and Fig.~\ref{fig:analysis_mi}.    }

\subsection{Empirical Mean Error}\label{sec:empirical_mean_error}
In this section, we discuss the error introduced by using a finite empirical mean in \eqref{eq:mean_fully}, \eqref{eq:mean_efim_likd}, and \eqref{eq:bprior_mean} instead of the true expectation. We 
assume that the 
learned score function is bounded for any sample in the dataset $\ds$, namely:
\begin{assumption}[
{Score Function Bounds}]\label{assume:bounded_score}
For $\Omega^*$, $\paramp^*$ and $\paramf^*$ we assume that:
\begin{enumerate}[label=\text{\ref{assume:bounded_score}}.\arabic*,labelsep=*, leftmargin=*]
    \item\label{assume:bound_post} $\norm{\postscores{\p_n}{\xsetr_n}}_2^2\leq \cb \quad \forall \xsetr_n,\p_n \in \ds$.
    \item \label{assume:bound_fisher}$\norm{\frac{1}{\niiddata}\sum_{\x\in\xsetr_n}\lscores{\x}{\p_n}\lscores{\x}{\p_n}^T}_2\leq \cm \quad \forall  \xsetr_n,\p_n  \in \ds $ .
\item\label{assume:bound_prior} $\norm{\priorscores{\p_n}}_2^2\leq \cp \quad \forall  \p_n  \in \ds$.
\end{enumerate}
\end{assumption}

{\begin{remark}
The bounds $\cb, \cm$, and $\cp$ are determined for the specific realization in $\ds$, \emph{after it has been observed.} {Hence these bounds are functions of $\ds$, $\cb=\cb(\ds), \cm=\cm(\ds), \cp=\cp(\ds)$, but this dependence will not be shown explicitly.}
\end{remark}
}

Assumptions~\ref{assume:bounded_score} can be inherently satisfied by selecting an appropriate neural network architecture and nonlinear activation function and by the assumption that all samples in finite size set $\ds$ are bounded as well. {Subject to these assumption, we obtain the following results.}

\begin{theorem}[LBCRB Empirical Mean Error: Posterior Approach]\label{thm:sampling_post}
{Suppose that Assumptions~\ref{assum:bcrb_reg}, \ref{ass:score_cond_reg}, and the score bound 
condition~\ref{assume:bound_post} hold.} 
    Let $\dbb=\mathrm{intdim}\brackets{\lbfimb}$,
    %
    {and define 
    $\nbe\triangleq\frac{4}{3}\brackets{u+\log\brackets{8 \dbb}}\brackets{\frac{\cb}{\trace{\lbfimb}}\cdot \dbb+1}$. 
    }
    Then for any $u>0$ 
    {and $\nds \geq \nbe$} the following bound holds with probability %
    {of at least } $1-\exp{\brackets{-u}}$:
    \begin{align}\label{eq:re_error_post_mean}
        {\mathrm{RE}_{B}^{(e)}}=&\tfrac{\norm{\lbfimbs-\lbfimb}_2}{\norm{\lbfimb}_2}\leq \bfse\triangleq 
        {1.5\sqrt{\frac{\nbe}{\nds}}}.
    \end{align}
    Furthermore, %
    {define
    $\nbet=\brackets{\log\brackets{1+2\dbb} +{0.52}} \cdot\brackets{\dbb\frac{\cb}{\trace{\lbfimb}}+1}$. Then for any $\nds\geq\nbet$ the relative empirical mean error is bounded in expectation by
    }
     \begin{align}\label{eq:emprical_mean_error_post_expection}
        \expectation{\mathrm{RE}_{B}^{(e)}}{\ds}\leq
        {\frac{6+1.5\sqrt{2\log\brackets{1+2\dbb}}}{\sqrt{\log\brackets{1+2\dbb} +{0.52}}
        }}\sqrt{\frac{\nbet}{\nds}}.
    \end{align}
\end{theorem}

\begin{theorem}[LBCRB Empirical Mean Error: Measurement-Prior Approach]\label{thm:sampling_mp}
    Suppose that  Assumptions~\ref{assum:bcrb_reg}, \ref{ass:score_reg_prior}, \ref{ass:score_reg} and the bounded score conditions~\ref{assume:bound_fisher} and \ref{assume:bound_prior} hold.
    Let %
    { $\dmpb=\mathrm{intdim}\brackets{\lbfimlp}$, and define $\nmpe\triangleq \frac{4}{3}\brackets{u+\log\brackets{8 \dmpb}}\brackets{\frac{ \cm+\frac{\cp}{\niideval}}{\trace{\tlmfim}+\frac{\trace{\tlpfim}}{\niideval}}\cdot \dmpb+1}$. }
     Then for any $u>0$ and $\nds\geq \nmpe$ the following bound holds with probability {of at least } $1-\exp{\brackets{-u}}$:
    \begin{align}\label{eq:re_error_mp_mean}
        {\mathrm{RE}_{MP}^{(e)}}=&\frac{\norm{\lbfimlps-\lbfimlp}_2}{\norm{\lbfimlp}_2}\leq \mpfse\triangleq {1.5}\sqrt{\frac{\nmpe}{\nds}}. 
    \end{align}
    Furthermore, define $\nmpet\triangleq\brackets{\log\brackets{1+2\dmpb} +{0.52}}\cdot\brackets{\dmpb\frac{ \cm+\frac{\cp}{\niideval}}{\trace{\tlmfim+\frac{\tlpfim}{\niideval}}}+1}$. Then %
    {for any $\nds\geq \nmpet$ the relative empirical mean error is bounded in expectation by}
     \begin{align}\label{eq:emprical_mean_error_mp_expection}
        &\expectation{\mathrm{RE}_{MP}^{(e)}}{\ds}\leq \frac{6+1.5\sqrt{2\log\brackets{1+2\dmpb}}}{\sqrt{\log\brackets{1+2\dmpb} +{0.52}}}\sqrt{\frac{\nmpet}{\nds}}.
    \end{align}
\end{theorem}
{Theorems \ref{thm:sampling_post} and \ref{thm:sampling_mp}, proved in Appendix~\ref{apx:proof_sample_error}, quantify, with explicit constants (and for sufficiently large $u$, with overwhelming probability) the effect of the size $\nds$ of the training set $\ds$ on the error in the estimated FIMs due to replacing expected values by finite means over $\ds$.  
We also observe
the effect of the intrinsic dimension of the respective FIM (which is upper bounded by $\np$):  as the effective number of 
parameters to estimate grows, 
more training samples $\nds$ are required 
to obtain the same FIM error, with the dependence being (up to log factors ) roughly linear in $\dbb$ or $\dmpb$.

{
Recall from Sec.~\ref{subsec:MP} that an advantage of the Measurement-Prior Approach 
over the Posterior Approach is that it is able to determine the LBCRB for any number $\niideval$ of i.i.d. samples using the same $\ds$, without any additional effort or data.
Another advantage of the Measurement-Prior Approach is revealed by a close comparison
of Thm.~\ref{thm:sampling_post} with Thm.~\ref{thm:sampling_mp}:
the Measurement-Prior Approach enjoys
a reduced 
finite mean}
error for datasets $\ds$ consisting of measurements with multiple $\niiddata > 1$ i.i.d. samples per parameter value.
}

{The reason for this behavior is that, as we argue below, with high probability, $\cb\geq \niiddata \cm +\cp$. Because $\lbfimlp\approx \fb \approx \lbfimb $, we also have $\dbb \approx \dmpb$. Toghether, this implies that $\nbe \geq \niiddata \nmpe$, and in turn
$\bfse \geq \sqrt{\niiddata}\mpfse$, i.e, the upper bound in \eqref{eq:re_error_mp_mean} on the empirical-mean error for the Measurement-Posterior Approach, is, with high probability, smaller by a factor of $\sqrt{\niiddata}$ than the corresponding bound in \eqref{eq:re_error_post_mean} for the Posterior Approach. 

To justify the claim that with high probability, $\cb\geq \niiddata \cm +\cp$, 
recall the definitions of $\cb, \cm$, and $\cp$ in Assumption~\ref{assume:bounded_score} as the maxima over $\ds$ of the expressions on the left hand sides of the respective definitions. To compare these quantities, we first compare their expected values by} proving in Appendix~\ref{apx:remark_c_relation_proof} the following relation between the true score functions.
\begin{prop} \label{prop:expected_score_ineq}
    \begin{align}
&\expectation{\norm{\nabla_{\p}\log\probt{\p|\xsetr}{\p|\xset}}_2^2}{}\geq\expectation{\norm{\nabla_{\p}\log\probt{\p}{\pr}}_2^2}{}+\nonumber\\
    &\niiddata \expectation{\norm{\frac{1}{\niiddata}\sum_{i=1}^{\niiddata}\nabla_{\p}\log\probt{\x_i|\p}{\X|\pr}\nabla_{\p}\log\probt{\x_i|\p}{\X|\pr}^T}_2}{}.\nonumber
\end{align}
\end{prop}
{Now, assuming that the learning of the score functions was successful and the score mismatches $\mathrm{sRE}_B, \mathrm{sRE}_M$ and $\mathrm{sRE}_P$ defined in Theorem~\ref{thm:lrn:direct} and Theorem~\ref{thm:lrn:lik_prior} are small, the same inequality as in Proposition~\ref{prop:expected_score_ineq} will apply to the respective learned scores $\postscores{\p_n}{\xsetr_n},\lscores{\x}{\p_n}$ and $\priorscores{\p_n}$. For $\niideval>1$, the large gap between the expected values suggests also a similar large gap between the maxima, implying the claim that $\cb\geq \niiddata \cm +\cp$ with high probability. 

This analysis is supported by the empirical results in Fig.~\ref{fig:sample_error_vs_n_samples}, which show a consistently lower empirical-mean errors for the Measurement-Prior Approach than for the Posterior Approach, but in terms of the bounds, and the actual errors.
 }

{In 
Section~\ref{subsec:consistency} we 
show, using Theorems~\ref{thm:sampling_post} and \ref{thm:sampling_mp},
that the empirical mean error vanishes as the size $\nds$ of the training data set grows.

\subsection{{\name{}} Learning Error
}
\label{subsec:Conv_Invert}
{
}
{
To obtain bounds on the deviation $\hat{\bcrb} -\bcrb $ of the \name{} from the BCRB due to learning error, we combine the effects of the approximation error (Theorems \ref{thm:lrn:direct} and \ref{thm:lrn:lik_prior}) and the empirical-mean error (Theorems \ref{thm:sampling_post} and \ref{thm:sampling_mp}, respectively) on the learned FIM, and propagate them through the inverse relation $\lbcrbbs = \lbfimbs^{-1} $ and $\lbcrblps = \lbfimlps^{-1}$. For the latter, we include conditions for the learned FIMs $\lbfimbs$ and $\lbfimlps$ to remain positive definite in the presence of learning errors, to ensure that its inverse, the \name, exists. This is expressed in the following corollaries, proved in Appendix~\ref{apx:proof_inv_re}.
}
\begin{corollary}\label{corr:bound_inv_post}
Suppose that the assumptions of Theorems  {\ref{thm:lrn:direct} and} \ref{thm:sampling_post}  hold {and 
\begin{equation}\label{eq:inv_cond_post}
    \frac{\norm{\lbfimb}_2}{\norm{\fb}_2}\bfse+\bfle <\frac{1}{\kappa\brackets{\fb}}.
\end{equation}
Then }
{$\lbfimbs\succ 0$} {(i.e., 
$\lbcrbbs$ exists)}, 
{and} the following holds with probability {of at least} $1-\exp\brackets{-u}$:
\begin{align}\label{eq:inv_bound_post}
    \mathrm{RE}_{B} {\triangleq
    \frac{\norm{\lbcrbbs-\bcrb}_2}{\norm{
    \bcrb
    }_2}}
\leq\kappa\brackets{\lbfimbs}\brackets{\frac{\norm{\lbfimb}_2}{\norm{\lbfimbs}_2}\bfse+\frac{\norm{\fb}_2}{\norm{\lbfimbs}_2}\bfle}
\end{align}
\end{corollary}

\begin{corollary}\label{corr:bound_inv_split}
Suppose that 
the assumptions of Theorems {\ref{thm:lrn:lik_prior} and} \ref{thm:sampling_mp} hold,
{and 
\begin{equation}\label{eq:inv_cond_mp}
    \frac{\norm{\lbfimlp}_2}{\norm{\fb}_2} \mpfse +\mpfle<\frac{1}{\kappa\brackets{\fb}}.
\end{equation}
Then,  
}
{$\lmfim\succ 0$ and $\lpfim\succ 0$} {(i.e., 
$\lbcrblps$ exists)}, {and} the following
holds with probability {of at least} $1-\exp\brackets{-u}$:
\begin{align}\label{eq:inv_bound_mp}
\mathrm{RE}_{MP} & \triangleq\frac{\norm{\lbcrblps-\bcrb}_2}{\norm{\bcrb}_2} \nonumber \\
&\leq\kappa\brackets{\lbfimlps}\brackets{\frac{\norm{\lbfimlp}_2}{\norm{\lbfimlps}_2} \mpfse+\frac{\norm{\fb}_2}{\norm{\lbfimlps}_2}\mpfle}.
\end{align}
\end{corollary}
%
{Corollaries \ref{corr:bound_inv_post} and \ref{corr:bound_inv_split} admit a simplified interpretation to first order in $\eta^{(e)}$ and $\eta^{(a)}$. Assuming small learning errors, we have $\lbfimlps \approx \lbfimlp \approx\fb \approx\lbfimb  \approx  \lbfimbs$. The bounds in \eqref{eq:inv_bound_post} and \eqref{eq:inv_bound_mp} then simplify, respectively, to
\begin{equation}
   \mathrm{RE}_{B} \leq\kappa\brackets{\bcrb}\brackets{
    \bfse+\bfle} 
\end{equation}
\begin{equation}
   \mathrm{RE}_{MP} \leq\kappa\brackets{\bcrb}\brackets{
    \mpfse+\mpfle}, 
\end{equation}
showing that for both methods, the respective relative empirical-mean and approximation errors essentially add up, with scaling by the condition number $\kappa\brackets{\bcrb} = \kappa\brackets{\fb}$ of the BCRB. 
}

{Conditions \eqref{eq:inv_cond_post} and \eqref{eq:inv_cond_mp} for positive-definiteness of the 
learned FIMs in Corollaries~\ref{corr:bound_inv_post} and \ref{corr:bound_inv_split} are sufficient conditions, but they are the weakest such conditions: if violated, then there may exist learning errors such that the learned FIM is not invertible. In this sense, they are necessary to be able to guarantee the existence of the \name{}.}

{These conditions 
can be further detailed into 
specific criteria concerning the empirical-mean error and the 
approximation error}.
{Assuming small approximation errors, $\lbfimb \approx \fb \approx \lbfimlp$}
\eqref{eq:inv_cond_post} and \eqref{eq:inv_cond_mp} imply that the minimum required sample size for the Posterior Approach is :
{
\begin{equation*}
    {\nds}>  2.25\kappa^2\brackets{\fb}
    %
    %
    \nbe,
\end{equation*}
}
}
and for the Measurement-Prior Approach:
\begin{equation*}
    \nds>   2.25\kappa^2\brackets{\fb}\nmpe.
\end{equation*}
{
}
Note that both requirements on $\nds$ are quite sensitive to the condition number of the BCRB $\bcrb$: larger condition number will require more training samples.
If these conditions on the training set size are not met, $\lbfimb$ and $\lbfimlp$ 
{may} 
be non-invertible. 
{Furthermore, by the discussion comparing Theorems Theorems \ref{thm:sampling_post} and \ref{thm:sampling_mp} in Section~\ref{sec:empirical_mean_error}, we  typically have $\nbe \geq \niiddata \nmpe$. Combining with the conditions above implies that the {Measurement-Prior} Approach will require smaller $\nds$ when $\niiddata>1$.}

{Turning to the approximation error, Conditions \eqref{eq:inv_cond_post} and \eqref{eq:inv_cond_mp} imply
$\bfle, \mpfle\leq \frac{1}{\kappa\brackets{\fb}} = \frac{1}{\kappa\brackets{\bcrb}}$,
i.e., that the relative approximation errors must be kept below a threshold determined by $\kappa\brackets{\bcrb}$. 
}



\subsection{ Consistency \& Convergence}
\label{subsec:consistency}
{We consider}  
the convergence of \name{} to the BCRB as number {$\nds$ of training} samples goes to infinity. 
{
First we}
show that the empirical-mean error converges to zero as {$\nds \rightarrow \infty$.} Then we prove that a neural network trained using score matching is a consistent estimator of 
the true score function. Finally, 
{combining these results with} Corollaries \ref{corr:bound_inv_post} and \ref{corr:bound_inv_split}, we show the convergence of the \name{} .


%
{The convergence of the empirical-mean error follows} directly from
Theorems~\ref{thm:sampling_post} and \ref{thm:sampling_mp} subject to the additional assumption $\frac{\cb}{\nds}, \frac{\cm}{\nds}, \frac{\cp}{\nds}\xrightarrow{\nds\rightarrow\infty}0
{\quad \text{a.s.}}$, as summarized in the following results.

\begin{corollary}[LBCRB Empirical-Mean Error Convergence: Posterior Approach]\label{corr:sampling_post} 
Suppose
that the assumptions of Theorem~\ref{thm:sampling_post} hold and that $\frac{\cb}{\nds}\xrightarrow{\nds\rightarrow\infty}0 
{\quad \text{a.s.}}$, then:
\begin{equation}
\mathrm{RE}_B^{(e)}\xrightarrow{\nds\rightarrow\infty}0 \quad\text{a.s.}
\end{equation}
    
\end{corollary}

\begin{corollary}[LBCRB Empirical-Mean Error Convergence: Measurement-Prior Approach]\label{corr:sampling_mp} 
Suppose that the assumptions of Theorem~\ref{thm:sampling_mp} hold and that
$
{\frac{\cm}{\nds}, \frac{\cp}{\nds}}\xrightarrow{\nds\rightarrow\infty}0 %
{\quad \text{a.s.}}$, then:
\begin{equation}
\mathrm{RE}_{MP}^{(e)}\xrightarrow{\nds\rightarrow\infty}0 \quad\text{a.s.}
\end{equation}
    
\end{corollary}

From Corollaries~\ref{corr:sampling_post} and \ref{corr:sampling_mp}, we observe that as the number of samples $\nds$
increases, the %
{empirical}-mean error is reduced%
{, vanishing with probability 1 in the limit of infinite training data}.
 \begin{remark}\label{remark:mean_error_convergance}
    The assumption  $\frac{\cb}{\nds}\xrightarrow{\nds\rightarrow\infty}0 {\quad \text{a.s.}}$ in Corrolary~\ref{corr:sampling_post} is 
    {rather mild. 
    It means that the maximum over the data set $\ds$ of the squared norm of the learned score grows at a slower rate than $|\ds|=\nds$. For example, 
    {as we show in Appendix~\ref{apx:remark_convergance_proof}}, it holds if $\postscores{\pr_n}{\xsetr_n}$ has finite moments up to 6th order. In particular, this requirement is less demanding than the assumption made in \cite{crafts2023bayesian} that the score vector is
        Sub-Gaussian. Indeed, a Sub-Gaussian r.v. has finite moments of any order, rather than just up to 6th order.} Similar comments apply to the assumption about $\cp$ and $\cm$ in Corollary~\ref{corr:sampling_mp}.
\end{remark}

{Next we address the approximation error. We will need the following assumptions about the score neural networks. The first set of assumptions, of realizability of the true score functions by the chosen NN model architectures, is the more fundamental.}
\begin{assumption}[{NN-}Realizable Score Function]\label{assume:realizable}
Let {$\mathcal{S}_B$, $\mathcal{S}_F$ and $\mathcal{S}_P$} be the sets of all score functions representable by the  architectures of the neural networks chosen for the Bayesian, Fisher and prior scores, respectively. {Define the shorthand notation for the true score functions $\vectorsym{s}^{0}_{B}\brackets{\p|\xsetr} \triangleq\nabla_{\p}\log\probt{\p|\xsetr}{ \p | \xset}$,}  $\vectorsym{s}^{0}_{F}\brackets{\x|\p} \triangleq\nabla_{\p}\log\probt{\x|\p}{\X|\p}$ and $\vectorsym{s}^{0}_{P}\brackets{\p}\triangleq\nabla_{\p}\log\probt{\p}{\p}$. Then assume that {all three true scores are realizable {by the chosen NN architectures,} i.e.}:
{$\vectorsym{s}^{0}_\alpha \in \mathcal{S}_\alpha$, $\alpha= B, F, P$.}
\end{assumption}

We argue that Assumption~\ref{assume:realizable} holds {(or can be approximated arbitrarily well)} in the regime of a neural networks with sufficient capacity. {In terms of our parametrized score neural network notation, this means, e.g., for the Bayesian score, that $\vectorsym{s}_{B}\brackets{\p|\xsetr; \Omega} = \vectorsym{s}^{0}_{B}\brackets{\p|\xsetr}$ for some $\Omega=\Omega^{*}$. (Unlike in \cite{hyvarinen2005estimation}, the presence of multiple such alternative $\Omega^*$ is not a concern in our application, and may in fact facilitate the optimization step of score matching.)} 

{The second set of assumptions, of Lipschitz continuity, are more technical regularity assumptions.}
They can be readily satisfied by choice of neural network architecture and training strategy (e.g. $\ell_2$ regularization).

\begin{assumption}[Lipschitz Continuity]\label{assume:lip_cont}
For each of the following three alternative settings, 
\begin{enumerate}[label=\text{\ref{assume:lip_cont}}.\arabic*,labelsep=*, leftmargin=*]
    \item $\alpha=B$, $\, f\brackets{\z;\Omega}=\postscore{\p}{\x;\Omega}$, $\, \z=[\p,\x]$, 
    $\, \randomvec{Z}=[\pr,\X]$, \label{assume:lip_cont_post}
    
    \item  $\alpha=P$, $\, f\brackets{\z;\Omega}=\priorscore{\p;\Omega}$, $\, \z=\p$, 
    $\, \randomvec{Z}=\pr$, \label{assume:lip_cont_prior}
    
    \item  $\alpha=F$, $f\brackets{\z;\Omega}=\lscore{\x}{\p;\Omega}$, $\z=[\p,\x]$, 
    $\randomvec{Z}=[\pr,\X]$.\label{assume:lip_cont_fisher}
    
\end{enumerate}

assume the following holds.\\
Functions $f_{\alpha}\brackets{\z;\Omega_{}}$, $\divc{f_{\alpha}\brackets{\z;\Omega_{}}}{\p}$ and $f_{\alpha}\brackets{\z;\Omega_{}}f_{\alpha}\brackets{\z;\Omega_{}}^T$ are Lipschitz continuous {with Lipschitz constants finite in expectation,} 
namely, for any {$\Omega_{1}, \Omega_{2} \in \mathcal{S}_{\alpha}$ {with $\Delta\Omega=\norm{\Omega_{1}-\Omega_{2}}_2$}:} \\
   1) 
  $\norm{f_{\alpha}\brackets{\z;\Omega_{1}}-f_{\alpha}\brackets{\z;\Omega_{2}}}_2\leq \xi_{\alpha}\brackets{\z}\Delta\Omega$,\\
    2) $\norm{\divc{f_{\alpha}\brackets{\z;\Omega_{1}}}{\p}-\divc{f_{\alpha}\brackets{\z;\Omega_{2}}}{\p}}_2\leq \tau_{\alpha}\brackets{\z}\Delta\Omega$,\\
    3) $\norm{f_{\alpha}\brackets{\z;\Omega_{1}}f_{\alpha}\brackets{\z;\Omega_{1}}^T-f_{\alpha}\brackets{\z;\Omega_{2}}f_{\alpha}\brackets{\z;\Omega_{2}}^T}_F\leq \zeta_{\alpha}\brackets{\z}\Delta\Omega$.\\
{with} $\expectation{\xi_\alpha\brackets{\randomvec{Z}}{^2}}{\randomvec{Z}},
\expectation{\tau_\alpha\brackets{\randomvec{Z}}}{\randomvec{Z}},
\expectation{ \zeta_{\alpha}\brackets{\randomvec{Z}}}{\randomvec{Z}}<\infty$. 

\end{assumption}


Then, {we obtain the following results.  }

\begin{theorem}[Posterior Score Consistency] \label{thm:post_consist}
    Suppose that Assumptions~\ref{assum:bcrb_reg}, \ref{ass:score_cond_reg}, {\ref{assume:lip_cont},} and {\ref{assume:realizable}}~ hold, {and that the minimization algorithm applied to the objective in \eqref{eq:score_post_mean} succeeds in finding a global minimizer $\Omega^*={\arg\min\limits_{\Omega}\lossbsm\brackets{\Omega}}$.
    }  Then,
    \begin{equation*}
        \lossbs\brackets{\Omega^{*}}\xrightarrow{a.s} 0\quad\text{as} \quad\abs{\ds}=\nds\rightarrow \infty.
    \end{equation*}
\end{theorem}

\begin{theorem}[Measurement and Prior Score Consistency]\label{thm:mp_consist}
    Suppose that Assumptions~\ref{assum:bcrb_reg}, \ref{ass:score_reg_prior}, \ref{ass:score_reg}, {\ref{assume:lip_cont},} {and \ref{assume:realizable}} hold, 
    {and that the minimization algorithms applied to the objectives in \eqref{eq:score_prior_mean} and \eqref{eq:score_lik_mean} succeed in finding respective global minimizers {$\paramf^{*}=\arg\min\limits_{\Omega}\lossfsm\brackets{\Omega}$ and $\paramp^{*}=\arg\min\limits_{\Omega}\losspsm\brackets{\Omega}$.}
    }
    Then,
        \begin{equation*}
        \lossfs\brackets{\paramf^{*}}\xrightarrow{a.s} 0\mkern9mu \text{and}\mkern9mu\lossps\brackets{\paramp^{*}}\xrightarrow{a.s} 0, \mkern9mu\text{as} \mkern9mu \abs{\ds}=\nds\rightarrow \infty. 
    \end{equation*}
\end{theorem}

Theorems~\ref{thm:post_consist} and \ref{thm:mp_consist} (proved in Appendix~\ref{apx:proof:consistency_post}) demonstrate that {the score neural networks optimized via score matching are strongly consistent estimators of the true score. }



{From a practical point of view, {the dependence of}
Theorems~\ref{thm:post_consist} and \ref{thm:mp_consist} {on the  realizability Assumption~\ref{assume:realizable}}
highlights the advantages of \peac{} that uses domain knowledge {in the Measurement-Prior Approach.}  \peac{} enables {the realization and learning of accurate score functions models}: (i) using less complex neural networks (fewer parameters, nonlinearity, etc.); (ii)  {using} less training data; and (iii) {with more reliable training 
(
minimization of the loss function).
}




 }


{Having established the a.s convergence of the empirical-mean  and the strong consistency of score matching, we are ready for our final results for the \name{} for the two methods.}

\begin{corollary}[Posterior Approach Convergence] \label{corr:re_zero}
Suppose that {the assumptions in Theorem~\ref{thm:post_consist}} {and Corollary~\ref{corr:sampling_post}} hold. Then,
    $\lim\limits_{\nds\rightarrow \infty } {\lbcrbbs =\bcrb}. \quad\text{a.s}$
\end{corollary}

\begin{corollary}[Measurement-Prior Approach Convergence] \label{corr:re_zero_mp}
Suppose that {the assumptions in Theorem~\ref{thm:mp_consist}} {and Corollary~\ref{corr:sampling_mp}} hold. Then,
  $  \lim\limits_{\nds\rightarrow \infty } {\lbcrblps=\bcrb}. \quad\text{a.s}$
\end{corollary}
{
We prove Corollary~\ref{corr:re_zero_mp}. The proof of Corollary~\ref{corr:re_zero} is exactly analogous, with  Corollary~\ref{corr:sampling_post} and Theorems~\ref{thm:lrn:direct} and \ref{thm:post_consist}  playing the role of  Corollary~\ref{corr:sampling_mp} and Theorems
\ref{thm:lrn:lik_prior}, and \ref{thm:mp_consist}, respectively. 
}

{\begin{proof}
   Observe that $\expectation{\lbfimlps}{\X,\pr}=\lbfimlp$.
   Using  Corollary~\ref{corr:sampling_mp} 
   we have that, $\lbfimlps$ 
   from 
   \eqref{eq:bfim_apx_final} converges almost surely to 
   its 
   expected value,
   i.e., $\lim\limits_{\nds\rightarrow\infty}\lbfimlps=\lbfimlp$ a.s.  
   Furthermore, combining
   Theorem~\ref{thm:lrn:lik_prior} 
   with Theorem~\ref{thm:mp_consist}
   , it follows that $\lim\limits_{\nds\rightarrow\infty} \lbfimlp=\fb$ a.s.
 Hence, $\lim\limits_{\nds\rightarrow\infty}\lbfimlps=\fb$ a.s.  
 Inverting yields the result. 
\end{proof}
}

{In summary, } as stated in Corollarys~\ref{corr:re_zero} and \ref{corr:re_zero_mp}, with sufficient training data and a properly trained neural networks, the\name{} {will provide the BCRB accurately}. 

%% file: files/model_examples.tex
\section{MEASUREMENTS MODEL EXAMPLES}\label{sec:example_models}
We present three measurement models to illustrate the benefits of the \name.

\subsection{Linear Observation Model}\label{sec:linear_model} 
Let $\pr\in\mathbb{R}^{\np}$ be the parameter {we wish to estimate} and $\matsym{A}\in\mathbb{R}^{{\nx}\times \np}$  a fixed mixing matrix, 
and define the  ${n}^{th}$ snapshot measurement ${\X}_{n}\in\mathbb{R}^{{\nx}}$ 
\begin{equation}\label{eq:linear-model}
    {\randomvec{X}}_{{n}}=\matsym{A}\pr+\randomvec{W}_{n},
\end{equation}
where $\randomvec{W}_{{n}}\sim\normaldis{0}{\matsym{\Sigma}}$ is  additive Gaussian noise with zero mean and covariance matrix $\matsym{\Sigma}$. 
We are given a measurement set $\xset=\set{{\X}_{n}}_{{n}=1}^{\niideval}$,
that consists of $\niideval$ snapshots of the measurement in \eqref{eq:linear-model}.  
For the prior of $\p$, we assume a Gaussian distribution,  $\pr\sim\normaldis{0}{\matsym{I}\sigma_P^2}.$  The linear model in \eqref{eq:linear-model}  with the Gaussian prior enables a detailed {study} of the suggested methods and also ensures that Assumption ~\ref{assume:realizable} %
{is satisfied by simple neural network realizations of the score functions}. 

The BCRB for %
the estimation of $\p$ from the measurement {set $\xset$} in this example is given by:
\begin{align}\label{eq:bcrb_linear}
\bcrb=\brackets{{\niideval} \cdot\matsym{A}^T\Sigma^{-1}\matsym{A}+\frac{1}{\sigma_P^2}}^{-1},
\end{align}
and %
{(as expected for the linear Gaussian observation model)} coincides with the covariance of LMMSE. In addition, the {\emph{true}} score vectors of the measurement model  \eqref{eq:linear-model} are given by:
\begin{equation}\label{eq:score_opt_linear}
\vectorsym{s}^{0}_F\brackets{{\x}|\p}=  \matsym{A}^T\Sigma^{-1}\brackets{{\x}-\matsym{A}\p}\textbf{ and } \vectorsym{s}^{0}_P\brackets{\p}=-\frac{\p}{\sigma_p^2}.
\end{equation}
{The true posterior score is given by $\vectorsym{s}^{0}_B\brackets{\p|\xsetr}=\sum_{\x\in\xsetr}\vectorsym{s}^{0}_F\brackets{{\x}|\p}+\vectorsym{s}^{0}_P$, where $\xsetr$ represents the set of $\niideval$ i.i.d. measurements. }

{The goal is to compute the LBCRB (a close approximation to the BCRB), when the matrices $\matsym{A}$ and $\matsym{\Sigma}$ are \emph{unknown}, by using a training data set 
$\ds=\set{\xset_n=\set{{\x}_m}_{m=1}^{\niiddata},\p_n}_{n=1}^{\nds}$. 
 It is important to observe that the number $\niiddata$ of i.i.d. samples present in the dataset for each value of $\p$ may differ from the desired number $\niideval$ for obtaining the LBCRB,
 i.e., $\niideval \neq \niiddata$.

\subsection{Frequency Estimation }\label{sec:freq_est}
A simple frequency estimation problem  illustrates the benefits of the \name{}. 
The observation model is  
\begin{equation}\label{eq:freq_est_model}
    \squareb{\X}_n=\cos{\brackets{\pr n}}+\squareb{\randomvec{W}}_n \quad\forall n \in [0,N-1],
\end{equation}
where $\p$ is the digital  frequency  parameter to be estimated, and $\randomvec{W}$ is an additive noise. 
We consider two types of noise: 1)  Gaussian noise $\randomvec{W}\sim \mathcal{N}\brackets{0,\Sigma}$ to validate the \name{} using closed-form results from \cite{van2007bayesian} (Example 3); and 2) a real-world underwater ambient noise taken from \cite{msg0-ag12-22} to simulate frequency estimation with  non-Gaussian noise. 
We consider a four-parameter Beta prior  $\p\sim\mathrm{Beta}\brackets{\alpha_{\p},\alpha_{\p},0,\pi}$. 
Recall that a random variable $\randomvec{z}\sim\mathrm{Beta}\brackets{\alpha,\beta,l,u}$ has PDF
\begin{equation}
    \probt{z}{\randomvec{z}}\triangleq \frac{\brackets{z-l}^{\alpha-1}\brackets{u-z}^{\beta-1}}{\brackets{u-l}^{\alpha+\beta+1}B\brackets{\alpha,\beta}},
\end{equation}
where  $l<u$ are the lower and upper support parameters, respectively, $\alpha, \beta >0$ are the shape parameters, and $B\brackets{\alpha,\beta}=\frac{\Gamma\brackets{\alpha}\Gamma\brackets{\beta}}{\Gamma\brackets{\beta+\alpha}}$ is the normalization constant, where $\Gamma\brackets{\alpha}$ is the Gamma function.

In the case of Gaussian noise, the observation model in \eqref{eq:freq_est_model} has the following true score functions:
\begin{subequations}
\begin{equation}\label{eq:score_opt_freq_lik}
\vectorsym{s}^{0}_F\brackets{\x|\p}=\matsym{\frac{\partial \mathcal{M}\brackets{\p} }{\partial \p}}^{T}\Sigma^{-1}\brackets{\x-\mathcal{M}\brackets{\p}},
\end{equation}
\begin{equation}
\vectorsym{s}^{0}_{P}\brackets{\p}=\frac{\alpha_{{\p}}-1}{{\p}}-\frac{\alpha_{{\p}}-1}{\pi-{\p}}
\end{equation}
\end{subequations}
where $\squareb{\mathcal{M}\brackets{\p}}_n=\cos{\brackets{\p n}}$. 
Here, we show the use of a \pe{} score neural network and its effect on the structure of the neural network. 
We assume that $\squareb{\mathcal{M}\brackets{\p}}_n=\cos{\brackets{\p n}}$ 
{has the known form of a sinusoidal function of the unknown parameter $\p$,}
whereas the distribution of $\matsym{W}$ is unknown. 

Applying $\squareb{\mathcal{M}\brackets{\p}}_n=\cos{\brackets{\p n}}$ to \eqref{eq:model_base_score} 
and comparing with \eqref{eq:score_opt_freq_lik}, we find that the  
 $\vectorsym{s}_I$ %
 {that provides the true score}
is given by $\Sigma^{-1}\brackets{\x-\vectorsym{\tau}}$. %
{In this example, the score neural network only 
needs to learn (using $\ds$ and score matching) the score function of the noise distribution, 
since the modeling of the dependence of the measurement on the frequency parameter is known. }
{Furthermore, $\vectorsym{s}_I$ } can be represented by a single layer of a fully connected neural network that needs to learn only the inverse of the covariance matrix. %
{
Importantly, as explained in Remark~\ref{remark:opt_mi}, $\vectorsym{s}_{I}$ is only indirectly {optimized} through the application of $\mathcal{M}$ using \eqref{eq:model_base_score}.
}

In contrast, without leveraging the model information, we 
have to learn the cosine function for the score -- a rather complicated nonlinear function. This 
leads to a significantly more complex neural network that 
requires more training samples  to train properly, %
{as demonstrated later in Figure~\ref{fig:analysis_mi}}. 

\subsection{%
{One Bit }Quantization}\label{sec:quantization_example}
Consider the linear observation model with quantization. %
{Specifically, let Let $\pr\in\mathbb{R}^{\np}$ be the parameter %
{we wish to estimate} and $\matsym{A}\in\mathbb{R}^{{\nx}\times \np}$  a fixed mixing matrix, 
and define the  ${n}^{th}$ snapshot measurement ${\X}_{n}\in\mathbb{R}^{{\nx}}$:} 
\begin{equation}\label{eq:lin_q}
{\X}_{n}=\mathrm{sign}\brackets{\matsym{A}\pr+\randomvec{W}_{n}+\boldsymbol{1}_{\nx}\cdot b },
\end{equation}
where %
{$$\squareb{\mathrm{sign}\brackets{\vectorsym{x}}}_i=\begin{cases}
    1 & \squareb{\vectorsym{x}}_i\geq 0\\
    -1 & \squareb{\vectorsym{x}}_i< 0
\end{cases}$$ is an element-wise sign function,} 
{ $\pr \in \mathbb{R}^{\np}$,}
$\randomvec{W}_{n}\sim\normaldis{0}{\matsym{\Sigma}}$ is additive zero-mean Gaussian noise  with covariance 
$\matsym{\Sigma}$, $\boldsymbol{1}_{\nx} \in \mathbb{R}^{\nx}$ is a vector of ones, 
and $b$ is a constant shift. Finally, for the prior
we assume $\pr\sim\normaldis{0}{\matsym{I}\sigma_p^2}$. 
Importantly, even for this simple observation model
\eqref{eq:lin_q}, there is no closed-form solution for the BCRB. However, in 
 the special case 
of a diagonal 
$\matsym{\Sigma}$, 
the 
{true} {Fisher} score 
has a closed-form solution (derived in Appendix~\ref{apx:derivation_q_score})
\begin{align}\label{eq:score_quantization}
    \vectorsym{s}^{0}_{F}\brackets{\x|\p}&=\frac{\partial\vectorsym{u}\brackets{\p}}{\partial\p}^T\vectorsym{\rho}\brackets{\x,\p}, \\
    \text{where } \squareb{\vectorsym{\rho}\brackets{\vectorsym{x},\p}}_i&=
    \textstyle{\frac{\squareb{\x}_i\exp\brackets{-\frac{\squareb{\vectorsym{u}\brackets{\p}}_i^2}{2\sigma_i^2}}}{\sqrt{2\pi\sigma^2_i}\Phi\brackets{\squareb{\x}_i\frac{\squareb{\vectorsym{u}\brackets{\p}}_i}{\sigma_i} }} } , \nonumber
\end{align}
$\vectorsym{u}\brackets{\p}=\matsym{A}\p+\boldsymbol{1}_{\nx}\cdot b$ is the un-quantized observation function, $\sigma_i^2=\squareb{\Sigma}_{i,i}$ is the $i^{th}$ element in the diagonal matrix $\Sigma$, and
$\Phi$ is the 
standard Gaussian CDF. %
{Note that the true prior and posterior scores are 
given by the same expressions as for the linear observation model in section~\ref{sec:linear_model}.}

Using \eqref{eq:score_quantization} in \eqref{eq:mean_fully}, we can determine, {for known $\matsym{A}, \matsym{\Sigma}$ and $b$}, the BCRB numerically.
However, this is a very limited special case, and a more general bound cannot be obtained since the CDF of the multivariate Gaussian distribution does not have a close form\cite{genz2009computation}, preventing  the computation of the PMF
$\pmft{\X|\p}{\x|\pr}$. %
{Needless to say, for unknown $\matsym{\Sigma}$, the BCRB cannot be determined conventionally at all.}

 {Similar to the linear problem of Section~\ref{sec:linear_model}, our goal is to compute the LBCRB (a close approximation to the BCRB), when $\matsym{A}$, $\matsym{\Sigma}$ and $b$ are \emph{unknown}, by using a training data set $\ds=\set{\xset_n=\set{{\x}_m}_{m=1}^{\niiddata},\p_n}_{n=1}^{\nds}$.

%% file: files/results.tex
\section{{Experiments}}\label{sec:experimental}
\subsection{{Overview}}
We provide experimental results to show the benefits and {evaluate} the suggested \name{}. 
We begin with a %
{study of the \name{} errors}  to show the effects of different errors on the learned bound quality using the simple linear Gaussian measurement model 
of Section~\ref{sec:linear_model}. This is followed by an {evaluation} of the \pe{} score neural network, where we present its benefits in terms of sample complexity, usability, and accuracy. Finally, we use the \name{} on two applications: 1) linear observation model with quantization; 2) the frequency estimation problem in underwater acoustics.  In these two problems, we obtain a learned BCRB that could not be obtained otherwise, showing the main benefit of the \name{}.    

\subsection{Setup}

{In all experiment we follow the same process. {To represent a general case,} we  begin the experiment by randomly generating once the static parameters of the measurement model (such as the mixing matrix $\matsym{A}$ and the covariance matrix $\matsym{\Sigma}$) as described in Appendix~\ref{apx:init_sp}. {These parameters of the measurement model are then kept fixed for all experiments.} 

With the measurement model in place, {we simulate the acquisition of a training data set $\ds$ by} randomly drawing $\nds$ 
independent and identically distributed measurement set-parameter pairs for each SNR. Specifically, 
for each SNR, data generation involves initially sampling $\p_i$ from the prior distribution $\probt{\p}{\pr}$. Then, $\niiddata$ independent snapshots ${\x}^{(i)}$ are drawn from 
$\probt{\x|\p_i}{\X{|}\pr}$ for every $i \in [1,\niiddata]$ using the measurement model. This is done $\nds$ times to create the data for a specific SNR $\mathcal{D}$. Combining the data for all {$n_c$} SNR values 
produces the dataset $\mathcal{D}_C=\set{\mathcal{D}_i,\mathrm{SNR}_i}_{i=1}^{n_c}$
of size %
{$N_T=n_c \cdot \nds$. 

}
\begin{figure*}[t]
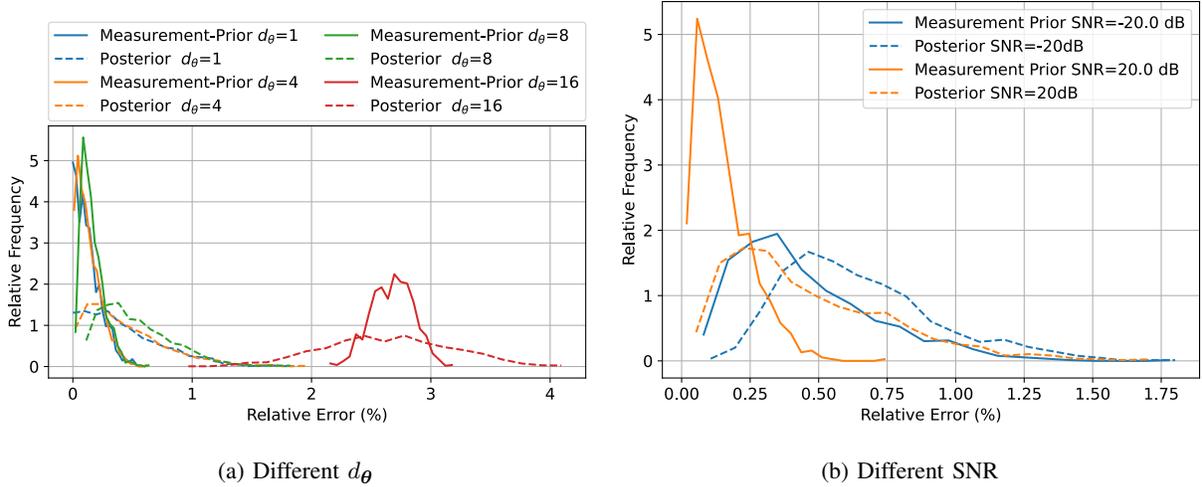

    \centering
    \begin{subfigure}[b]{0.45\textwidth}
        \centering
        \includesvg[width=1.0\textwidth]{files/results/linear_v2/sampling_error_analysis_different_m_fim_high_snr.svg}
        \caption{Different $\np$}\label{fig:m_re_hist}
    \end{subfigure}%
    \begin{subfigure}[b]{0.45\textwidth}
        \centering
        \includesvg[width=1.0\textwidth]{files/results/linear_v2/sampling_error_analysis_different_snr.svg}
        \caption{Different SNR}\label{fig:snr_re_hist}
    \end{subfigure}
    \caption{Histograms %
    {(using 1000 Monte-Carlo trials)} 
    of the \name{} relative errors 
    $\mathrm{RE}_{B}$ 
    \eqref{eq:inv_bound_post} and $\mathrm{RE}_{MP}$ \eqref{eq:inv_bound_mp} for a linear Gaussian example. }
    \label{fig:hist_sample}
\end{figure*}
{Instead of training a single score NN for each SNR, we use {the SNR as an additional conditioning variable.} Conditioning variables are used extensively  in various generative models \cite{mirza2014conditional,abdelhamed2019noise,liu2019conditional,ho2021classifier} and enable the training of a single generative model to operate in several conditions. In this study, we apply the SNR as a conditioning variable in the following manner.  }
During the training phase, %
{(similar to \cite{habi2024learning})} we optimize a single score
NN across $n_c$ signal-to-noise ratios, utilizing the SNR as a conditioning variable. 
{In inference, we evaluate the LBCRB for each SNR individually, {providing the value of the SNR as a conditioning input to the trained score NN.} This approach enables to improve the sample complexity {of learning the score NN}, since {the score functions for} different SNRs can be very similar.  }

Denoting this conditioning variable as $\vectorsym{c}$,  the score models are expressed as $\postscore{\p}{\xsetr,\vectorsym{c}}$ and $\lscore{\x}{\p,\vectorsym{c}}$.
Note that the condition $\vectorsym{c}$ does not effect the prior score neural network. 
In this study, {we use SNR as the conditioning variable $\vectorsym{c}$,
but it can vary depending on the specific problem. }
Then, to {learn} the score models with the conditional variable, we use 
{the data set $\ds_C$ that also includes the conditioning variables, 
with the objective functions \eqref{eq:score_post_mean} and} \eqref{eq:score_lik_mean} replaced, respectively, by the following.
}

\begin{equation*}
    \lossbsm\brackets{\Omega}=\frac{1}{\nds}\sum_{\p,\xsetr,\vectorsym{c}\in\ds}\norm{\postscore{\p}{\xsetr,\vectorsym{c};\Omega}}_2^2+2\trace{\overline{\matsym{J}}_B},
\end{equation*}
\begin{align*}
    &\lossfsm\brackets{\paramf; \paramp}=\frac{1}{\nds\cdot \niiddata}\sum_{\xsetr,\p,\vectorsym{c}\in\mathcal{D}}\sum_{\x\in\xsetr}\norm{\lscore{\x}{\p,\vectorsym{c};\paramf}}_2^2\nonumber \\
    &+2\sum_{\xsetr,\p,\vectorsym{c}\in\mathcal{D}}\sum_{\x\in\xsetr}\frac{\lscore{\x}{\p,\vectorsym{c};\paramf}^T\priorscore{\p;\paramp}}{\nds\cdot \niiddata}+2\trace{\overline{\matsym{J}}_{F}},
\end{align*}
where $\overline{\matsym{J}}_{F}\brackets{\paramf}\triangleq\frac{1}{\nds\cdot \niiddata}\sum_{\xsetr,\p,\vectorsym{c}\in\mathcal{D}}\sum_{\x\in\xsetr}{\frac{\partial \lscore{\x}{\p,\vectorsym{c};\paramf}}{\partial\p}}$ 
and $\overline{\matsym{J}}_B\brackets{\Omega}\triangleq\frac{1}{\nds}\sum_{\p,\xsetr,\vectorsym{c}\in\ds}{\frac{\partial \postscore{\p}{\xsetr,\vectorsym{c};\Omega}}{\partial\p}}$ are the average Jacobian matrices. 

Once training is completed, we evaluate 
the \name{} %
{at the $i^{th}$ SNR as follows}
\begin{equation*}
    \lbcrbbs^{(i)}=\brackets{\frac{1}{\nds}\sum_{\p,\xsetr\in\mathcal{D}_i}\postscores{\p}{\xsetr,\mathrm{SNR}_i}\postscores{\p}{\xsetr,\mathrm{SNR}_i}^T}^{-1}
\end{equation*}
\begin{equation}
    \lbcrblps^{(i)}{(\niideval)} =\brackets{\niideval\cdot  \lmfim^{(i)}+ \lpfim}^{-1},
\end{equation}
where 
{
        \begin{equation*}
     \lmfim^{(i)}\triangleq     \tfrac{1}{\nds\cdot \niiddata} \sum_{\p,\xsetr\in\ds_i} \sum_{\x\in\xsetr}\lscores{\x}{\p, \mathrm{SNR}_i}\lscores{\x}{\p, \mathrm{SNR}_i}^T   
        \end{equation*} 
        }
is the LBFIM of the $i^{th}$ SNR. It is important to note that we evaluate the \name{} under the same SNR conditions seen during training %
{i.e., we don't interpolate to unseen SNR values}. 

In all experiments, except where otherwise specified, we 
assume that the only information available about the observation model is the data set $\ds$, used during the training and evaluation phases. %
{In addition, in all experiments, unless otherwise specified, we use the same set of hyperparameters. 
In all experiments, the score neural networks are trained on 20 different SNR values with $\nds=60K$ samples 
per SNR.  } The 
neural network architectures and
hyperparameters 
are detailed in Appendix~\ref{apx:param}.


We 
evaluate the performance of the \name{} method in Sections~\ref{sec:error_anaylsis_exp} and \ref{sec:model_score_analysis}, where we study the  relative errors %
{$\mathrm{RE}_{B}$ and $\mathrm{RE}_{MP}$ defined in \eqref{eq:inv_bound_post} and \eqref{eq:inv_bound_mp}} under various conditions such as SNR, parameter dimension, etc., and the mean ($\mathrm{mRE}$) relative error 
\begin{equation}\label{eq:mean_re}
    \mathrm{mRE}\triangleq\frac{1}{n_c}\sum_{j=1}^{n_c}\mathrm{RE}\brackets{\mathrm{SNR}_j}.
\end{equation}


\subsection{{
Empirical-Mean Error with Linear Gaussian Observations}}\label{sec:error_anaylsis_exp}
We begin by examining the empirical-mean error %
{and the corresponding theoretical results Theorem~\ref{thm:sampling_post} and \ref{thm:sampling_mp}.}
To eliminate the {approximation error} in this experiment, we employ the Gaussian observation model 
of Section~\ref{sec:linear_model} (with $\nx=16$ and a prior variance $\sigma_p=2.5$) and use the {true}
scores ($\vectorsym{s}_{F}^{0}$, $\vectorsym{s}_{P}^{0}$) in \eqref{eq:score_opt_linear}.  

We evaluate the \name{} with a dataset size of $\nds^{{c}}=64k$ with $\niiddata=\niideval=10$ 
and compute the relative errors 
$\mathrm{RE}_{B}$ \eqref{eq:inv_bound_post} and $\mathrm{RE}_{MP}$ \eqref{eq:inv_bound_mp}
 compared to the BFIM expression. This procedure is repeated 1000 times, and the {histogram} is shown in Figure~\ref{fig:hist_sample} 
 for various sizes of the parameter vector $\np$ (at SNR $=0dB$) and different SNR values (with $\np=4$).

From Figure~\ref{fig:m_re_hist}, we observe that a large $\np$ leads to {increased} error, {in agreement with the predictions} 
in Theorem~\ref{thm:sampling_mp} and Theorem~\ref{thm:sampling_post}. In the case of SNR, Figure~\ref{fig:snr_re_hist} shows that a low SNR results in a high empirical-mean error. 
Furthermore, both Figures~\ref{fig:m_re_hist} and \ref{fig:snr_re_hist} demonstrate that the Measurement-Prior
Approach yields smaller errors than
the Posterior Approach. 


To further validate the theoretical results on the empirical-mean error, 
we evaluated the \name{} with various data set sizes from {$\nds=$}32k to  {$\nds=$}256k using the following parameters: $\np=4$, $\nx=16$, {$\niiddata=\niideval=10$} and $SNR=20dB$. We repeat this process 1000 times for each size of the data set and present the mean and standard divination in Figure~\ref{fig:sample_error_vs_n_samples}. In addition to the empirical standard deviation, we 
{plot the} theoretical bounds 
in \eqref{eq:re_error_post_mean} and \eqref{eq:re_error_mp_mean} {(which can be computed in this case\footnote{ When analytical expression {are not available} the bound can be approximated by replacing the expectation with an empirical mean.})}.
\begin{figure}
    \centering
    \includesvg[width=0.8\linewidth]{files/results/linear_v2/sampling_error_analysis_vs_n_samples_fim_high_snr.svg}
    \caption{{
    Empirical-mean errors  
    $\mathrm{RE}_{B}^{(e)}$ and $\mathrm{RE}_{MP}^{(e)}$ {vs. $\nds$}
    for the linear Gaussian estimation problem {with $\niiddata=10$}. 
    Upper part: 
    theoretical bounds \eqref{eq:re_error_post_mean} and \eqref{eq:re_error_mp_mean}. 
    Bottom part: 
    average and standard deviation of the actual empirical-mean error over 1000 Monte-Carlo trials. 
    }
    }
    
    \label{fig:sample_error_vs_n_samples}
\end{figure}

\begin{figure}
    \centering
    \includesvg[width=0.8\linewidth]{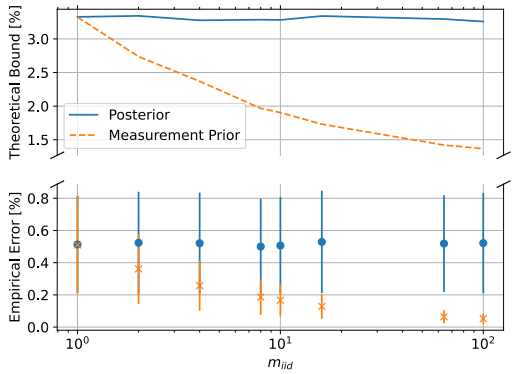}
    \caption{
    Empirical-mean errors $\mathrm{RE}_{B}^{(e)}$ and $\mathrm{RE}_{MP}^{(e)}$ vs $\niiddata$
    for the linear Gaussian estimation problem. 
    Upper part: 
    theoretical bounds \eqref{eq:re_error_post_mean} and \eqref{eq:re_error_mp_mean}. 
    Bottom part: 
    average and standard deviation of the actual empirical-mean error over 1000 Monte-Carlo trials. 
    }
    \label{fig:sample_error_vs_m_iid_samples}
\end{figure}

From Figure~\ref{fig:sample_error_vs_n_samples} we observe that the error decreases with increasing number of samples $\nds$. In addition, comparing to the theoretical bound, 
we observe that both the theoretical and empirical results decrease with the same rate. %
{Additionally, this result shows that the Measurement-Prior Approach surpasses the Posterior Approach  in terms empirical-mean error in these scenarios. 
To study this further, we performed a similar experiment with 
$\nds^{{c}}=64k$ 
using various values of $\niiddata=\niideval$, 
with results 
shown in Figure~\ref{fig:sample_error_vs_m_iid_samples}. 

Figure~\ref{fig:sample_error_vs_m_iid_samples} reveals that, as $\niiddata$ grows, the empirical-mean error in the Measurement-Prior Approach diminishes, whereas in the Posterior Approach the error remains unchanged. This observation 
{is predicted by} the bounds in Theorems~\ref{thm:sampling_post} and \ref{thm:sampling_mp}, and the discussion that followed them. {With reference to Theorems~\ref{thm:sampling_post} and \ref{thm:sampling_mp}, Figure~\ref{fig:sample_error_vs_m_iid_samples} shows that $\frac{\cb}{\trace{\fb}}$ stays constant as $\niiddata$ increases, whereas $\frac{\niiddata\cb +\cp}{\niiddata\trace{\fm}+\trace{\fp}}$ decreases.  } 
Moreover, this highlights an additional benefit of using the Measurement-Prior Approach to 
reduce the empirical-mean error.

{Finally, we explore the effect of different $\niideval$ on the empirical-mean error, further validating the theoretical results. To this end,
we evaluate the LBCRB with various of $\niideval$ values using the following parameters 
(
$\nds=64k$, $\np=4$, $\nx=16$, %
{$\niiddata=10$} and $SNR=-50dB$\footnote{
Note that at  higher SNRs $\niideval$ has little  effect on the empirical-mean error, 
{
as predicted by
} Theorem~\ref{thm:sampling_mp}, 
where the upper bound on {$\mathrm{RE}_{MP}^{(e)}$ depends on 
 $(\niideval c_m+c_p)\Big/ \left(\niideval\mathrm{Tr}(\tlmfim)+\mathrm{Tr}(\tlpfim) \right) $. Now, $\cm$ and $\mathrm{Tr}(\tlmfim)$ both scale up identically with SNR, so that at high SNR they dominate, and the effect of $\niideval$ 
 in the numerator and denominator cancels.
} 
}  
). 
%
We repeat this process 1000 times for each size of the data set and present the mean and standard divination in Figure~\ref{fig:sample_error_vs_n_iid}. In addition to the empirical standard deviation, we plot
\eqref{eq:re_error_mp_mean}.  }
\begin{figure}
    \centering
    \includesvg[width=0.8\linewidth]{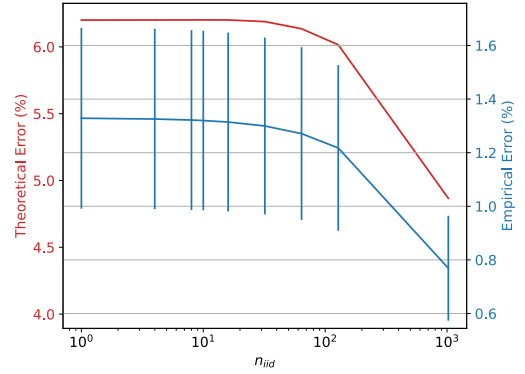}
    \caption{
    {
    Empirical-mean \name{} error $\mathrm{RE}_{MP}^{(e)}$ \eqref{eq:re_error_mp_mean} 
     for the linear Gaussian estimation problem,  vs. the assumed number $\niideval$ of i.i.d measurement samples.
        The Measurement-Prior Approach is able to compute the \name{} for 
    any  $\niideval$ 
    without more data or retraining.   
    Left y-axis:  theoretical bound (red). 
    Right y-axis:  actual $\mathrm{RE}_{MP}^{(e)}$ 
    (blue); average and standard deviation over 1000 Monte-Carlo trials.
    }
    }
    \label{fig:sample_error_vs_n_iid}
\end{figure}

The results in Figures~\ref{fig:hist_sample}, \ref{fig:sample_error_vs_n_samples}, \ref{fig:sample_error_vs_m_iid_samples} and  \ref{fig:sample_error_vs_n_iid}
{(which do not include approximation error)} validate the theoretical findings in Theorems~\ref{thm:sampling_post} and \ref{thm:sampling_mp}. 

\subsection{%
{
Approximation Error}}
{This section examines the approximation error in two scenarios: (i) 
{multiple i.i.d samples $\niideval>1$ (to compare between the two approaches we require that dataset $\ds$ includes multiple i.i.d. samples  $\niiddata=\niideval$); } 
and (ii) with the use of a \pe{} score NN.
}

\subsubsection{%
{
Linear Gaussian Observations 
with $\niideval>1$}}
%
To illustrate the advantages 
of 
{the Measurement-Prior Approach when the LBCRB is computed for} {$\niideval>1$} i.i.d. samples, %
%
we performed the following experiment on the linear Gaussian observation model of Section~\ref{sec:linear_model} (%
{with parameters $\np=4$, $\nx=10$,  and SNR values uniformly spaced between -20dB to 20dB}). We trained two different score models: one employing the Posterior Approach {with $|\xset|=\niideval=\niiddata$,  
}
and the other {employing} 
the Measurement-Prior Approach 
{that computes the measurement score and FIM for a single sample, but evaluates the BCRB for $\niideval=\niiddata$. This was repeated} 
with various numbers (1, 2, 3, 4, and 5) of i.i.d. samples 
$\niideval$, with
results 
presented in Figure~\ref{fig:k_study}. 

\begin{figure}
    \centering
    \includesvg[width=0.8\linewidth]{files/results/linear_v2/k_study.svg}
    \caption{Mean Relative Error 
    {
    $\mathrm{mRE}$} 
    \eqref{eq:mean_re} 
    {
    vs.} number $\niideval$ of i.i.d. samples { per $\p$ value in the training data set $\ds$.
    }
    } \label{fig:k_study}
\end{figure}

The results indicate that as $\niideval$  increases, the mean relative error %
{$\mathrm{mRE}$} 
{defined in \eqref{eq:mean_re}} {mostly} decreases for the Measurement-Prior Approach, 
whereas it increases for the Posterior Approach. %
{We emphasize that the empirical-mean error does not influence this outcome in Figure~\ref{fig:k_study}. 
We can conclude this from 
Figure~\ref{fig:sample_error_vs_m_iid_samples}, which shows that as $\niiddata$ increases, the Posterior Approach empirical-mean error is essentially constant,
whereas Figure~\ref{fig:k_study} shows an increase in combined error.  }

The cause of this behavior of {the approximation error} lies in the Posterior Approach requirement to incorporate {all measurement samples, whether 
i.i.d. or not, 
jointly as a combined measurement $\xset$ }through a score NN
with a larger input dimension. This increases the NN's capacity, %
and as a consequence, the sample complexity for its training increases too. {(This is trivially evident in the example of this experiment, where} the score NN is a fully-connected linear layer, with number of parameters determined by the input and output dimensions.)

Conversely, in the Measurement-Prior Approach, the NN maintains a constant input dimension and operates on each i.i.d. sample independently, ensuring that the model capacity remains unchanged. A further advantage of the Measurement-Prior Approach is that the NN does not need to adapt to the presence of $\niideval$ i.i.d. samples, while in the Posterior Approach it must infer {the fact that they are i.i.d.} from the dataset. Consequently, in the Measurement-Prior Approach, increasing $\niideval$ has a similar effect to increasing the size of the dataset $\nds$.

\subsubsection{
\pe{}
Score NN
with 
the Frequency Estimation Model}\label{sec:model_score_analysis}
\begin{figure}
    \centering
    \includesvg[width=0.4\textwidth]{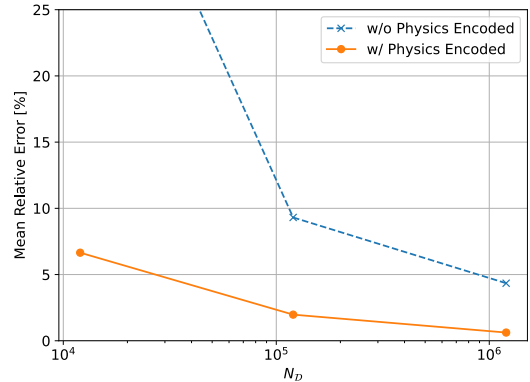}
    \caption{\name{} Mean relative error $\mathrm{mRE}$ \eqref{eq:mean_re} vs.  dataset set size $\nds$ with $\niiddata=\niideval=1$  using the Measurement-Prior Approach with a score NN with (solid line) and without (dashed line) {physics encoding}.    }
    \label{fig:analysis_mi}
\end{figure}
To 
study the benefits of the \pe{} score neural network for the \name{}, {we use the Measurement-Prior Approach} on the frequency estimation problem model of Section~\ref{sec:freq_est} (with parameters $\nx=16$, $\niiddata=1$ and $\alpha_\omega=100$).
{For the Fisher Score we} train, for comparison,  two different score neural networks: 
(i) {
a regular NN, without 
Physics-encoding;
and (ii) {
a NN with  
Physics-encoding.}
{In the context of this example,} by 
\pe{} by we mean that the structure of the problem is known, but the noise covariance is unknown.} Then the Fisher score function has the form of \eqref{eq:model_base_score} with the model function $\squareb{\mathcal{M}\brackets{\p}}_n=\cos{\brackets{\omega n }}$.

We train the two score NNs
using datasets of varying sizes $\nds$: 1200K, 120K, and 12K samples. 
The Fisher {score} 
NNs
 {without Physics-encoding} consists of a three-layer multilayer perceptron (MLP) that uses Swish\cite{ramachandran2017searching} activation functions between layers, incorporating conditional inputs such as SNR and $\p$. In contrast, the \pe{} Fisher score 
NN {has} 
just one layer of MLP without any nonlinear activation function. 

We perform an {evaluation of}  the \pe{} {score neural network} to highlight its advantages, particularly in terms of sample complexity.
The results are shown in Figure ~\ref{fig:analysis_mi}. 
    

Figure~\ref{fig:analysis_mi}  
shows
the mean relative error $\mathrm{mRE}$ \eqref{eq:mean_re}
{of the \name{}}  vs. dataset size $\nds$. These results reveal that the \pe{} {score} neural network consistently surpasses the regular model, particularly with the smaller dataset size of $N_T=12K$ (comprising $\nds=600$ samples per SNR value). 


\subsection{Example Use Cases of the \name{} }
In this section, we consider
two signal processing problems in which calculating the BCRB is impractical. These examples underscore the primary advantage of \name{}, which 
enables the determination of a lower bound on Bayesian estimation when both prior and measurement distributions are unknown or only partially known. 

\subsubsection{One-Bit Quantization}
{We consider the observation model of Section~\ref{sec:quantization_example} with correlated noise and one-bit quantization. }Despite its seemingly straightforward nature,  the score function for this observation model \emph{cannot be calculated analytically,} {even when the measurement model (matrices $\matsym{A}$ and $\matsym{\Sigma}$ and quantizer model) are known. Furthermore, we consider this problem when both measurement model and prior on $\p$ are  \emph{completely unknown}.
}
This emphasizes the main benefit of the \name{}, which can be used without any knowledge about measurement or prior distributions.

{For the noise correlation, we set $\matsym{\Sigma}$ in the measurement model such that
}
the noise components in $\randomvec{W}$ are correlated pairwise with correlation factor $\rho$, namely $\squareb{\matsym{\Sigma}}_{i,j}=\rho \sigma_i\sigma_j \quad\forall i\neq j$ and $\squareb{\matsym{\Sigma}}_{i}= \sigma_i^2\quad \forall i$  . 
{
The LBCRB  computed at 
two {representative values of $\rho$, $\rho=0$ (uncorrelated noise), and  $\rho=0.9$}
is shown}
in Figure~\ref{fig:q_res}.

{To assess the learning error of the LBCRB in this (highly) non-linear problem,} we evaluate the LBCRB using the true score functions  $\lscore{\x}{\p}=\vectorsym{s}_F^0$ and $\priorscore{\p}=\vectorsym{s}_P^0$ for uncorrelated noise $\rho=0$. {(Recall that 
 for $\rho=0$ and known model parameters, the true score function $\vectorsym{s}_F^0$ is available analytically.)} 
  It is evident {in Figure~\ref{fig:q_res}} that the LBCRB with the true score {essentially} coincides with the LBCRB with the learned score, indicating {
  the high accuracy of the learned LBCRB. 
  }

We observe that the bound plateaues at low and high SNRs where the bound is dominated by the prior FIM (marked in Figure~\ref{fig:q_res} as the "no information region}). This is because at very high SNRs {the information conveyed by the quantized measurements is greatly reduced} as shown in \cite{9664619} in non-Bayesian cases. {(Intuitively, the noise serves as dithering before quantization, which is known to help overcome quantization error.)} Here, since we are investigating the Bayesian bound, in this region the prior FIM becomes the dominant part, thus the bound plateaues. By examining the mid SNR range (between -8dB to 8dB) we observe that increasing the correlation factor $\rho$ reduces the bound (this was validated over several $\rho$ values). 

{To gain further insight and demonstrate the effectiveness of the \name{}}, we compare to the performance of an approximation of the MMSE estimator, implemented using a neural network.
Specifically, we train a neural network to estimate $\p$ from $\X$ 
{
using the mean square error as the loss function for training}. We find that {with uncorrelated noise}, the MMSE estimator attains the LBCRB across all SNRs. Conversely, in the presence of correlated noise, the MMSE meets the bound in the no-information region, but in the mid SNR range it does not always achieve the bound.
{This suggests that in the latter case, there is room for improvement over our approximate MMSE estimator.}

\begin{figure}
    \centering
    \includesvg[width=0.8\linewidth]{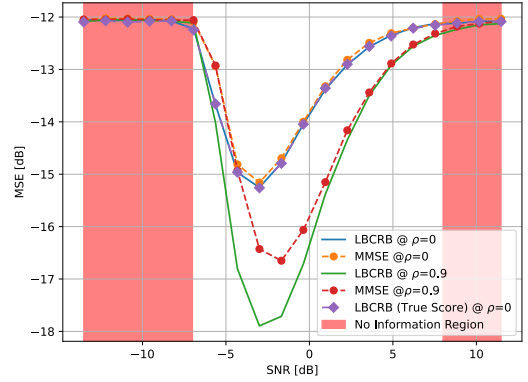}
    \caption{\name{} for the linear Gaussian model with 1-bit quantized measurement in 
    two case:(i) with correlated noise ($\rho=0.9$); and (ii) uncorrelated noises ($\rho=0$). 
    The two no-information region are marked in red. "True score" stands for evaluating the LBCRB without any {approximation error}.  
    }
    \label{fig:q_res}
\end{figure}
In conclusion, we  highlight the advantages of employing Fisher score matching {in quantized measurement scenarios such as the one considered in this experiment.} Previous research \cite{habi2022generative} 
obtains a non-Bayesian learned bound for quantized measurement by using normalizing flows. However, this approach requires an extra approximation via dequantization of the measurement distribution. This requirement arises because normalizing flows depend on an invertible mapping between the data distribution and a manageable base distribution, often a standard Gaussian. Instead, by learning the Fisher score function, we eliminate the need for this additional approximation.
In Bayesian contexts, normalizing flows can be employed to learn the posterior distribution and subsequently derive the posterior score. However, as demonstrated  in {the next subsection,} this method does not take advantage of  additional side information on the measurement model.
\subsubsection{Frequency Estimation }
\begin{figure}
    \centering
    \includesvg[width=0.8\linewidth]{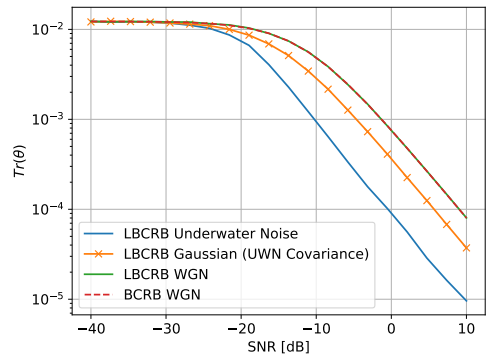}
    \caption{{
    Bounds on frequency estimation with real ocean underwater noise (UWN) vs. Gaussian noises. The \name{} for white Gaussian noise (WGN) coincides with the analytical BCRB for WGN.
    }
    %
    }
    \label{fig:under_weahter_freq}
\end{figure}

We consider the frequency estimation problem of Section~\ref{sec:freq_est}. We train a \pe{} score neural network on a frequency estimation problem (with parameters $\nx=16$, $\niiddata=1$ and $\alpha=100$)
with  ocean underwater ambient noise. We assume that the cosine dependence $\squareb{\mathcal{M}\brackets{\p}}_n=\cos\brackets{{\p}\cdot n}$ is known, and the 
NN only requires to learn the underwater ambient noise. The NN
is a five-layer multilayer perceptron (MLP) that uses Swish\cite{ramachandran2017searching} activation functions between layers, incorporating conditioning inputs such as SNR and $\p$. We compare the \name{} with the underwater ambient noise to the BCRB with white Gaussian noise (WGN) when both noises have the same variance. The various bounds  on the {frequency estimate of $\p$} are shown in Figure~\ref{fig:under_weahter_freq}.

We observe that {the \name{}}
with underwater ambient noise {is uniformly lower than the BCRB for 
WGN of the same variance}. 
This is expected, since WGN yields the largest CRB for a given noise variance \cite{stoica2011gaussian}. 

%
For further comparison, 
we 
compute the LBCRB for the same problem,
but with Gaussian noise having the same covariance matrix as the underwater noise. 
{The resulting bound (denoted in Figure~\ref{fig:under_weahter_freq} by "Gaussian (UWN Covariance)" lies below the BCRB for WGN, but above the \name{} for underwater noise.}  

{First, we observe that the LBCRB and BCRB with WGN coincide, 
demonstrating the good accuracy of the LBCRB. }
{Moreover, } {the comparison between the \name{} with the Gaussian (UWN Covariance) noise and the \name{} with the underwater noise} demonstrates a situation where the LBCRB {with Underwater noise} 
provides a more accurate bound {than would be obtained under the Gaussian assumption.} This potentially highlights an area for improvement in estimation {algorithms in such scenarios.}

%% file: files/appendix/implamentation_deatiles.tex
\subsection{Implementation Details}
{
We 
use a conditional NN for the score NN
and (similar to \cite{habi2024learning}) train a single NN  on multiple S-s, which are used as a conditioning variable. 
}
\subsubsection{Neural Network Structure}
\begin{figure}
    \centering
    \input{files/tikz/inject_mlp}
    \caption{Basic NN Block}
    \label{fig:basic_block}
\end{figure}
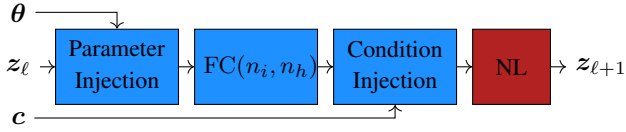

basic blocks {
illustrated in Figure~\ref{fig:basic_block}, each performing the following sequence of operations:}
(i) parameter injection, 
where the input is scaled and shifted based on the parameter vector $\p$; 
(ii) a fully connected operation; 
(iii) condition injection (not used in the prior score model $\priorscore{\p}$ ) involving the application of a scale factor to the features; and finally (iv) 
point-wise Swish \cite{ramachandran2017searching} non-linearity.
{The parameters of the NN structures used for the various examples are listed in Table~\ref{tab:net-param}.}
\subsubsection{Hyper-parameters {and Training}}\label{apx:param}
In all experiments, we use the following hyperparameters. We train the score neural network for  200 epochs using the AdamW \cite{loshchilov2018decoupled} optimizer with learning rate 4e-4 and weight decay $1e-4$, 
and a data set of $60k$ samples for each S- condition.
If a smaller dataset is used, then the number of epochs is adjusted to keep the same
number of gradient updates.
The batch size is set to 512. We employ exponentially decayed averages (EMA) of the weights with a decay rate of 0.999 as suggested in \cite{song2020improved}. At the end of the training we take the weights from the last EMA update. 
\begin{table}[]
\caption{Neural Network Configuration and Parameters for Each Example. The Score Model specifies the type of score function approximated by the NN.
"N Blocks" indicates the number of basic blocks. 
"Non-linearity" {indicates the presence/absence of a nonlinearity in the NN.} {"True Score Realizable" indicates whether the true score function} can be represented by the NN. 
{{"\pe{}"} {indicates whether the NN is \pe{}.} }. 
}
\label{tab:net-param}
\resizebox{\columnwidth}{!}{%
\begin{tabular}{ccccccc}
\hline
Example                                                                                          & \begin{tabular}[c]{@{}c@{}}Score\\ Model\end{tabular} & \begin{tabular}[c]{@{}c@{}}N\\ Blocks\end{tabular} & Width & \begin{tabular}[c]{@{}c@{}}Non\\ Linearity\end{tabular} & \begin{tabular}[c]{@{}c@{}}True \\ Score \\ Realizable\end{tabular} & \begin{tabular}[c]{@{}c@{}}Physics \\ Encoded\end{tabular} \\ \hline
Linear                                                                                           & Posterior                                             & 1                                                  & -    & -                                                      & $\cmark$                                                           & -                                                \\
\multirow{2}{*}{Linear}                                                                          & Fisher                                            & 1                                                  & -    & -                                                      & $\cmark$                                                           & -                                                 \\
                                                                                                 & Prior                                                 & 1                                                  & -    & -                                                      & $\cmark$                                                           & -                                                 \\
\multirow{2}{*}{Quantization}                                                                    & Fisher                                            & 3                                                  & 96    & $\cmark$                                                        & -                                                         & -                                              \\
                                                                                                 & Prior                                                 & 1                                                  & -    & -                                                      & $\cmark$                                                           & -                                                 \\
\multirow{2}{*}{\begin{tabular}[c]{@{}c@{}}Frequency Estimation\\ Gaussian Noise \end{tabular}}   & Fisher                                            & 3                                                  & 96    & $\cmark$                                                        & -                                                          & -                                                 \\
                                                                                                 & Prior                                                 & 3                                                  & 96    &$\cmark$                                                        & -                                                          & -                                                  \\
\multirow{2}{*}{\begin{tabular}[c]{@{}c@{}}Frequency Estimation\\ Gaussian Noise\end{tabular}}   & Fisher                                            & 1                                                  & -    & -                                                      & $\cmark$                                                           & $\cmark$                                                  \\
                                                                                                 & Prior                                                 & 3                                                  & 96    & $\cmark$                                                        & -                                                          &-                                                \\
\multirow{2}{*}{\begin{tabular}[c]{@{}c@{}}Frequency Estimation\\ Underwater Noise\end{tabular}} & Fisher                                            & 2                                                  & 96    & $\cmark$                                                        & -                                                          & $\cmark$                                                  \\
                                                                                                 & Prior                                                 & 3                                                  & 96    & $\cmark$                                                       & -                                                          & -                                              \\ \hline
\end{tabular}%
}
\end{table}
\subsubsection{Measurement Models {Setup}}\label{apx:init_sp}
The linear measurement model \eqref{eq:linear-model} or the quantized linear measurement model \eqref{eq:lin_q} are {set up as follows.} Each element in $\matsym{A}$ is generated by $\squareb{\matsym{A}}_{ij}\sim\normaldis{0}{1}$. For the covariance matrix, we first generate $\matsym{U}\in\mathbb{R}^{d_x\times d_x}$   as 
$\squareb{\matsym{U}}_{ij}\sim\normaldis{0}{1}$ and then $\matsym{\Sigma}=\frac{\matsym{U}\matsym{U}^T}{\trace{\matsym{U}\matsym{U}^T}}$. 

%% file: files/tikz/inject_mlp.tex
\begin{tikzpicture}[node distance=2cm]
    \node (ry) {$\vectorsym{z}_\ell$};
    \node[below=0.25 cm of ry] (rc) {$\vectorsym{c}$};
    \node[above=0.25 cm of ry] (rp) {$\p$};
    \node[fc,text width=1.4cm,minimum width=1.4cm,right=0.2cm of ry,align=center] (fc2)  {\small Parameter \\
    Injection};
    \node[fc,text width=1.4cm,minimum width=1.5cm,right =0.2cm of fc2,align=center] (fc1)  {\small $\mathrm{FC}(n_{i},n_{h})$};

    \node[fc,text width=1.4cm,minimum width=1.4cm,right=0.2cm of fc1,align=center] (fc_n)  {\small Condition \\
    Injection};
        \node[pool,text width=0.8cm,minimum width=0.5cm,right=0.2cm of fc_n,align=center] (nl)  {\small NL};
    \node[right =0.2cm of nl] (out) {$\vectorsym{z}_{\ell+1}$};
    \draw[->] (ry) -- (fc2);
    \draw[->] (fc2) -- (fc1);
    \draw[->] (fc1) -- (fc_n);
    \draw[->] (fc_n) -- (nl);
    \draw[->] (nl) -- (out);
    \draw[->] (rc) -| (fc_n);
    \draw[->] (rp) -| (fc2);
  \end{tikzpicture}

%% file: files/proofs/proofs.tex
\section{Proofs}\label{sec:proofs}
\input{files/proofs/liklihood_score_proof}

\input{files/proofs/bound_error}

\input{files/proofs/consistency}

%% file: files/proofs/liklihood_score_proof.tex
\subsection{Proof Fisher Score Matching Theorem~\ref{thm:liklihood}}\label{sec:lik_score_proof}
We will need the following minor extension of
Lemma 4 from \cite{hyvarinen2005estimation} to include, in addition to infinite limits, also the case of a finite boundary as in \cite{liu2022estimating}: 
\begin{lemma}\label{lemma:int_by_parts} Let $\z\in\mathcal{Z}\subset\mathbb{R}^n$ {
and assume that 
${h}:\mathcal{Z}\rightarrow\mathbb{R}$ and $g:\mathcal{Z}\rightarrow\mathbb{R}$ are differentiable functions. T}hen: 

    \begin{align}
        &\lim\limits_{a\rightarrow \partial z_1^{-},b\rightarrow \partial z_1^{+}} {h}\brackets{a,z_2,\hdots,z_n}g\brackets{a,z_2,\hdots,z_n}\\
        &-{h}\brackets{b,z_2,\hdots,z_n}g\brackets{b,z_2,\hdots,z_n}\\
        &=\int_{\zeta_1^{-}}^{z_1^{+}}{h}\brackets{\z}\frac{\partial g\brackets{\z}}{\partial z_1}dz_1+\int_{z_1^{-}}^{z_1^{+}}g\brackets{\z}\frac{\partial {h}\brackets{\z}}{\partial z_1}dz_1,\nonumber
    \end{align}
    where $\partial z_1^{-}$ and $\partial z_1^{+}$ represent the lower and upper limits of $z_1$, respectively{, which may be finite or infinite.}
\end{lemma}
\begin{proof}
    \begin{equation}
        \frac{\partial {h}\brackets{\z}g\brackets{\z}}{\partial z_1}={h}\brackets{\z}\frac{\partial g\brackets{\z}}{\partial z_1}+g\brackets{\z}\frac{\partial {h}\brackets{\z}}{\partial z_1}.
    \end{equation}
    Now integrate w.r.t. $z_1$, 
    keeping 
    the rest of the elements in $\z$ 
    fixed.
\end{proof}
Note that {although we only wrote the case of $i=1$,}
Lemma~\ref{lemma:int_by_parts} can be applied to any component $\z_i$ of $\z$. 

Now we proceed to prove the {Fisher Score Matching} Theorem.  
\begin{proof}
{
Subject to Assumption~\ref{ass:bound_expection} 
}
\begin{align}\label{eq:step_one_proof_alt}
        \lossfs\brackets{\paramf}&=\expectation{\norm{\lscore{\X}{\p;\paramf}-\nabla_{\p}\log\probt{\randomvec{X}|\p}{\X|\p}}_2^2}{\randomvec{X},\p},\nonumber\\
        &=\expectation{\norm{\lscore{\X}{\p;\paramf}}_2^2}{\randomvec{X},\p}\nonumber\\
        &-2\expectation{\lscore{\X}{\p;\paramf}^T\nabla_{\p}\log\probt{\randomvec{X}|\p}{\X|\p}}{\randomvec{X},\p}+C.
    \end{align}
    where $C=\expectation{\norm{\nabla_{\p}\log\probt{\randomvec{X}|\p}{\X|\p}}_2^2}{\randomvec{X},\p}$  is constant independent of $\paramf$. %
     {
    For convenience, we drop $\paramf$ 
    in the sequel, 
    denoting
    $\lscore{\X}{\p}=\lscore{\X}{\p;\paramf}$. }
    Define
\begin{align}\label{eq:direct_r_define_v2}
    r&\triangleq2\expectation{\nabla_{\p}\log\probt{\vectorsym{x}|\p}{\X|\p}^T\lscore{\X}{\p}}{\randomvec{X},\p}\nonumber\\
    &=2\sum_i\expectation{\squareb{\nabla_{\p}\log\probt{\vectorsym{x}|\p}{\X|\p}}_i\squareb{\lscore{\X}{\p}}_i}{\randomvec{X},\p}.
\end{align}
$w_i\triangleq\expectation{\squareb{\nabla_{\p}\log\probt{\vectorsym{x}|\p}{\X|\p}}_i\squareb{\lscore{\X}{\p}}_i}{\randomvec{X},\p}$, then:
\begin{align}\label{eq:w_before_egt}
    &w_i\\
    &=\int_{\x\in\Upsilon}\int_{\p\in\Theta} \squareb{\lscore{\X}{\p}}_i\divc{\log\probt{\vectorsym{x}|\p}{\X|\p}}{\squareb{\p}_i} \probt{\x,\p}{\X,\p}d\x d\p\nonumber\\
    &=\int_{\x\in\Upsilon}\int_{\p\in\Theta} \squareb{\lscore{\X}{\p}}_i\divc{\probt{\vectorsym{x}|\p}{\X|\p}}{\squareb{\p}_i} \probt{\p}{\p}d\x d\p.\nonumber
\end{align}
     {
    Thanks to the boundary 
     condition \eqref{eq:boundary_condtions_gen_direct} 
     and Assumptions~\ref{assum:diff_prob} and ~\ref{assum:diff_net} we can 
     apply} Lemma ~\ref{lemma:int_by_parts} by setting $g\brackets{\x,\p}=\divc{\probt{\vectorsym{x}|\p}{\X|\p}}{\squareb{\p}_i}$ and ${h}\brackets{\x,\p}=\squareb{\lscore{\X}{\p}}_i\probt{\p}{\p}$. 
     This yields
\begin{align}\label{eq:finaly_step_direct}
    &w_i\nonumber\\
    &=-\int_{\x\in\Upsilon}\int_{\p\in\Theta}\probt{\vectorsym{x}|\p}{\X|\p} \divc{\squareb{\lscore{\X}{\p}}_i\probt{\p}{\p}}{\squareb{\p}_i}d\x d\p\nonumber\\
        &=-\expectation{\divc{\squareb{\lscore{\X}{\p}}_i}{\squareb{\p}_i}}{\X,\p}
        \nonumber\\
        &-\expectation{\squareb{\lscore{\X}{\p}}_i\squareb{\nabla_{\p}\log\probt{\p}{\p}}_i}{\X,\p}.
\end{align}
Finally,  combining \eqref{eq:step_one_proof_alt}, \eqref{eq:direct_r_define_v2} and \eqref{eq:finaly_step_direct} we obtain 
\eqref{eq:score_match_param_gen_direct}.
\end{proof}

%% file: files/proofs/bound_error.tex
\subsection{Proofs of 
the \name{} {Approximation Error}}
\newcommand{\vDelta}{\vectorsym{\Delta}}
\newcommand{\deltasvec}[0]{\vectorsym{\Delta}_{s}}
\newcommand{\shat}[0]{\hat{\vectorsym{s}}}
\newcommand{\svec}[0]{\vectorsym{s}}
{We will need the following technical result.}
\begin{lemma}[{From vector Difference to Covariance Difference}]\label{lemma:cov_vec_error}
    Let {$\Xr,\Yr \in \mathbb{R}^n$ be random vectors on a common probability space, and define \begin{equation*}
        \mathcal{J}=\expectation{\norm{\Xr-\Yr}_2^2}{}, \quad \quad  \matsym{F}=\expectation{\Yr\Yr^T}{}.
    \end{equation*}
    Then:}
    {
\begin{align}\label{eq:vector_error}
\frac{\norm{\expectation{\Xr\Xr^T -\Yr\Yr^T}{}}_2}{\norm{\matsym{F}}_2} \leq \frac{\mathcal{J}}{\norm{\matsym{F}}_2}+2\sqrt{\frac{\mathcal{J}}{\norm{\matsym{F}}_2}}.
\end{align}
Furthermore, if $\frac{\mathcal{J}}{\expectation{\norm{\Yr}_2^2}{}} < {0.16}$, then the RHS of \eqref{eq:vector_error} is upper bounded by
\begin{equation} \label{eq:vector_error-simplified}
2.4\sqrt{\frac{\mathcal{J}}{\norm{\matsym{F}}_2}} =
2.4\sqrt{\intdim\brackets{\F}\frac{\mathcal{J}}{\expectation{\norm{\Yr}_2^2}{}}}.
\end{equation}
    }    
\end{lemma}
\begin{proof}
    {Let $\vDelta \triangleq \Xr-\Yr$. Then
    \begin{align}\label{eq:bound_proof_eq_delta_cov}
 \matsym{P}&=\norm{\expectation{\Xr\Xr^T -\Yr\Yr^T}{}}_2 \nonumber\\
 &=
 \norm{\expectation{\vDelta\vDelta^T +\vDelta\Yr^T + \Yr \vDelta^T}{}}_2
    \end{align}
    }
    Considering
    the first term in \eqref{eq:bound_proof_eq_delta_cov} yields
    {\begin{equation}\label{eq:part_one}
        \norm{\expectation{\vDelta\vDelta^T}{}}_2 \leq 
        \trace{\expectation{{\vDelta\vDelta^T}}{}} =  \expectation{\norm{\vDelta}_2^2}{}=\mathcal{J}
        \end{equation}
    }
    Next, {considering}  the second term in \eqref{eq:bound_proof_eq_delta_cov}, yields:
    \begin{align}\label{eq:part_two}
        &\norm{\expectation{\vDelta\Yr^T+\Yr\vDelta^T}{}}_2 
        {=} 2 \norm{\expectation{\vDelta\Yr^T}{}}_2\nonumber\\
        &\leq 2\sqrt{\norm{\expectation{\vDelta\vDelta^T}{}}_2\norm{\expectation{\Yr\Yr^T}{\z}}_2}\nonumber\\
        &\leq 2\sqrt{\expectation{\norm{\vDelta}_2^2}{}\norm{\expectation{\Yr\Yr^T}{\z}}_2}
    \end{align}
    where the first equality holds by the identity $\norm{\matsym{A} + \matsym{A}^T}_2 = 2\norm{\matsym{A}}_2$, the inequality on the second line follows by Lemma XIII.1 from \cite{habi2023learning}, and the third line by \eqref{eq:part_one}.
    Combining \eqref{eq:part_one}, \eqref{eq:part_two} with \eqref{eq:bound_proof_eq_delta_cov}, {applying the triangle inequality,} and dividing by $\norm{\matsym{F}}_2$ {establishes \eqref{eq:vector_error}}:
    \begin{equation}\label{eq:covar_error_full}
        \frac{\matsym{P}}{\norm{\F}_2}\leq\frac{\mathcal{J}}{\norm{\matsym{F}}_2}+2\sqrt{\frac{\mathcal{J}}{\norm{\matsym{F}}_2}}.
    \end{equation}
   {Now, it is easily verified that $x + 2 \sqrt{x} \leq 2.4 \sqrt{x} $ holds for any $0 \leq x \leq 0.16$. 
This leads immediately to the upper bound on the LHS of \eqref{eq:vector_error-simplified}.  
   } 
    Finally, replacing 
    \begin{align}
        \frac{\mathcal{J}}{\norm{\matsym{F}}_2}&=\frac{\trace{\matsym{\F}}\mathcal{J}}{\trace{\matsym{\F}}\norm{\matsym{F}}_2}=\mathrm{intdim}\brackets{\F}\frac{\mathcal{J}}{\trace{\matsym{\F}}}\nonumber\\
                                              &=\mathrm{intdim}\brackets{\F}\frac{\mathcal{J}}{\expectation{\norm{\Yr}_2^2}{}}
    \end{align}
    {yields the alternative form of the upper bound in \eqref{eq:vector_error-simplified}.}
\end{proof}
{With} Lemma~\ref{lemma:cov_vec_error} {in hand,} we {proceed to prove the various bounds on the {Approximation error}}.
\subsubsection{Posterior {FIM } {Approximation Error} {Thm.~\ref{thm:lrn:direct}}}\label{proof:post:lrn}
\begin{proof}
    We set $\Xr=\postscores{\p}{\xset}$ and $\Yr=\nabla_{\p}\log\probt{\p|\xset}{\p|\xset}$, then we apply Lemma~\ref{lemma:cov_vec_error}, 
    {establishing \eqref{eq:error_learn_post_approch}}. 
\end{proof}

\subsubsection{Measurement-Prior {FIM } {Approximation error} {Thm.~\ref{thm:lrn:lik_prior}}}\label{proof:lik_prior:lrn}
\begin{proof}
{
We apply Lemma~\ref{lemma:cov_vec_error} twice.
First, for} the measurement FIM  
we set $\Xr=\lscores{\x}{\p}$ and $\Yr=\nabla_{\p}\log\probt{\X|\p}{\x|\p}$
{in} Lemma~\ref{lemma:cov_vec_error}, which results in
\begin{equation}\label{eq:mfim_error}
    \frac{\norm{\tlmfim-\fm}_2}{\norm{\fm}_2}\leq \mfle\triangleq {2.4}\sqrt{\mathrm{intdim}\brackets{\fm}\cdot \frac{\lossfs}{\trace{\fm}}}.
\end{equation}
{Second, for} the 
prior FIM, we set $\Xr=\priorscores{\p}$ and $\Yr=\nabla_{\p}\log\probt{\p}{\p}$
{in} Lemma~\ref{lemma:cov_vec_error}, yielding
\begin{equation}\label{eq:pfim_error}
    \frac{\norm{\tlpfim-\fp}_2}{\norm{\fp}_2}  \leq\eta_{p}^{(l)}\triangleq {2.4}\sqrt{\mathrm{intdim}\brackets{\fp}\cdot \frac{\lossps}{\trace{\fp}}}.
\end{equation}
Finally, we use the BFIM decomposition $\fb=\niideval\cdot\fm+\fp$ with the triangular inequality. Then, we employ equations \eqref{eq:mfim_error} and \eqref{eq:pfim_error}, to obtain Theorem~\ref{thm:lrn:lik_prior}.
\end{proof}



\input{files/proofs/sample_error_v2}
\input{files/proofs/c_relation}

\input{files/proofs/empircal_mean_error_convergance}

\subsection{
LBCRB Relative Error Bound}\label{apx:proof_inv_re}
\subsubsection{
Relative Error Posterior Approach {Corr.~\ref{corr:bound_inv_post}}}\label{sec:proof_post_inv}
\begin{proof}
    If $\lbfimbs \succ 0$ it is  invertible and
    \begin{align}\label{eq:mat_inv_post}
        \mathrm{RE}_{B}&=\frac{\norm{\lbcrbbs-\bcrb}_2}{\norm{\bcrb}_2}=\frac{\norm{\lbfimbs^{-1}\brackets{\fb-\lbfimbs}\fb^{-1}}_2}{\norm{\bcrb}_2}\nonumber\\
        &\leq\norm{\lbcrbbs}_2\norm{\fb-\lbfimbs}_2\nonumber\\
        &\leq\norm{\lbcrbbs}_2\norm{\lbfimb-\lbfimbs}_2+\norm{\lbcrbbs}_2\norm{\fb-\lbfimb}_2 .
    \end{align}
    By combining \eqref{eq:error_learn_post_approch} and \eqref{eq:re_error_post_mean}  with \eqref{eq:mat_inv_post} we have:
    \begin{align}\label{eq:mat_inv_post_step2}
        \mathrm{RE}_{B}\leq\norm{\lbcrbbs}_2\brackets{\norm{\lbfimb}_2\bfse+\norm{\fb}\bfle}.
    \end{align}    
    Multiplying and dividing\eqref{eq:mat_inv_post_step2}  by $\norm{\lbfimbs}_2$  results in \eqref{eq:inv_bound_post}. Next, we provide the {
    condition 
    for $\lbfimbs\succ 0$.} {From \eqref{eq:error_learn_post_approch}, \eqref{eq:re_error_post_mean} and using the triangular  inequality we have
    \begin{align}\label{eq:fimb_bound_joint}
        \norm{\lbfimbs-\fb}_2&\leq\norm{\lbfimbs-\lbfimb}_2+\norm{\lbfimb-\fb}_2\nonumber\\
                              &\leq \norm{\lbfimb}_2\bfse+\norm{\fb}\bfle 
    \end{align}
    
    }
    
    Using Assumption~\ref{assume:non_singular}  we have that $\fb\succ 0$, and by definition we have that $\lbfimb$ {is symmetric}. 
    Now by \eqref{eq:fimb_bound_joint} we have with probability $1-\exp\brackets{-u}$, $\lbfimbs \succ0$ 
    if
        \begin{equation*}
       \norm{\lbfimb}_2\bfse+\norm{\fb}\bfle <\eigmin{\fb},
    \end{equation*}
    {where the inequality follows from a standard} matrix perturbation result, cf. \cite[Theorem 2.2]{stewart1977perturbation}.
    Finally, dividing both sides by $\norm{\fb}_2$ 
    yields \eqref{eq:inv_cond_post}.
\end{proof}
\subsubsection{
Relative Error Measurement-Prior Approach {Corr.~\ref{corr:bound_inv_split}}}\label{sec:proof_lik_prior_inv}
\begin{proof}
If $\lbfimlps \succ 0$ then it is invertible, 
and
\begin{align}\label{eq:mat_inv_mp}
    \mathrm{RE}_{MP}&\triangleq\frac{\norm{\lbcrblps-\bcrb}_2}{\norm{\bcrb}_2}\nonumber\\
                    &=\frac{\norm{\lbcrblps\brackets{\fb-\lbfimlps}\bcrb}_2}{\norm{\bcrb}_2}\nonumber\\
                    &\leq \norm{\lbcrblps}_2\norm{\fb-\lbfimlps}_2\nonumber\\
                    &\leq  \norm{\lbcrblps}_2\norm{\lbfimlp-\lbfimlps}_2\nonumber\\
                    &+\norm{\lbcrblps}_2\norm{\lbfimlp-\fb}_2
\end{align}
    By combining \eqref{eq:error_learn_lik_prior_approch} and \eqref{eq:re_error_mp_mean}  with \eqref{eq:mat_inv_mp} we have:
    \begin{align}\label{eq:mat_inv_mp_step2}
        \mathrm{RE}_{MP}&= \norm{\lbcrblps}_2 \brackets{\norm{\fm}_2\mfle+\norm{\fp}_2\pfle}\nonumber\\
                        &+\norm{\lbcrblps}_2\norm{\lbfimlp}_2 \mpfse
    \end{align}
    Multiplying and dividing \eqref{eq:mat_inv_mp_step2}  by $\norm{\lbfimlps}_2$  results in \eqref{eq:inv_bound_mp}.
    Next, we provide the 
    condition for
    {$\lbfimlps \succ 0$}. {From \eqref{eq:error_learn_lik_prior_approch}, \eqref{eq:re_error_mp_mean} and using the triangular  inequality we have
    \begin{align}\label{eq:fim_mp_bound_joint}
        \norm{\lbfimlps-\fb}_2&\leq\norm{\lbfimlps-\lbfimlp}_2+\norm{\lbfimlp-\fb}_2\nonumber\\
                              &\leq \norm{\lbfimb}_2\bfse+\norm{\fb}\bfle 
    \end{align}}

    By definition we have $\lbcrblps\succeq 0$ and $\fb\succ 0$, using \eqref{eq:error_learn_lik_prior_approch} and \eqref{eq:re_error_mp_mean} we have with probability $1-\exp\brackets{-u}$, $\lbfimbs \succ 0$ 
    if:
    \begin{equation*}
       \norm{\fm}_2\mfle+\norm{\fp}_2\pfle+\norm{\lbfimlp}_2 \mpfse <\eigmin{\fb},
    \end{equation*}
    {where the inequality is by the matrix perturbation result as above. }
    Dividing both sides by $\norm{\fb}_2$ 
    yields\eqref{eq:inv_cond_post}.
    \end{proof}

%% file: files/proofs/sample_error_v2.tex
\subsection{Proofs of 
Empirical-Mean Error}\label{apx:proof_sample_error}
{We will use the following result, which is an easy corollary of} "Matrix Bernstein: Hermitian Case with Intrinsic Dimension" \cite[Chapter 7]{tropp2015introduction}.
\begin{prop}
    [Matrix Bernstein {for Spectral Norm}: Symmetric Case with Intrinsic Dimension]\label{prop:bernstein}
    Let $\matsym{X}_k$ be a finite set of random symmetric matrices, $\matsym{Y}=\sum_{k}\matsym{X}_k$ be a sum over the finite set,  $$\matsym{V}\succeq \expectation{\matsym{Y}^2}{}=\sum_k\expectation{\matsym{X}_k^2}{},$$ 
    be a semi-definite upper bound, and $\nu=\norm{\matsym{V}}_2$ its {spectral} norm.     Assume that:
    $$\expectation{\matsym{X}_k}{\matsym{X}_k}=0\quad\mathrm{and} \quad \norm{\matsym{X}_k}_2\leq L \quad\forall k,$$
     then for $t\geq \sqrt{\nu}+\frac{L}{3}$,
   \begin{equation}\label{eq:lemma_bernstein_prob}
        \mathbb{P}\brackets{\norm{\matsym{Y}}\geq t}\leq 8 d_i
        \cdot\exp{\brackets{-\frac{0.5t^2}{\nu+Lt/3}}},
    \end{equation}
    where $d_i=\mathrm{intdim}\brackets{\matsym{V}}\triangleq\frac{\trace{\matsym{V}}}{\norm{\matsym{V}}_2}$ is the Intrinsic Dimension of matrix $\matsym{V}$. In addition the Expectation Bound is given by:
    \begin{align}\label{eq:lemma_bound_expection}
        \expectation{\norm{\matsym{Y}}_2}{\ds}&\leq \sqrt{\nu}\brackets{4+\sqrt{2\log\brackets{1+2d_i}}} \nonumber\\ 
        &+\frac{2L}{3}\brackets{4+\log\brackets{
        1+2d_i}}
    \end{align}
    
\end{prop}
\begin{proof}
    {The relation} 
    $\norm{\matsym{Y}}_2=\max\brackets{\lambda_{max}\brackets{\matsym{Y}},\lambda_{max}\brackets{-\matsym{Y}}}$ 
    together with the union bound {yield}
    \begin{align*}
        \mathbb{P}\brackets{\norm{\matsym{Y}}_2\geq t}\leq\mathbb{P}\brackets{\lambda_{max}\brackets{\matsym{Y}}\geq t}+\mathbb{P}\brackets{\lambda_{max}\brackets{-\matsym{Y}}\geq t}.
    \end{align*}
    {Applying} Matrix Bernstein: Hermitian Case with Intrinsic Dimension from \cite[Chapter 7]{tropp2015introduction} twice {establishes}  Proposition~\ref{prop:bernstein}. 
\end{proof}
{Given some $u>0$,} we wish to bound  $\norm{\matsym{Y}}<t$ with probability at least $1-\exp\brackets{-u}$. {This is provided by the following corollary of Proposition~\ref{prop:bernstein}}.
\begin{corollary} \label{cor:bernstein}
Let
\begin{flalign} 
t_u \triangleq &{\frac{bL}{3}}\brackets{1+\sqrt{\frac{{18}\nu}{bL^2}+1}}, 
\text{ where }  b= {u+\log\brackets{8d_i}}. \nonumber
\end{flalign}
Then, subject to the assumptions of Proposition~\ref{prop:bernstein},
    \begin{equation} \label{eq:t_bound}
    \mathbb{P}\brackets{\norm{\matsym{Y}}\leq t_u} \geq 1-\exp\brackets{-u} 
    .
    \end{equation}
\end{corollary}
\begin{proof}
    Consider the event complementary to that in \eqref{eq:t_bound},
%
\begin{align}\label{eq:apply_bound_one}
   {\mathbb{P}\brackets{\norm{\matsym{Y}}\geq  t} \leq \exp\brackets{-u}} .
\end{align}
To ensue that \eqref{eq:apply_bound_one} holds, we 
{
require 
}
\begin{align}\label{eq:apply_bound}
        {8 d_i
        \cdot\exp{\brackets{-\frac{0.5t^2}{\nu+Lt/3}}} \leq \exp\brackets{-u},}
\end{align}
which, {by \eqref{eq:lemma_bernstein_prob} from Proposition~\ref{prop:bernstein},} implies ~\eqref{eq:apply_bound_one}.
{
Solving \eqref{eq:apply_bound} for $t\geq 0$ yields $t\geq t_u$.

}
Now, any $t\geq t_u$
also satisfies $t\geq \sqrt{\nu}+\frac{L}{3}$ as required by Proposition~\ref{prop:bernstein}, and therefore yields a valid bound in \eqref{eq:apply_bound_one}.
%
{Selecting $t=t_u$ yields 
 the tightest bound in \eqref{eq:t_bound}, and establishes the corollary. 
}
\end{proof}

%
%

\subsubsection{Posterior Approach Empirical Mean Error {Thm. ~\ref{thm:sampling_post}}}
\begin{proof}
Denote $\vectorsym{s}_k=\postscores{\p_k}{\x_k}$ {and} $\matsym{X}_k=\frac{1}{\nds}\brackets{\vectorsym{s}_k\vectorsym{s}_k^T-\lbfimb}$. 
 First we validate the assumptions of Proposition~\ref{prop:bernstein}.
    \begin{align*}
        \expectation{\matsym{X}_k}{\ds}=\frac{1}{\nds}\brackets{\expectation{\vectorsym{s}_k\vectorsym{s}_k^T}{\ds}-\lbfimb}=0,
    \end{align*}
    and 
    \begin{align}\label{eq:l_value_post}
        \norm{\matsym{X}_k}&\leq\frac{1}{\nds}\brackets{\norm{\vectorsym{s}_k\vectorsym{s}_k^T}+\norm{\lbfimb}}\nonumber\\
        &\leq \frac{1}{\nds}\brackets{\cb+\norm{\lbfimb}_2} \triangleq L.
    \end{align}
    Next, we calculate $\matsym{V}$:
    \begin{align*}
        \expectation{\matsym{X}_k^2}{\ds}&=\frac{1}{\nds^2}\expectation{{\vectorsym{s}_k\vectorsym{s}_k^T}\vectorsym{s}_k\vectorsym{s}_k^T -\lbfimb^2}{\ds}\\
        &\preceq \frac{1}{\nds^2}\brackets{\cb\expectation{\vectorsym{s}_k\vectorsym{s}_k^T}{\ds} -\lbfimb^2}\preceq  \frac{\cb\lbfimb}{\nds^2}.
    \end{align*}
    {The first inequality is obtained by $\vectorsym{s}_k\vectorsym{s}_k^T\preceq \norm{\vectorsym{s}_k\vectorsym{s}_k^T}_2\matsym{I}$}.
    Then
    \begin{equation}\label{eq:v_value_post}
    \matsym{V}=\frac{\cb \lbfimb}{\nds} \quad\mathrm{and} \quad \nu=\frac{\cb}{\nds}\norm{\lbfimb}_2.
    \end{equation}
           {Now, using} \eqref{eq:l_value_post}, \eqref{eq:v_value_post} in Corollary ~\ref{cor:bernstein} we obtain:
            \begin{align}
              &\frac{t_u}{\norm{\lbfimb}_2}=\frac{\nbe}{4\nds}\brackets{1+\sqrt{1+24\frac{\nds}{\nbe}\frac{\cb}{\cb+\norm{\fb}_2}}}\nonumber\\
              &=\frac{\nbe}{4\nds}\brackets{1+\sqrt{1+24\frac{\nds}{\nbe}\frac{1}{1+\frac{\norm{\fb}_2}{\cb}}}}\nonumber\\
              &\leq \frac{\nbe}{4\nds}\brackets{1+\sqrt{1+24\frac{\nds}{\nbe}}} \label{eq:tBe_bound}
            \end{align}
            where $\nbe {\triangleq}\frac{4}{3}\brackets{u+\log\brackets{8\dbb}}\brackets{\frac{\cb}{\norm{\lbfimb}_2}+1}$. Using the 
            relation {$\trace{\lbfimb}=\dbb\norm{\lbfimb}_2$} 
provides the alternative form
            $\nbe=\frac{4}{3}\brackets{u+\log\brackets{8\dbb}}\brackets{\dbb\frac{\cb}{\trace{\lbfimb}}+1}$.  {Finally, using the easiliy verified inequality $1+\sqrt{1+x} \leq \sqrt{3x/2}$ {for $x\geq 24$,} in \eqref{eq:tBe_bound} yields }
            \begin{equation}\label{eq:bound_t_post_v2}
                \frac{t_u}{\norm{\lbfimb}_2}\leq { 1.5\sqrt{\frac{\nbe}{\nds}}},
            \end{equation}
            for any {$\nds \geq \nbe$}. {Applying the  upper bound of \eqref{eq:bound_t_post_v2}
 in Corollary~\ref{cor:bernstein} yields \eqref{eq:re_error_post_mean} of Theorem~\ref{thm:sampling_post}.}
 
             
             Next, {defining $\phi_B{\triangleq}\frac{\cb}{\norm{\lbfimb}_2}+1$,
             we use \eqref{eq:l_value_post} and  \eqref{eq:v_value_post} in 
        \eqref{eq:lemma_bound_expection} 
        to obtain a bound on the normalized expected value 
             }
    \begin{align*}
        &\expectation{\frac{\norm{\lbfimbs-\lbfimb}_2}{\norm{\lbfimb}_2}}{\ds}\leq 
         \sqrt{\frac{\cb}{\norm{\lbfimb}_2\nds}}\brackets{4+\sqrt{2
         {\psi_B}
         }}\\
        &+\frac{2}{3\nds}\phi_B\brackets{4+
       {\psi_B}
        } \text{ where } \psi_B= \log\brackets{1+2\dbb} 
    \end{align*}
      Using  $\frac{\cb}{\norm{\lbfimb}_2}\leq \phi_B$ and defining $\alpha\triangleq \frac{2\brackets{4+\psi_B}}{3\brackets{4+\sqrt{2\psi_B}}}$ yields 
    {\begin{align*}
        \expectation{\frac{\norm{\lbfimbs-\lbfimb}_2}{\norm{\lbfimb}_2}}{\ds}\leq \brackets{4+\sqrt{2\psi_B}}\brackets{\alpha\frac{\phi_B}{\nds}+\sqrt{\frac{\phi_B}{\nds}} }.
    \end{align*}
    }
     {Now, it is easily verified that $\alpha x+\sqrt{x}\leq 1.5\sqrt{x}$ holds for any $0\leq x \leq \frac{1}{4\alpha^2}$. It  follows that:}
    {\begin{flalign}
      \label{eq:expection_bound_post}
        \textstyle{\expectation{\frac{\norm{\lbfimbs-\lbfimb}_2}{\norm{\lbfimb}_2}}{\ds}} \leq \textstyle{\brackets{6+1.5\sqrt{2\log\brackets{1+2\dbb}}}}\sqrt{\frac{\phi_B}{\nds}}
    \end{flalign}
    for any $\nds \geq 4\alpha^2 \phi_B$.  Since $\psi_B=\log\brackets{1+2\dbb}\geq 1$ it is easy to verify that $4\alpha^2\leq \psi_B +{0.52}$. Hence the error bound \eqref{eq:expection_bound_post} 
    holds for $\nds \geq \brackets{\psi_B +{0.52}}\phi_B$.  Using the relation $\trace{\lbfimb}=\dbb\norm{\lbfimb}_2$ we define $\nbet\triangleq\brackets{\log\brackets{1+2d_B} +{0.52}} \cdot\brackets{\dbb\frac{\cb}{\trace{\lbfimb}}+1}$, producing the form used in Thm. ~\ref{thm:sampling_post}.
    }
\end{proof}
\subsubsection{
Measurement-Prior Empirical Mean Error {Thm.~\ref{thm:sampling_mp}}}
\begin{proof}
  Denote {$\vectorsym{s}_{k,i}=\lscores{\tilde{\x}_{k,i}}{\p_k}$} as the Fisher score  vector  and $\vectorsym{p}_k=\priorscores{\p_k}$ as the prior score of the $k^{th}$ sample in $\ds$.
    Now, let $$\matsym{X}_k=\frac{1}{\nds}\brackets{\frac{\niideval}{\niiddata}\cdot\sum_{i=1}^{\niiddata} \vectorsym{s}_{k,i}\vectorsym{s}_{k,i}^T+ \vectorsym{p}_k\vectorsym{p}_k^T-\lbfimlp}=\frac{1}{\nds}\matsym{R}_k.$$
    First we validate the assumptions of {Proposition~\ref{prop:bernstein}}:
    \begin{align*}
        &\expectation{\matsym{X}_k}{\ds}\nonumber\\
        &=\frac{1}{\nds}\brackets{\niideval\expectation{\vectorsym{s}_{k,i}\vectorsym{s}_{k,i}^T}{\ds}+\expectation{\vectorsym{p}_k\vectorsym{p}_k^T}{\ds}-\lbfimlp}=0,
    \end{align*}
    {for every $i^{th}$ i.i.d sample} and 
    \begin{align}\label{eq:l_value_lik_prior}
        \norm{\matsym{X}_k}&\leq\frac{1}{\nds}\brackets{\frac{\niideval}{\niiddata}\cdot{\norm{\sum_{i=1}^{\niiddata}\vectorsym{s}_{k,i}\vectorsym{s}_{k,i}^T}}+\norm{\vectorsym{p}_k\vectorsym{p}_k^T}+\norm{\lbfimlp}}\nonumber\\
        &\leq \frac{1}{\nds}\brackets{\niideval \cm+\cp+\norm{\lbfimlp}_2}=L.
    \end{align}
   Next, we calculate $\matsym{V}$:
    \begin{align*}
        &\matsym{R}^2_k={\frac{\niideval}{\niiddata}\sum_{i=1}^{\niiddata}\vectorsym{s}_{k,i}\vectorsym{s}_{k,i}^T}{\frac{\niideval}{\niiddata}\sum_{i=1}^{\niiddata}\vectorsym{s}_{k,i}\vectorsym{s}_{k,i}^T}+\vectorsym{p}_k\vectorsym{p}_k^T\vectorsym{p}_k\vectorsym{p}_k^T\nonumber\\
        &+\lbfimlp^2+{\frac{\niideval}{\niiddata}\sum_{i=1}^{\niiddata}\vectorsym{s}_{k,i}\vectorsym{s}_{k,i}^T}\vectorsym{p}_k\vectorsym{p}_k^T+\vectorsym{p}_k\vectorsym{p}_k^T{\frac{\niideval}{\niiddata}\sum_{i=1}^{\niiddata}\vectorsym{s}_{k,i}\vectorsym{s}_{k,i}^T}\\
        &-\brackets{{\frac{\niideval}{\niiddata}\sum_{i=1}^{\niiddata}\vectorsym{s}_{k,i}\vectorsym{s}_{k,i}^T}+\vectorsym{p}_k\vectorsym{p}_k^T}\lbfimlp\\
            &-\lbfimlp \brackets{{\frac{\niideval}{\niiddata}\sum_{i=1}^{\niiddata}\vectorsym{s}_{k,i}\vectorsym{s}_{k,i}^T}+\vectorsym{p}_k\vectorsym{p}_k^T}
    \end{align*}
    Taking the expectation and using that $\expectation{\frac{\niideval}{\niiddata}\sum_{i=1}^{\niiddata}\vectorsym{s}_{k,i}\vectorsym{s}_{k,i}^T+\vectorsym{p}_k\vectorsym{p}_k^T}{\ds}=\lbfimlp$ yields
    \begin{align*}
        &\expectation{\matsym{R}_k^2}{\ds}= \frac{\niideval^2}{\niiddata}\cm\expectation{\sum_{i=1}^{\niiddata}\vectorsym{s}_{k,i}\vectorsym{s}_{k,i}^T} {\ds}+\cp\expectation{\vectorsym{p}_k\vectorsym{p}_k^T}{\ds}\\
        &+\lbfimlp^2+ \frac{\niideval}{\niiddata}\cp\expectation{\sum_{i=1}^{\niiddata}\vectorsym{s}_{k,i}\vectorsym{s}_{k,i}^T} {\ds}+\cm\niideval\expectation{\vectorsym{p}_k\vectorsym{p}_k^T}{\ds}\\
        &-2\lbfimlp^2.
    \end{align*}
    Reordering and using $\expectation{\frac{\niideval}{\niiddata}\sum_{i=1}^{\niiddata}\vectorsym{s}_{k,i}\vectorsym{s}_{k,i}^T+\vectorsym{p}_k\vectorsym{p}_k^T}{\ds}=\lbfimlp$ we have:
    
    \begin{align*}
        \expectation{\matsym{R}_k^2}{\ds}&=\brackets{\niideval \cm+\cp}\lbfimlp-\lbfimlp^2 \\
        &\preceq\brackets{\niideval \cm+\cp}\lbfimlp.
    \end{align*}
    Then,
    \begin{equation}\label{eq:v_value_mp}
\matsym{V}=\frac{\expectation{\matsym{R}_k^2}{\ds}}{\nds}\preceq\frac{\brackets{\frac{\niideval}{\niiddata}\cm+\cp}\lbfimlp}{\nds} 
    \end{equation}
    Finally, {using}  equations \eqref{eq:l_value_lik_prior}, \eqref{eq:v_value_mp} {and} { Corollary~\ref{cor:bernstein} 
    {and following the same steps as  in the proof of Theorem~\ref{thm:sampling_post}},
    we establish Theorem~\ref{thm:sampling_mp}.}

\end{proof}

%% file: files/proofs/c_relation.tex
\subsection{
  Score Norms Squared - Relation (Proposition~\ref{prop:expected_score_ineq})}\label{apx:remark_c_relation_proof}

\begin{proof}
   Define:
    \begin{align*}
        \matsym{A}&\triangleq \norm{\nabla_{\p}\log\probt{\p|\xsetr}{\p|\xset}}_2^2\\
        \matsym{B}&\triangleq\norm{\frac{1}{\niiddata}\sum_{i=1}^{\niiddata}\nabla_{\p}\log\probt{\x_i|\p}{\X|\pr}\nabla_{\p}\log\probt{\x_i|\p}{\X|\pr}^T}_2
    \end{align*}
Then, 
\begin{align}
    &\expectation{\matsym{A}}{\X,\pr}=\expectation{\norm{\sum_{i=1}^{\niiddata}\nabla_{\p}\log\probt{\x_i|\p}{\X|\pr}}_2^2}{\X,\pr}\nonumber\\
    &+\expectation{\norm{\nabla_{\p}\log\probt{\p}{\pr}}_2^2}{\X,\pr}\nonumber\\
    &+\expectation{2\nabla_{\p}\log\probt{\p}{\pr}^T\sum_{i=1}^{\niiddata}\nabla_{\p}\log\probt{\x_i|\p}{\X|\pr}}{\X,\pr} \label{eq:expect_post_score_squared}
\end{align}
Now, by  assumption ~\ref{assume:common_support} we have $\expectation{\sum_{i=1}^{\niiddata}\nabla_{\p}\log\probt{\x_i|\p}{\X|\pr}}{\xset|\pr}$ $=0$, {thus 
eliminating the third term on the RHS of \eqref{eq:expect_post_score_squared}.} Turning to the first term on the RHS of \eqref{eq:expect_post_score_squared} and expanding yields
\begin{align*}
{
    } 
    &\sum_{i=1}^{\niiddata}\expectation{\norm{\nabla_{\p}\log\probt{\x_i|\p}{\X|\pr}}_2^2}{\X,\pr}\nonumber\\
    &+\sum_{\substack{i=1\\
    j=1\\ 
    j\neq i} }\expectation{\nabla_{\p}\log\probt{\x_i|\p}{\X|\pr}^T\nabla_{\p}\log\probt{\x_{j}|\p}{\X|\pr}}{\X,\pr}\nonumber
\end{align*}
Using assumption ~\ref{assume:common_support} the cross-terms zero out resulting in:
\begin{align*}
    &\expectation{\norm{\sum_{i=1}^{\niiddata}\nabla_{\p}\log\probt{\x_i|\p}{\X|\pr}}_2^2}{\X,\pr}
    =\nonumber\\
    &\sum_{i=1}^{\niiddata}\expectation{\norm{\nabla_{\p}\log\probt{\x_i|\p}{\X|\pr}\nabla_{\p}\log\probt{\x_i|\p}{\X|\pr}^T}_2}{\X,\pr}\nonumber\\
    &\geq \niiddata \expectation{\matsym{B}}{\X,\pr}
\end{align*}
Where the inequality  follows by Jensen's inequality. {Combining with \eqref{eq:expect_post_score_squared} (without the third term) yields the result.}
\end{proof}

%% file: files/proofs/empircal_mean_error_convergance.tex
\subsection{
Empirical-Mean Error Convergence Condition (Remark~\ref{remark:mean_error_convergance})}\label{apx:remark_convergance_proof}

\begin{prop}
    \label{prop:nonneg_z}
    Let $\randomvec{Z}_n\sim\probt{\z}{\Z}$ be a sequence of $N$ nonegative {i.i.d} random variables. Assume that the 3rd moment  {of $\probt{\z}{\Z}$} is finite, i.e.,
    ${\expectation{\randomvec{Z}^3}{}}\leq \infty$.  Then
     {
     \begin{equation*}
       \lim\limits_{N \rightarrow \infty } 
       \tfrac{1}{N} \max\limits_{1\leq n\leq N} \randomvec{Z}_n
       = 0 \quad \text{a.s.}
    \end{equation*}
    }
\end{prop}
\begin{proof}
By 
the Markov inequality, for any nonegative r.v.
$Z$,
\begin{equation} \label{eq:Markov3}
\probP(Z > t) = \probP(Z^3 > t^3) \leq \frac{\expectation{Z^3}{}}{t^3}.
\end{equation}
Now, {using first} the union bound {and then \eqref{eq:Markov3} we have}
{
\begin{align}\label{eq:prob_max_union_N}
    \probP\brackets{
   \tfrac{1}{N} \max\limits_{1\leq n\leq N} \randomvec{Z}_n
    >N^{-\tfrac{1}{4}}}
    &\leq N \probP\brackets{\randomvec{Z}>N \cdot N^{-\tfrac{1}{4}}
    } 
     \nonumber\\
      &\leq N \frac{\expectation{Z^3}{}}{(N^{\tfrac{3}{4}})^3 }
          = \frac{\expectation{Z^3}{}}{N^{1.25}}.
\end{align}
}
Next, it is easily established that 
$$ \sum_{N=1}^\infty \frac{\expectation{Z^3}{}}{N^{1.25}} < \infty .$$
Therefore, by the Borel-Cantelli Lemma, the event on the LHS of \eqref{eq:prob_max_union_N} cannot happen infinitely often as $N\rightarrow \infty$.
\end{proof}
\begin{corollary}
    Let $\randomvec{S}_n \in \mathbb{R}^k$, {$\randomvec{S}_n \sim \probt{\boldsymbol{s}}{\randomvec{s}} $} be a sequence of $\nds$ {i.i.d} random vectors. Assume that 
    {
    $\probt{\boldsymbol{s}}{\randomvec{s}}$ has a finite 6$^{th}$ moment
    i.e., $\expectation{\squareb{\randomvec{S}}_i^l\squareb{\randomvec{S}}_j^m}{}<\infty$ 
    for all  $i,j\in[1,k]$, and $1 \leq l+m\leq 6$. 
    }  Then,
    \begin{equation*}
      \lim\limits_{N \rightarrow \infty }   \tfrac{1}{N}\max\limits_{1\leq n\leq N} \norm{\randomvec{S}_n}_2^2
        =0 \quad \text{a.s.}
    \end{equation*}
\end{corollary}
\begin{proof}
   Consider the sequence of nonnegative random variables $\randomvec{Z}_n=\norm{\randomvec{S}_n}^2_2=\sum_i\squareb{\randomvec{S}_n}_i^2$. 
   Because $\randomvec{S}_n$ has  finite moments up to 6$^{th}$ order,
   it follows that $\randomvec{Z}_n$ has a 
   finite 
   3rd moment. Applying Proposition ~\ref{prop:nonneg_z} yields desired result.  
\end{proof}

%% file: files/proofs/consistency.tex
\subsection{Strong Consistency Proofs, Thm.~\ref{thm:post_consist} and Thm.~\ref{thm:mp_consist}}
{The following theorem is a key tool in our proofs. The version presented is  adapted slightly to the case where the domain $\mathcal{S}$ of the parameters $\Omega$ is a subset of Euclidean space.} 
\begin{theorem}[Strong Uniform Law of Large Numbers \cite{andrews1992generic}]\label{lemma:sulln}
    Let  $\Z_n\in\mathbb{R}^n$ be a sequence of random variables and $f:(\mathbb{R}^n,\mathcal{S})\rightarrow \mathbb{R}$ 
    a function.
     Assume that:
    \begin{enumerate}[label=\text{\ref{lemma:sulln}}.\arabic*,labelsep=*, leftmargin=*]
    \item $\mathcal{S}$ is compact.\label{assume:sulln_compact} 
    \item  Lipschitz Condition: $\abs{f\brackets{\Z_n,\Omega_1}-f\brackets{\Z_n,\Omega_2}}\leq B\brackets{\Z_n}h\brackets{
    {\norm{\Omega_1-\Omega_2}_2}}\quad\forall\Omega_1, \Omega_2\in\mathcal{S} \qquad \text{a.s}$
    for some measurable function $B$ and some nonrandom function $h$ that  satisfies {$\lim_{x\rightarrow 0}h\brackets{x} =0$}.\label{assume:lip_cond}
    
    \item $\expectation{B\brackets{\Z_n}}{\Z_n}\leq \infty$\label{assume:bound_b}
    
    \item Pointwise strong convergence 
    $\forall\Omega\in\mathcal{S}$:\label{assume:sulln_conv}
    $$\frac{1}{N}\sum_{n=1}^N \brackets{f\brackets{\Z_n,\Omega}- \expectation{f\brackets{\Z_n,\Omega}}{\Z_n}}\xrightarrow{N\rightarrow\infty}0 \quad \text{a.s.}$$
\end{enumerate}
Then:
\begin{equation*}
    \sup\limits_{\Omega}\abs{\frac{1}{N}\sum_{n=1}^N \brackets{f\brackets{\Z_n,\Omega}- \expectation{f\brackets{\Z_n,\Omega}}{\Z_n}}}\xrightarrow{N\rightarrow\infty}0 \quad \text{a.s.}
\end{equation*}
\end{theorem}

\subsubsection{Posterior Score Consistency Proof (Thm~\ref{thm:post_consist})}
\label{apx:proof:consistency_post}
\begin{lemma}[Posterior Score Loss is Lipschitz continuous]\label{lemma:lip_score}
 Define the per-sample loss of the posterior score matching $\ell_B\brackets{\p,\xsetr;\Omega}\triangleq{\norm{\postscore{\p}{\xsetr;\Omega}}_2^2}+2\trace{\frac{\partial \postscore{\p}{\xsetr;\Omega}{}}{\partial\p}} $, and $\Delta_\Omega \triangleq  \norm{\Omega_{1}-\Omega_{2}}_2$. 
 Assume that:
  \begin{itemize}
      \item {The posterior} score {Jacobian matrix} is Lipschitz continuous 
      $$\norm{\nabla_{\p}\postscore{\p}{\xsetr;\Omega_1}-\nabla_{\p}\postscore{\p}{\xsetr;\Omega_2}}_2\leq \tau_B\brackets{\p,\xsetr}\Delta_\Omega$$
      \item {The score} outer product is Lipschitz continuous
      $$\norm{\matsym{R}\brackets{\p|\xsetr;\Omega_1}-\matsym{R}\brackets{\p|\xsetr;\Omega_2}}_F\leq \zeta_B\brackets{\p,\xsetr}\Delta_\Omega$$
  \end{itemize}
where $\matsym{R}\brackets{\p|\xsetr;\Omega} \triangleq\postscore{\p}{\xsetr;\Omega}\postscore{\p}{\xsetr;\Omega}^T$.  {In addition assume that the following expectations: $\expectation{\tau_B\brackets{\pr,\xset}}{\pr,\xset}<\infty$ and $\expectation{\zeta_B\brackets{\pr,\xset}}{\pr,\xset}<\infty$ are finite.} Then
    \begin{equation}
        \abs{\ell_B\brackets{\p,\xsetr;\Omega_1}-\ell_B\brackets{\p,\xsetr;\Omega_2}}\leq L_B\brackets{\p,\xsetr}\Delta_\Omega,
    \end{equation}
where 
$L_B\brackets{\p,\xsetr}=\sqrt{\np}\brackets{2\tau_B\brackets{\p,\xsetr}+\zeta_B\brackets{\p,\xsetr}}$ is the Liphschitz constant. {Furthermore, $\expectation{L_B\brackets{\pr,\xset}}{\pr,\xset}<\infty$. }

\end{lemma}
{By concatenating $\p$ and $\xsetr$ into a single vector $\z$,
the proof of Lemma~\ref{lemma:lip_score} 
is similar to that of 
Lemma 10 in \cite{crafts2023bayesian}.
}


{Next using Theorem~\ref{lemma:sulln} and Lemma~\ref{lemma:lip_score} we prove the Consistency of the Posterior Score (Theorem~\ref{thm:post_consist}) }
\begin{proof}
{Let $\tilde{\Omega}=\arg\min\limits_{\Omega}\lossbst\brackets{\Omega}$. Then}
    \begin{align}
        &\lossbs\brackets{\Omega^*}-\lossbs\brackets{\tilde{\Omega}}=\lossbst\brackets{\Omega^*}-\lossbst\brackets{\tilde{\Omega}}\nonumber\\
        &=\lossbst\brackets{\Omega^*}-\lossbsm\brackets{\Omega^*}+\lossbsm\brackets{\Omega^*}-\lossbsm\brackets{\tilde{\Omega}} \nonumber\\
        &+\lossbsm\brackets{\tilde{\Omega}}-\lossbst\brackets{\tilde{\Omega}}\nonumber\\
        &{\leq
        \lossbst\brackets{\Omega^*}-\lossbsm\brackets{\Omega^*}
        +\lossbsm\brackets{\tilde{\Omega}}-\lossbst\brackets{\tilde{\Omega}}}\nonumber\\
        &\leq 
        2\sup\limits_{\Omega}\norm{\lossbsm\brackets{\Omega}-\lossbst\brackets{\Omega}}_2,
    \end{align}
    where {the first inequality holds because 
    }
    $\lossbsm\brackets{\Omega^*}-\lossbsm\brackets{\tilde{\Omega}}\leq0$ by definition of $\Omega^*$ and $\tilde{\Omega}$.
    
    {Next we 
    apply Theorem ~\ref{lemma:sulln}, by setting 
    {$\Z_n=[\pr_n,\xset_n]$ as the $n^{th}$ sample from the dataset $\ds$ and}
    $f\brackets{\Z_n;\Omega}= {\ell_B\brackets{\p_n,\xsetr_n;\Omega}} $ as the per-sample loss of the posterior score matching {defined in Lemma~\ref{lemma:lip_score}.  
    }
    As first step we validate that all assumptions in Theorem~\ref{lemma:sulln} hold. Assumptions~\ref{assume:lip_cond} and ~\ref{assume:bound_b} hold using Assumption ~\ref{assume:lip_cont_post} and Lemma~\ref{lemma:lip_score}. Assumption~\ref{assume:sulln_compact} holds by choosing the right the parameter space for the neural network (e.g., using {so-called} weight decay{, i.e., $\ell_2$ regularization}).}

    Next, we show pointwise strong convergence (Assumption~\ref{assume:sulln_conv} ) hold by showing {that 
    $\expectation{\ell_B\brackets{\pr,\xset;\Omega}}{\xset,\pr}$ is finite.   }
    {
    \begin{align}\label{eq:conv_post_exp_finie_p1}
        &\expectation{\ell_B\brackets{\pr,\xset;\Omega}}{\xset,\pr}=
        \expectation{\norm{\postscore{\pr}{\xset;\Omega}}_2^2}{\xset,\pr}\nonumber\\
        &+2\expectation{{\trace{\frac{\partial \postscore{\pr}{\xset;\Omega}}{\partial\pr}}}}{\xset,\pr}.
    \end{align}
    The first term in \eqref{eq:conv_post_exp_finie_p1} is finite using Assumption~\ref{assum:expected_cond_score}. As for the second term we have:
    \begin{align}\label{eq:conv_post_exp_finie_p2}
        &\expectation{{\trace{\frac{\partial \postscore{\pr}{\xset;\Omega}}{\partial\pr}}}}{\xset,\pr}=\nonumber\\
        &\sum_i \expectation{\squareb{\frac{\partial \postscore{\pr}{\xset;\Omega}}{\partial\pr}}_{i,i}}{\xset,\pr}
    \end{align}
    Assuming Boundary {condition} \eqref{eq:boundary_conditions_post} {and} assumptions~\ref{assum:diff_prob_post} and ~\ref{assum:net_cond_score}{ all hold, allows to use 
    }
    Lemma ~\ref{lemma:int_by_parts} by setting $g\brackets{\xset,\pr}=\squareb{\postscore{\pr}{\xset;\Omega}}_i$ and $h\brackets{\xset,\pr}=\probt{\pr|\xset}{\pr|\xset}$ yielding
    \begin{align}\label{eq:conv_post_exp_finie_p3}
       &w_i\triangleq\expectation{\squareb{\frac{\partial \postscore{\pr}{\xset;\Omega}}{\partial\pr}}_{i,i}}{\pr,\xset}\nonumber\\
       &=-\int_{\xset,\pr} \squareb{\postscore{\pr}{\xset;\Omega}}_i \squareb{\frac{\partial\probt{\pr|\xset}{\pr|\xset}}{\partial\pr}}_i\ d\xset d\pr\nonumber\\
       &=-\expectation{\squareb{\postscore{\pr}{\xset;\Omega}}_i \squareb{\nabla_{\pr}\log\probt{\pr|\xset}{\pr|\xset}}_i}{\pr,\xset}\nonumber\\
       &\leq \abs{\expectation{\squareb{\postscore{\pr}{\xset;\Omega}}_i \squareb{\nabla_{\pr}\log\probt{\pr|\xset}{\pr|\xset}}_i}{\pr,\xset}}
    \end{align}
    Next using Cauchy–Schwarz inequality we have:
    \begin{align}\label{eq:conv_post_exp_finie_p2_cs}
        &\abs{w_i}\\
        &\leq \sqrt{\expectation{\squareb{\postscore{\pr}{\xset;\Omega}}_i^2}{\pr,\xset}\expectation{\squareb{\nabla_{\pr}\log\probt{\pr|\xset}{\pr|\xset}}_i^2}{\pr,\xset}}\nonumber
    \end{align}
    Combining \eqref{eq:conv_post_exp_finie_p2} and \eqref{eq:conv_post_exp_finie_p3} and \eqref{eq:conv_post_exp_finie_p2_cs} along with Assumption ~\ref{assum:bound_expection_post} and ~\ref{assum:net_cond_score} 
    showing that $\expectation{{\trace{\frac{\partial \postscore{\pr}{\xset;\Omega}}{\partial\pr}}}}{\xset,\pr}$ is finite and SLLN holds.
    }    
    Then we have pointwise strong convergence that results in
    \begin{equation}\label{eq:conv_probability}
        \lossbs\brackets{\Omega^*}-\lossbs\brackets{\tilde{\Omega}}\leq 2\sup\limits_{\Omega}\norm{\lossbsm\brackets{\Omega}-\lossbst\brackets{\Omega}}_2\xrightarrow{a.s.} 0,
    \end{equation}
    as $\nds\rightarrow\infty$.  In last part, we show that $\lossbs\brackets{\tilde{\Omega}}=0$:
    \begin{align}\label{eq:loss_zero}
        &\lossbs\brackets{\arg\min\limits_{\Omega}\lossbst\brackets{\Omega}}=\lossbs\brackets{\arg\min\limits_{\Omega}\lossbst\brackets{\Omega}+C}\nonumber\\
        &=\lossbs\brackets{\arg\min\limits_{\Omega}\lossbs\brackets{\Omega}}=0,
    \end{align}
    where $C$ is a constant independent of $\Omega$. The last step follows by the
    realizable score function assumption~\ref{assume:realizable}.  Finally, combining \eqref{eq:conv_probability} and \eqref{eq:loss_zero} yields Thm~\ref{thm:post_consist}. 
\end{proof}
\subsubsection{Prior and Fisher Score Consistency Proof (Thm~\ref{thm:mp_consist})}
First we 
establish 
that the Lipschit continuouity requirements are met. 
\begin{lemma}[Fisher Score is Lipschitz continuous]\label{lemma:lip_score_fisher}
        {Define the per-sample loss of the Fisher score matching $\ell_F\brackets{\x,\p;\Omega}\triangleq \norm{\lscore{{\x}}{\p;\paramf}}_2^2
    +2\lscore{{\x}}{\p;\Omega}^T\priorscore{\p;\paramp{^*}}+2\trace{\frac{\partial \lscore{{\x}}{\p;\Omega}}{\partial\p}} $, and $\Delta_\Omega \triangleq  \norm{\Omega_{1}-\Omega_{2}}_2$. }
    Assume that: 
    \begin{itemize}
        \item  The Fisher score is  Lipschitz continuous:
        $\norm{\lscore{\x}{\p;\Omega_1}-\lscore{\x}{\p;\Omega_2}}_2\leq \xi_{F}\brackets{\x,\p}\Delta\Omega$
      \item {The Fisher} score {Jacobian matrix} is Lipschitz continuous 
      $$\norm{\nabla_{\p}\lscore{\x}{\p;\Omega_1}-\nabla_{\p}\lscore{\x}{\p;\Omega_2}}_2\leq \tau_F\brackets{\x,\p}\Delta_\Omega$$
      \item {The score} outer product is Lipschitz continuous
      $$\norm{\matsym{R}\brackets{\x,\p;\Omega_1}-\matsym{R}\brackets{\x,\p;\Omega_2}}_F\leq \zeta_F\brackets{\x,\p}\Delta_\Omega$$
  \end{itemize}
  In addition assume that the following expectations: $\expectation{\xi_F\brackets{\x,\p}^2}{\X,\pr}<\infty$, $\expectation{\tau_F\brackets{\x,\p}}{\X,\pr}<\infty$,  and  $\expectation{\zeta_F\brackets{\x,\p}}{\X,\pr}<\infty$ are finite.
    Then
    \begin{equation}
        \abs{\ell_F\brackets{\x,\p;\Omega_{1}}-\ell_F\brackets{\x,\p;\Omega_{2}}}\leq L_F\brackets{\x,\p}\Delta\Omega,
    \end{equation}
    where $L_F\brackets{\x,\p}=\sqrt{\np}\brackets{2\tau_F\brackets{\x,\p}+\zeta_F\brackets{\x,\p}}+2\xi\brackets{\x,\p}\norm{\priorscore{\p;\paramp{^*}}}_2$ is the Liphschitz constant. {Furthermore, $\expectation{L_F\brackets{\X,\pr}}{\X,\pr}<\infty$. }
\end{lemma}
\begin{proof}
    \begin{align}
        &\abs{\ell_F\brackets{\x,\p;\Omega_1}-\ell_F\brackets{\x,\p;\Omega_2}}\leq \abs{p_1}+2\abs{p_2}+2\abs{p_3}\nonumber
    \end{align}
    where $p_1=\norm{\lscore{{\X}}{\p;\Omega_1}}_2^2-\norm{\lscore{{\X}}{\p;\Omega_2}}_2^2$, $p_2=\trace{\frac{\partial \lscore{{\X;\Omega_1}}{\p}}{\partial\p}}-\trace{\frac{\partial \lscore{{\X;\Omega_2}}{\p}}{\partial\p}}$ and
    \begin{equation*}
        p_3= \brackets{\lscore{{\X}}{\p;\Omega_1}-\lscore{{\X}}{\p;\Omega_2}}^T\priorscore{\p;\paramp{^*}}
    \end{equation*}
    To obtain the bound on $p_1$ and $p_2$, we use the same derivation as of Lemma 10 from \cite{crafts2023bayesian}. Next we left with bounding $p_3$:
    \begin{align*}
        p_3\leq \norm{\lscore{{\X}}{\p;\Omega_1}-\lscore{{\X}}{\p;\Omega_2}}_2\norm{\priorscore{\p;\paramp{^*}}}_2
    \end{align*}
    Using assumption ~\ref{assume:lip_cont} we have:
    \begin{align}\label{eq:p3_final}
        p_3\leq \xi_F\brackets{\x,\p}\norm{\priorscore{\p;\paramp{^*}}}_2\Delta\Omega
    \end{align}
    Combining the results of $p_1$ and $p_2$ with \eqref{eq:p3_final} results in Lemma ~\ref{lemma:lip_score_fisher}.
    Next, we proof that $\expectation{L_F\brackets{\X,\pr}}{\X,\pr}<\infty$. Using the Cauchy–Schwarz inequality we have that:
    \begin{align}
        \expectation{\xi_F\brackets{\X,\pr}\norm{\priorscore{\pr;\paramp{^*}}}_2}{\X,\pr}\leq \nonumber\\
        \sqrt{\expectation{\xi_F\brackets{\X,\pr}^2}{\X,\pr}\expectation{\priorscore{\pr;\paramp{^*}}^2}{\pr}}
    \end{align}
    which is finite using Assumption~\ref{assume:expecte_fine_score_matching} and the Lemma assumption $\expectation{\tau_F\brackets{\x,\p}^2}{\X,\pr}<\infty$. Combining with the rest of the Lemma assumption, we obtain that $\expectation{L_F\brackets{\X,\pr}}{\X,\pr}<\infty$

\end{proof}

\begin{lemma}[Prior Score is Lipschitz continuous]\label{lemma:lip_score_prior}

    {Define the per-sample loss of the prior score matching $\ell_P\brackets{\p;\Omega}\triangleq{\norm{\priorscore{\p;\Omega}}_2^2}+2\trace{\frac{\partial \priorscore{\p;\Omega}{}}{\partial\p}} $, and $\Delta_\Omega \triangleq  \norm{\Omega_{1}-\Omega_{2}}_2$. }
    Assume that: 
    \begin{itemize}
      \item {The prior} score {Jacobian matrix} is Lipschitz continuous 
      $$\norm{\nabla_{\p}\priorscore{\p;\Omega_1}-\nabla_{\p}\priorscore{\p;\Omega_2}}_2\leq \tau_P\brackets{\p}\Delta_\Omega$$
      \item {The score} outer product is Lipschitz continuous
      $$\norm{\matsym{R}\brackets{\p;\Omega_1}-\matsym{R}\brackets{\p;\Omega_2}}_F\leq \zeta_P\brackets{\p}\Delta_\Omega$$
  \end{itemize}
where $\matsym{R}\brackets{\p;\Omega} \triangleq\priorscore{\p;\Omega}\priorscore{\p;\Omega}^T$. {In addition assume that $\expectation{\tau_P\brackets{\pr}}{\pr}<\infty$ and $\expectation{\zeta_P\brackets{\pr}}{\pr}<\infty$.}
    Then
    \begin{equation}
        \abs{\ell_P\brackets{\p;\Omega_{1}}-\ell_P\brackets{\p;\Omega_{2}}}\leq L_P\brackets{\p}\Delta_\Omega,
    \end{equation}
    where $L_P\brackets{\z}=\sqrt{\np}\brackets{2\tau_P\brackets{\x,\p}+\zeta_P\brackets{\x,\p}}$ is the Liphschitz constant. {Furthermore, $\expectation{L_P\brackets{\pr}}{\pr}<\infty$. }
\end{lemma}
The proof of Lemma~\ref{lemma:lip_score_prior} is the same as the proof of Lemma 10 in \cite{crafts2023bayesian}. 
To establish consistency in Thm~\ref{thm:mp_consist}, we employ Lemmas~\ref{lemma:lip_score_fisher} and ~\ref{lemma:lip_score_prior} and proceed with a derivation analogous to that in ~\ref{apx:proof:consistency_post}.

%% file: files/appendix/examples.tex
\subsection{Quantized Measurement Score Function \eqref{eq:score_quantization}}\label{apx:derivation_q_score}
Here we derive the {Fisher} score function for the  
following quantized observation model,
\begin{equation}
\X=Q\brackets{u\brackets{\p}+\randomvec{W},t,n_b},
\end{equation}
where $\randomvec{W}\sim\normaldis{0}{\matsym{\Sigma}}$ is an additive  Gaussian noise with zero mean and diagonal covariance matrix $\matsym{\Sigma}$. Now, the PMF of $\X$ given $\p$ {
for $\Sigma$ 
a diagonal matrix 
} is given by:
\begin{align}
    &\pmft{\x=\vectorsym{q}|\p}{\X|\p}=\prod_{i=1}^{\nx}\pmft{\squareb{\x}_i=\squareb{\vectorsym{q}}_i|\p}{\X_i|\p}\nonumber\\
&=\prod_{i=1}^{\nx}\int_{b_l\brackets{\squareb{\vectorsym{q}}_i}}^{b_u\brackets{\squareb{\vectorsym{q}}_i}} \frac{1}{\sqrt{2\pi\sigma^2_i}}\exp{\brackets{-\tfrac{\brackets{x-\squareb{u\brackets{\p}}_i}^2}{2\sigma^2_i}}}dx,
\end{align}
where  $b_l$ and $b_u$ are functions that provide the lower  and upper quantization boundary, respectively.
Then the Fisher score vector is given by:
\begin{align}\label{eq:score_quant}
    &\nabla_{\p}\log\pmft{\x=\vectorsym{q}|\p}{\X|\p}=\nonumber\\
    &\sum_{i=1}^{\nx}\tfrac{\frac{\partial\squareb{\vectorsym{u}\brackets{\p}}_i}{\partial\p}   ^T\int_{b_l\brackets{\squareb{\vectorsym{q}}_i}}^{b_u\brackets{\squareb{\vectorsym{q}}_i}} \exp{\brackets{-\frac{\brackets{x-\squareb{\vectorsym{u}\brackets{\p}}_i}^2}{2\sigma^2_i}}}\brackets{x-\squareb{\vectorsym{u}\brackets{\p}}_i}dx}{\sigma^2_i\int_{b_l\brackets{\squareb{\vectorsym{q}}_i}}^{b_u\brackets{\squareb{\vectorsym{q}}_i}} \exp{\brackets{-\frac{\brackets{x-\squareb{\vectorsym{u}\brackets{\p}}_i}^2}{2\sigma^2_i}}}dx}\nonumber\\
    &=\sum_{i=1}^{\nx}\frac{\frac{\partial\squareb{\vectorsym{u}\brackets{\p}}_i}{\partial\p}^T}{\sigma^2_i}\brackets{\frac{\mathfrak{N}_i}{\mathfrak{D}_i} -\squareb{\vectorsym{u}\brackets{\p}}_i},
\end{align}
where 
\begin{align*}
    \mathfrak{N}_i&=\int_{b_l\brackets{\squareb{\vectorsym{q}}_i}}^{b_u\brackets{\squareb{\vectorsym{q}}_i}} x\exp{\brackets{-\tfrac{\brackets{x-\squareb{\vectorsym{u}\brackets{\p}}_i}^2}{2\sigma^2_i}}}dx,\\
    \mathfrak{D}_i&=\int_{b_l\brackets{\squareb{\vectorsym{q}}_i}}^{b_u\brackets{\squareb{\vectorsym{q}}_i}} \exp{\brackets{-\tfrac{\brackets{x-\squareb{\vectorsym{u}\brackets{\p}}_i}^2}{2\sigma^2_i}}}dx,
\end{align*}

Next, {define the distances 
$\Delta b_{u,i}\brackets{\x,\p}=\frac{b_u\brackets{[\x]_i}-\squareb{\vectorsym{u}\brackets{\p}}_i}{\sigma_i}$ and $\Delta b_{l,i}\brackets{\x,\p}=\frac{b_l\brackets{[\x]_i}-\squareb{\vectorsym{u}\brackets{\p}}_i}{\sigma_i}$ from the upper and lower quantization boundaries, $\Delta E_i\triangleq  \exp\brackets{-\frac{\Delta b_{l,i}\brackets{\x,\p}^2}{2}}-\exp\brackets{-\frac{\Delta b_{u,i}\brackets{\x,\p}^2}{2}}$, and $\Delta\Phi_i\triangleq\Phi\brackets{\Delta b_{u,i}\brackets{\x,\p} }-\Phi\brackets{\Delta b_{l,i}\brackets{\x,\p}}$, where
$\Phi$ is the cumulative distribution function (CDF) of the standard Gaussian. 
Then
}
\begin{align} \label{eq:score_quant_dem} 
    \mathfrak{D}_i&=\sqrt{2\pi\sigma_i^2}\Delta\Phi_i \\
    \mathfrak{N}_i&=
     \sigma_i^2\Delta E_i+\sqrt{2\pi\sigma^2_i}\squareb{\vectorsym{u}\brackets{\p}}_i\Delta\Phi_i.\label{eq:score_quant_num}
\end{align}
Combining \eqref{eq:score_quant_dem}, \eqref{eq:score_quant_num} and plugging into \eqref{eq:score_quant} results in:
\begin{align}\label{eq:score_quant_p2}
\nabla_{\p}\log\pmft{\X=\vectorsym{q}|\p}{\x|\p}
    &=\sum_{i=1}^{\nx}\frac{\frac{\partial\squareb{\vectorsym{u}\brackets{\p}}_i}{\partial\p}^T}{\sqrt{2\pi\sigma^2_i}}\frac{\Delta E_i}{\Delta\Phi_i}.
\end{align}

Using the fact that $\squareb{\x}_i\in \{-1,1\}\quad\forall i$ we can simplify $\Delta\Phi_i$ and $\Delta E_i$ for the two cases as follows.
\begin{itemize}
    \item $\squareb{\x}_i=-1$ in which case {$\Phi\brackets{\Delta b_{l,i}\brackets{\x,\p}=-\infty} =0$} and $\Delta b_{u,i}\brackets{\x,\p}=\frac{-\squareb{\vectorsym{u}\brackets{\p}}_i}{\sigma_i}$ resulting in:
    \begin{subequations}\label{eq:simply_one}
    \begin{align}
        \Delta\Phi_i&=\Phi\brackets{-\tfrac{\squareb{\vectorsym{u}\brackets{\p}}_i}{\sigma_i} }
        \Delta E_i &=-\exp\brackets{-\tfrac{\squareb{\vectorsym{u}\brackets{\p}}_i^2}{2\sigma_i^2}}.
    \end{align}
    \end{subequations}
    
    \item $\squareb{\x}_i=1$ in which case $\Delta b_{l,i}\brackets{\x,\p}=\frac{-\squareb{\vectorsym{u}\brackets{\p}}_i}{\sigma_i}$ and {$\Phi\brackets{\Delta b_{u,i}\brackets{\x,\p}=\infty}=1$} resulting in:
    \begin{subequations}\label{eq:simply_neg_one}
    \begin{align}
        \Delta\Phi_i&=
        {1} -\Phi\brackets{-\tfrac{-\squareb{\vectorsym{u}\brackets{\p}}_i}{\sigma_i}}=\Phi\brackets{\tfrac{\squareb{\vectorsym{u}\brackets{\p}}_i}{\sigma_i} } \\
        \Delta E_i &=\exp\brackets{-\tfrac{\squareb{\vectorsym{u}\brackets{\p}}_i^2}{2\sigma_i^2}},
    \end{align}
        \end{subequations}
\end{itemize}
Combing \eqref{eq:simply_one} and \eqref{eq:simply_neg_one} results in:
\begin{align}\label{eq:q_comb}
    \squareb{\vectorsym{\rho}\brackets{\vectorsym{x},\p}}_i&=\frac{\Delta E_i}{\sqrt{2\pi\sigma^2_i}\Delta\Phi_i}\nonumber\\
                                                           &=\frac{\squareb{\x}_i\exp\brackets{-\frac{\squareb{\vectorsym{u}\brackets{\p}}_i^2}{2\sigma_i^2}}}{\sqrt{2\pi\sigma^2_i}\Phi\brackets{\squareb{\x}_i\frac{\squareb{\vectorsym{u}\brackets{\p}}_i}{\sigma_i} }}.
\end{align}
Plugging back  \eqref{eq:q_comb}  into \eqref{eq:score_quant_p2} results in \eqref{eq:score_quantization}.

%% file: files/appendix/prior_fim.tex
\subsection{Prior FIMs}\label{apx:prior_fim}
In this section we derive the prior Fisher information matrix of Beta and Gaussian distributions.
\subsubsection{Prior FIM of Beta Distribution}
Let $\kappa\sim\mathrm{Beta}\brackets{\alpha_{\kappa},\beta_{\kappa},a_l,a_h}$ be a random variable that is distribution according to the four-parameter beta distribution. Then, the PDF of $\kappa$ is given by:
\begin{equation}
    \probt{\kappa}{\kappa}=\frac{\brackets{\kappa-a_l}^{\alpha_{\kappa}-1}\brackets{a_u-\kappa}^{\beta_{\kappa}-1}}{\brackets{a_h-a_l}^{\alpha_{\kappa}+\beta_{\kappa}-1}\mathrm{B}\brackets{\alpha_{\kappa},\beta_{\kappa}}}
\end{equation}
where  $a_l,a_u$ are the lower and upper support parameters ($a_u>a_l$), respectively, $\alpha_{\kappa}>0$, $\beta_{\kappa}>0$ are the shape parameters, $B\brackets{\alpha,\beta}=\frac{\Gamma\brackets{\alpha}\Gamma\brackets{\beta}}{\Gamma\brackets{\beta+\alpha}}$ is the normalization constant (Beta function) and $\Gamma\brackets{\alpha}$ is the Gamma function.
As a direct result, the score of $\kappa$ is given by:
\begin{equation}
    \frac{d \log\probt{\kappa}{\kappa}}{d\kappa}=\frac{\alpha_{\kappa}-1}{\kappa-a_l}-\frac{\beta_{\kappa}-1}{a_h-\kappa}
\end{equation}
Assuming that $\alpha_{\kappa}-2>0$
\begin{equation}
    \expectation{\frac{\brackets{\alpha_{\kappa}-1}^2}{\brackets{\kappa-a_l}^2}}{\kappa}=\frac{ \brackets{\alpha_{\kappa}-1}^2B\brackets{\alpha_{\kappa}-2,\beta_{\kappa}}}{\brackets{a_h-a_l}^{2} B\brackets{\alpha_{\kappa},\beta_{\kappa}}}
\end{equation}
Next, using the definition of Beta function we have:
\begin{equation}
    \expectation{\frac{\brackets{\alpha_{\kappa}-1}^2}{\brackets{\kappa-a_l}^2}}{\kappa}=\frac{ \brackets{\alpha_{\kappa}-1}^2\Gamma\brackets{\alpha_{\kappa}-2}\Gamma\brackets{\alpha_{\kappa}+\beta_{\kappa}}}{\brackets{a_h-a_l}^{2} \Gamma\brackets{\alpha_{\kappa}}\Gamma\brackets{\alpha_{\kappa}-2+\beta_{\kappa}}}
\end{equation}
Using the Gamma function properties we have:
\begin{equation}\label{eq:beta_prior_p1}
\expectation{\frac{\brackets{\alpha_{\kappa}-1}^2}{\brackets{\kappa-a_l}^2}}{\kappa}=\frac{ \brackets{\alpha_{\kappa}-1+\beta_{\kappa}}\brackets{\alpha_{\kappa}-2+\beta_{\kappa}}\brackets{\alpha_{\kappa}-1}}{\brackets{a_h-a_l}^{2} \brackets{\alpha_{\kappa}-2}}
\end{equation}
In a similar fashion we have that:
\begin{align}\label{eq:beta_prior_p2}
    &\expectation{\frac{\brackets{\beta_{\kappa}-1}^2}{\brackets{a_h-\kappa}^2}}{\kappa}=\frac{ \brackets{\beta_{\kappa}-1}^2B\brackets{\alpha_{\kappa},\beta_{\kappa}-2}}{\brackets{a_h-a_l}^{2} B\brackets{\alpha_{\kappa},\beta_{\kappa}}}\nonumber\\
    &=\frac{\brackets{\beta_{\kappa}-1}^2\Gamma\brackets{\beta_{\kappa}-2}\Gamma\brackets{\alpha_{\kappa}+\beta_{\kappa}}}{\brackets{a_h-a_l}^{2} \Gamma\brackets{\alpha_{\kappa}+\beta_{\kappa}-2} \Gamma\brackets{\beta_{\kappa}}}\nonumber\\
    &=\frac{ \brackets{\alpha_{\kappa}-1+\beta_{\kappa}}\brackets{\alpha_{\kappa}-2+\beta_{\kappa}}\brackets{\beta_{\kappa}-1}}{\brackets{a_h-a_l}^{2} \brackets{\beta_{\kappa}-2}}
\end{align}
\begin{align}\label{eq:beta_prior_p3}
    &\expectation{\frac{\brackets{\alpha_{\kappa}-1}\brackets{\beta_{\kappa}-1}}{\brackets{\kappa-a_l}\brackets{a_h-\kappa}}}{\kappa}=\frac{ \brackets{\alpha_{\kappa}-1}\brackets{\beta_{\kappa}-1}B\brackets{\alpha_{\kappa}-1,\beta_{\kappa}-1}}{\brackets{a_h-a_l}^{2} B\brackets{\alpha_{\kappa},\beta_{\kappa}}}\nonumber\\
    &=\frac{\brackets{\alpha_{\kappa}-1}\brackets{\beta_{\kappa}-1}\Gamma\brackets{\alpha_{\kappa}-1}\Gamma\brackets{\beta_{\kappa}-1}\Gamma\brackets{\beta_{\kappa}+\alpha_{\kappa}}}{\brackets{a_h-a_l}^{2}\Gamma\brackets{\alpha_{\kappa}}\Gamma\brackets{\beta_{\kappa}}\Gamma\brackets{\beta_{\kappa}+\alpha_{\kappa}-2}}\nonumber\\
    &=\frac{\brackets{\beta_{\kappa}+\alpha_{\kappa}-1}\brackets{\beta_{\kappa}+\alpha_{\kappa}-2}}{\brackets{a_h-a_l}^{2}}
\end{align}
Finally combining \eqref{eq:beta_prior_p1}, \eqref{eq:beta_prior_p2} and \eqref{eq:beta_prior_p3}  results in:
\begin{align}\label{eq:beta_prior_fim}
&\expectation{\brackets{\frac{d \log\probt{\kappa}{\kappa}}{d\kappa}}^2}{\kappa}\nonumber \\
&=\frac{\brackets{\alpha_{\kappa}-1+\beta_{\kappa}}\brackets{\alpha_{\kappa}-2+\beta_{\kappa}}}{\brackets{a_h-a_l}^{2}}\brackets{\frac{1}{\alpha_{\kappa}-2}+\frac{1}{\beta_{\kappa}-2}}
\end{align}



\subsubsection{Prior FIM of Gaussian Distribution}\label{apx:prior_fim_gaussian}
Let $\vectorsym{\kappa}\sim\normaldis{\vectorsym{\mu}}{\matsym{\Sigma}}$ be a random vector of size $k$ that distributed according to Gaussian distribution with mean $\vectorsym{\mu}$ and covariance $\matsym{\Sigma}$.
\begin{equation}
    \probt{\vectorsym{\kappa}}{\vectorsym{\kappa}}=\frac{1}{\sqrt{\brackets{2\pi}^k\det{\matsym{\Sigma}}}}\exp\brackets{-\frac{1}{2}\brackets{\vectorsym{\kappa}-\vectorsym{\mu}}^T\Sigma^{-1}\brackets{\vectorsym{\kappa}-\vectorsym{\mu}}}
\end{equation}
Then the score vector of $\vectorsym{\kappa}$ is given by:
\begin{equation}
    \nabla_{\vectorsym{\kappa}}\log\brackets{\probt{\vectorsym{\kappa}}{\vectorsym{\kappa}}}=-\matsym{\Sigma}^{-1}\brackets{\vectorsym{\kappa}-\vectorsym{\mu}}
\end{equation}
Finally the Prior FIM of Gaussian random vector is given by:
\begin{equation}
    \expectation{\nabla_{\vectorsym{\kappa}}\log\brackets{\probt{\vectorsym{\kappa}}{\vectorsym{\kappa}}}\nabla_{\vectorsym{\kappa}}\log\brackets{\probt{\vectorsym{\kappa}}{\vectorsym{\kappa}}}^T}{\vectorsym{\kappa}}=\matsym{\Sigma}^{-1}.
\end{equation}

%% file: main.bbl
\begin{thebibliography}{10}
\providecommand{\url}[1]{#1}
\csname url@samestyle\endcsname
\providecommand{\newblock}{\relax}
\providecommand{\bibinfo}[2]{#2}
\providecommand{\BIBentrySTDinterwordspacing}{\spaceskip=0pt\relax}
\providecommand{\BIBentryALTinterwordstretchfactor}{4}
\providecommand{\BIBentryALTinterwordspacing}{\spaceskip=\fontdimen2\font plus
\BIBentryALTinterwordstretchfactor\fontdimen3\font minus \fontdimen4\font\relax}
\providecommand{\BIBforeignlanguage}[2]{{%
\expandafter\ifx\csname l@#1\endcsname\relax
\typeout{** WARNING: IEEEtran.bst: No hyphenation pattern has been}%
\typeout{** loaded for the language `#1'. Using the pattern for}%
\typeout{** the default language instead.}%
\else
\language=\csname l@#1\endcsname
\fi
#2}}
\providecommand{\BIBdecl}{\relax}
\BIBdecl

\bibitem{van2004detection}
H.~L. Van~Trees, \emph{Detection, estimation, and modulation theory, part I: detection, estimation, and linear modulation theory}.\hskip 1em plus 0.5em minus 0.4em\relax John Wiley \& Sons, 2004.

\bibitem{10184105}
D.~A. Tubail and S.~Ikki, ``Range-direction tracking and {B}ayesian {C}ramer–{R}ao bound analysis in mmwave systems equipped with imperfect transceivers,'' \emph{IEEE Wireless Communications Letters}, vol.~12, no.~10, pp. 1806--1810, 2023.

\bibitem{10140073}
B.~Siebler, S.~Sand, and U.~D. Hanebeck, ``Bayesian {C}ramér-{R}ao lower bounds for magnetic field-based train localization,'' in \emph{2023 IEEE/ION Position, Location and Navigation Symposium (PLANS)}, 2023, pp. 814--820.

\bibitem{nasir2013phase}
A.~A. Nasir, H.~Mehrpouyan, R.~Schober, and Y.~Hua, ``Phase noise in mimo systems: {B}ayesian {C}ram{\'e}r--{R}ao bounds and soft-input estimation,'' \emph{IEEE transactions on signal processing}, vol.~61, no.~10, pp. 2675--2692, 2013.

\bibitem{xu2004bayesian}
W.~Xu, A.~B. Baggeroer, and C.~D. Richmond, ``Bayesian bounds for matched-field parameter estimation,'' \emph{IEEE Transactions on Signal Processing}, vol.~52, no.~12, pp. 3293--3305, 2004.

\bibitem{rosentha2024asymptotically}
N.~E. Rosentha and J.~Tabrikian, ``Asymptotically tight misspecified bayesian cram{\'e}r-rao bound,'' in \emph{ICASSP 2024-2024 IEEE International Conference on Acoustics, Speech and Signal Processing (ICASSP)}.\hskip 1em plus 0.5em minus 0.4em\relax IEEE, 2024, pp. 9916--9920.

\bibitem{mazor2024limitations}
Y.~Mazor, I.~E. Berman, and T.~Routtenberg, ``On the limitations of the bayesian cram{\'e}r-rao bound for mixed-resolution data,'' \emph{IEEE Signal Processing Letters}, 2024.

\bibitem{huleihel2013optimal}
W.~Huleihel, J.~Tabrikian, and R.~Shavit, ``Optimal adaptive waveform design for cognitive mimo radar,'' \emph{IEEE Transactions on Signal Processing}, vol.~61, no.~20, pp. 5075--5089, 2013.

\bibitem{turlapaty2014bayesian}
A.~Turlapaty and Y.~Jin, ``Bayesian sequential parameter estimation by cognitive radar with multiantenna arrays,'' \emph{IEEE Transactions on Signal Processing}, vol.~63, no.~4, pp. 974--987, 2014.

\bibitem{zuo2010conditional}
L.~Zuo, R.~Niu, and P.~K. Varshney, ``Conditional posterior {C}ram{\'e}r--{R}ao lower bounds for nonlinear sequential bayesian estimation,'' \emph{IEEE Transactions on Signal Processing}, vol.~59, no.~1, pp. 1--14, 2010.

\bibitem{sun2024optimal}
H.~Sun, J.~Tabrikian, H.~Messer, and H.~Gao, ``Optimal ratio between coherent and orthogonal signals in sparse mimo radar,'' in \emph{2024 IEEE 13rd Sensor Array and Multichannel Signal Processing Workshop (SAM)}.\hskip 1em plus 0.5em minus 0.4em\relax IEEE, 2024, pp. 1--5.

\bibitem{lutati22_interspeech}
S.~Lutati, E.~Nachmani, and L.~Wolf, ``Sepit: Approaching a single channel speech separation bound,'' in \emph{Interspeech 2022}, 2022, pp. 5323--5327.

\bibitem{duy2022fisher}
T.~T. Duy, L.~V. Nguyen, V.-D. Nguyen, N.~L. Trung, and K.~Abed-Meraim, ``Fisher information neural estimation,'' in \emph{2022 30th European Signal Processing Conference (EUSIPCO)}.\hskip 1em plus 0.5em minus 0.4em\relax IEEE, 2022, pp. 2111--2115.

\bibitem{6975144}
V.~Berisha and A.~O. Hero, ``Empirical non-parametric estimation of the fisher information,'' \emph{IEEE Signal Processing Letters}, vol.~22, no.~7, pp. 988--992, 2015.

\bibitem{song2019generative}
Y.~Song and S.~Ermon, ``Generative modeling by estimating gradients of the data distribution,'' \emph{Advances in neural information processing systems}, vol.~32, 2019.

\bibitem{kobyzev2020normalizing}
I.~Kobyzev, S.~Prince, and M.~Brubaker, ``Normalizing flows: An introduction and review of current methods,'' \emph{IEEE Transactions on Pattern Analysis and Machine Intelligence}, 2020.

\bibitem{habi2023learning}
H.~V. Habi, H.~Messer, and Y.~Bresler, ``Learning to bound: A generative {C}ram{\'e}r-{R}ao bound,'' \emph{IEEE Transactions on Signal Processing}, 2023.

\bibitem{habi2023learned}
H.~V. Habi, H.~Messer, and Y.~Bresler, ``Learned generative misspecified lower bound,'' in \emph{ICASSP 2023-2023 IEEE International Conference on Acoustics, Speech and Signal Processing (ICASSP)}.\hskip 1em plus 0.5em minus 0.4em\relax IEEE, 2023, pp. 1--5.

\bibitem{habi2024learning}
H.~V. Habi, H.~Messer, and Y.~Bresler, ``Learning the {B}arankin lower bound on doa estimation error,'' in \emph{ICASSP 2024-2024 IEEE International Conference on Acoustics, Speech and Signal Processing (ICASSP)}.\hskip 1em plus 0.5em minus 0.4em\relax IEEE, 2024, pp. 9906--9910.

\bibitem{papamakarios2021normalizing}
G.~Papamakarios, E.~Nalisnick, D.~J. Rezende, S.~Mohamed, and B.~Lakshminarayanan, ``Normalizing flows for probabilistic modeling and inference,'' \emph{Journal of Machine Learning Research}, vol.~22, no.~57, pp. 1--64, 2021.

\bibitem{habi2022generative}
H.~V. Habi, H.~Messer, and Y.~Bresler, ``A generative {C}ram{\'e}r-{R}ao bound on frequency estimation with learned measurement distribution,'' in \emph{2022 IEEE 12th Sensor Array and Multichannel Signal Processing Workshop (SAM)}.\hskip 1em plus 0.5em minus 0.4em\relax IEEE, 2022, pp. 176--180.

\bibitem{crafts2023bayesian}
E.~S. Crafts, X.~Zhang, and B.~Zhao, ``Bayesian {C}ramér-{R}ao bound estimation with score-based models,'' \emph{IEEE Transactions on Information Theory}, pp. 1--1, 2024.

\bibitem{hyvarinen2005estimation}
A.~Hyv{\"a}rinen and P.~Dayan, ``Estimation of non-normalized statistical models by score matching.'' \emph{Journal of Machine Learning Research}, vol.~6, no.~4, 2005.

\bibitem{abdelhamed2019noise}
A.~Abdelhamed, M.~A. Brubaker, and M.~S. Brown, ``Noise flow: Noise modeling with conditional normalizing flows,'' in \emph{Proceedings of the IEEE/CVF International Conference on Computer Vision}, 2019, pp. 3165--3173.

\bibitem{weiss2023towards}
A.~Weiss, A.~C. Singer, and G.~W. Wornell, ``Towards robust data-driven underwater acoustic localization: A deep cnn solution with performance guarantees for model mismatch,'' in \emph{ICASSP 2023-2023 IEEE International Conference on Acoustics, Speech and Signal Processing (ICASSP)}.\hskip 1em plus 0.5em minus 0.4em\relax IEEE, 2023, pp. 1--5.

\bibitem{msg0-ag12-22}
\BIBentryALTinterwordspacing
L.~Domingos, P.~Skelton, and P.~Santos, ``Vtuad: Vessel type underwater acoustic data,'' 2022. [Online]. Available: \url{https://dx.doi.org/10.21227/msg0-ag12}
\BIBentrySTDinterwordspacing

\bibitem{shlezinger2022model}
N.~Shlezinger, Y.~C. Eldar, and S.~P. Boyd, ``Model-based deep learning: On the intersection of deep learning and optimization,'' \emph{IEEE Access}, vol.~10, pp. 115\,384--115\,398, 2022.

\bibitem{shlezinger2023model}
N.~Shlezinger, J.~Whang, Y.~C. Eldar, and A.~G. Dimakis, ``Model-based deep learning,'' \emph{Proceedings of the IEEE}, vol. 111, no.~5, pp. 465--499, 2023.

\bibitem{banerjee2024physics}
C.~Banerjee, K.~Nguyen, C.~Fookes, and K.~George, ``Physics-informed computer vision: A review and perspectives,'' \emph{ACM Computing Surveys}, vol.~57, no.~1, pp. 1--38, 2024.

\bibitem{faroughi2024physics}
S.~A. Faroughi, N.~M. Pawar, C.~Fernandes, M.~Raissi, S.~Das, N.~K. Kalantari, and S.~Kourosh~Mahjour, ``Physics-guided, physics-informed, and physics-encoded neural networks and operators in scientific computing: Fluid and solid mechanics,'' \emph{Journal of Computing and Information Science in Engineering}, vol.~24, no.~4, p. 040802, 2024.

\bibitem{meinders2024application}
M.~B. Meinders, J.~Yang, and E.~v.~d. Linden, ``Application of physics encoded neural networks to improve predictability of properties of complex multi-scale systems,'' \emph{Scientific Reports}, vol.~14, no.~1, p. 15015, 2024.

\bibitem{liu2022estimating}
S.~Liu, T.~Kanamori, and D.~J. Williams, ``Estimating density models with truncation boundaries using score matching,'' \emph{Journal of Machine Learning Research}, vol.~23, no. 186, pp. 1--38, 2022.

\bibitem{yu2019generalized}
S.~Yu, M.~Drton, and A.~Shojaie, ``Generalized score matching for non-negative data,'' \emph{The Journal of Machine Learning Research}, vol.~20, no.~1, pp. 2779--2848, 2019.

\bibitem{yu2022generalized}
S.~Yu, M.~Drton, and A.~Shojaie, ``Generalized score matching for general domains,'' \emph{Information and Inference: A Journal of the IMA}, vol.~11, no.~2, pp. 739--780, 2022.

\bibitem{mirza2014conditional}
M.~Mirza and S.~Osindero, ``Conditional generative adversarial nets,'' \emph{arXiv preprint arXiv:1411.1784}, 2014.

\bibitem{liu2019conditional}
R.~Liu, Y.~Liu, X.~Gong, X.~Wang, and H.~Li, ``Conditional adversarial generative flow for controllable image synthesis,'' in \emph{Proceedings of the IEEE/CVF Conference on Computer Vision and Pattern Recognition}, 2019, pp. 7992--8001.

\bibitem{ho2021classifier}
J.~Ho and T.~Salimans, ``Classifier-free diffusion guidance,'' in \emph{NeurIPS 2021 Workshop on Deep Generative Models and Downstream Applications}, 2021.

\bibitem{erbe2019effects}
C.~Erbe, S.~A. Marley, R.~P. Schoeman, J.~N. Smith, L.~E. Trigg, and C.~B. Embling, ``The effects of ship noise on marine mammals—a review,'' \emph{Frontiers in Marine Science}, vol.~6, p. 606, 2019.

\bibitem{lbcrb_repo}
H.~V. Habi, ``Learned {B}ayesian e {C}ram\'er {R}ao bound,'' \url{https://github.com/haihabi/Learned-BCRB}, 2024.

\bibitem{van2007bayesian}
H.~L. Van~Trees and K.~L. Bell, ``Bayesian bounds for parameter estimation and nonlinear filtering/tracking,'' \emph{AMC}, vol.~10, p.~12, 2007.

\bibitem{weinstein1988general}
E.~Weinstein and A.~J. Weiss, ``A general class of lower bounds in parameter estimation,'' \emph{IEEE Transactions on Information Theory}, vol.~34, no.~2, pp. 338--342, 1988.

\bibitem{zeitler2012bayesian}
G.~Zeitler, G.~Kramer, and A.~C. Singer, ``Bayesian parameter estimation using single-bit dithered quantization,'' \emph{IEEE Transactions on Signal Processing}, vol.~60, no.~6, pp. 2713--2726, 2012.

\bibitem{hyvarinen2007some}
A.~Hyv{\"a}rinen, ``Some extensions of score matching,'' \emph{Computational statistics \& data analysis}, vol.~51, no.~5, pp. 2499--2512, 2007.

\bibitem{song2020score}
Y.~Song, J.~Sohl-Dickstein, D.~P. Kingma, A.~Kumar, S.~Ermon, and B.~Poole, ``Score-based generative modeling through stochastic differential equations,'' \emph{arXiv preprint arXiv:2011.13456}, 2020.

\bibitem{shalev2014understanding}
S.~Shalev-Shwartz and S.~Ben-David, \emph{Understanding machine learning: From theory to algorithms}.\hskip 1em plus 0.5em minus 0.4em\relax Cambridge university press, 2014.

\bibitem{tropp2015introduction}
J.~A. Tropp \emph{et~al.}, ``An introduction to matrix concentration inequalities,'' \emph{Foundations and Trends{\textregistered} in Machine Learning}, vol.~8, no. 1-2, pp. 1--230, 2015.

\bibitem{ipsen2024stable}
I.~C. Ipsen and A.~K. Saibaba, ``Stable rank and intrinsic dimension of real and complex matrices,'' \emph{arXiv preprint arXiv:2407.21594}, 2024.

\bibitem{genz2009computation}
A.~Genz and F.~Bretz, \emph{Computation of multivariate normal and t probabilities}.\hskip 1em plus 0.5em minus 0.4em\relax Springer Science \& Business Media, 2009, vol. 195.

\bibitem{ramachandran2017searching}
P.~Ramachandran, B.~Zoph, and Q.~V. Le, ``Searching for activation functions,'' \emph{arXiv preprint arXiv:1710.05941}, 2017.

\bibitem{9664619}
P.~Stoica, X.~Shang, and Y.~Cheng, ``The {C}ramér–{R}ao bound for signal parameter estimation from quantized data [lecture notes],'' \emph{IEEE Signal Processing Magazine}, vol.~39, no.~1, pp. 118--125, 2022.

\bibitem{stoica2011gaussian}
P.~Stoica and P.~Babu, ``The gaussian data assumption leads to the largest {C}ram{\'e}r-{R}ao bound [lecture notes],'' \emph{IEEE Signal Processing Magazine}, vol.~28, no.~3, pp. 132--133, 2011.

\bibitem{loshchilov2018decoupled}
\BIBentryALTinterwordspacing
I.~Loshchilov and F.~Hutter, ``Decoupled weight decay regularization,'' in \emph{International Conference on Learning Representations}, 2019. [Online]. Available: \url{https://openreview.net/forum?id=Bkg6RiCqY7}
\BIBentrySTDinterwordspacing

\bibitem{song2020improved}
Y.~Song and S.~Ermon, ``Improved techniques for training score-based generative models,'' \emph{Advances in neural information processing systems}, vol.~33, pp. 12\,438--12\,448, 2020.

\bibitem{stewart1977perturbation}
G.~W. Stewart, ``On the perturbation of pseudo-inverses, projections and linear least squares problems,'' \emph{SIAM review}, vol.~19, no.~4, pp. 634--662, 1977.

\bibitem{andrews1992generic}
D.~W. Andrews, ``Generic uniform convergence,'' \emph{Econometric theory}, vol.~8, no.~2, pp. 241--257, 1992.

\end{thebibliography}
